\documentclass{emulateapj}
\usepackage{apjfonts}

\usepackage{lscape,amsmath,graphicx,natbib,longtable,rotating,threeparttable}

\newcommand{\etal}{et al.}
\newcommand{\hbeta}{H{$\beta$}}
\newcommand{\halpha}{H{$\alpha$}}

\newcommand{\CIV}{C{\sevenrm IV}}

\def\FeII{Fe\,{\sc ii}}
\def\MgII{Mg\,{\sc ii}}

\def \OIII {[O\,{\sc iii}]}

\newcommand{\OIIIb}{[O{\sevenrm\,III}]\,$\lambda$5007}

\newcommand{\NIIab}{[N{\sevenrm\,II}]\,$\lambda\lambda$6548,6584}

\newcommand{\bracket}[1]{\left\langle#1\right\rangle}
   \font\sevenrm=cmr7 scaled 1000

\begin{document}

\title{Constraining sub-parsec Binary Supermassive Black Holes in Quasars with Multi-Epoch Spectroscopy. I. The General Quasar Population}

\shorttitle{SUB-PARSEC BINARY SMBHS IN QUASARS. I.}

%\slugcomment{Draft Version}

\shortauthors{SHEN ET~AL.}
\author{Yue Shen\altaffilmark{1,5}, Xin Liu\altaffilmark{2,5}, Abraham Loeb\altaffilmark{3}, Scott Tremaine\altaffilmark{4}
} 
\altaffiltext{1}{Carnegie Observatories, 813 Santa Barbara Street, Pasadena,
CA 91101, USA; yshen@obs.carnegiescience.edu}

\altaffiltext{2}{Department of Physics and Astronomy, University of California, Los Angeles, CA 90095, USA}

\altaffiltext{3}{Harvard-Smithsonian Center for Astrophysics, 60 Garden
Street, Cambridge, MA 02138, USA}

\altaffiltext{4}{Institute for Advanced Study, Princeton, NJ 08540, USA}

\altaffiltext{5}{Hubble Fellow}

\begin{abstract}

We perform a systematic search for sub-parsec binary supermassive black holes (BHs) in normal broad line quasars at $z<0.8$, using multi-epoch SDSS spectroscopy of the broad \hbeta\ line. Our working model is that (1) one and only one of the two BHs in the binary is active; (2) the active BH dynamically dominates its own broad line region (BLR) in the binary system, so that the mean velocity of the BLR reflects the mean velocity of its host BH; (3) the inactive companion BH is orbiting at a distance of a few $R_{\rm BLR}$, where $R_{\rm BLR}\sim0.01-0.1$ pc is the BLR size. We search for the expected line-of-sight acceleration of the broad line velocity from binary orbital motion by cross-correlating SDSS spectra from two epochs separated by up to several years in the quasar restframe. Out of $\sim 700$ pairs of spectra for which we have good measurements of the velocity shift between two epochs ($1\sigma$ error $\sim 40\ {\rm km\,s^{-1}}$), we detect $28$ systems with significant velocity shifts in broad \hbeta, among which seven are the best candidates for the hypothesized binaries, four are most likely due to broad line variability in single BHs, and the rest are ambiguous. Continued spectroscopic observations of these candidates will easily strengthen or disprove these claims. We use the distribution of the observed accelerations (mostly non-detections) to place constraints on the abundance of such binary systems among the general quasar population. Excess variance in the velocity shift is inferred for observations separated by longer than $0.4$ yr (quasar restframe). Attributing all the excess to binary motion would imply that most of the quasars in this sample must be in binaries; that the inactive BH must be on average more massive than the active one; and that the binary separation is at most a few times the size of the BLR. However, if this excess variance is partly or largely due to long-term broad line variability, the requirement of a large population of close binaries is much weakened or even disfavored for massive companions. Future time-domain spectroscopic surveys of normal quasars can provide the vital prior information on the structure function of stochastic velocity shifts induced by broad line variability in single BHs. Such surveys with improved spectral quality, increased time baseline, and more epochs, can greatly improve the statistical constraints of this method on the general binary population in broad line quasars, further shrink the allowed binary parameter space, and detect true sub-parsec binaries. 

%the scenario that $\gtrsim 50\%$ quasars are in equal mass binaries with separations less than twice the broad line region size can be ruled out at $\gtrsim 95\%$ confidence; more massive companions, smaller binary separations, and/or higher binary fractions, are increasingly less favored by the observed acceleration distribution. 
%The companion paper will focus on a specific class of quasars with offset
%broad line peak relative to the systemic redshift, for which we have been
%acquiring second-epoch spectra. These objects consist of a tiny fraction of
%all SDSS quasars.
\end{abstract}


\section{Introduction}

%\textbf{Read and cite lots of papers on the general topic of binary SMBHs here. We will also outline the difference between this paper and Paper II. }

The search for and characterization of the binary supermassive black hole (BBH) population are of great importance to understanding both galaxy formation and gravitational wave physics. Since most massive galaxies are believed to harbor central SMBHs, the formation of BBHs is an inevitable consequence of frequent galaxy mergers. The dynamical evolution of the BBH system within the merged galaxy, and its interaction with the host both dynamically and via possible feedback during baryonic accretion onto one or both BHs, encode crucial information about the assembly of the bulge and the central SMBH. If the BBH eventually coalesces, it will produce a gravitational siren that could be detected with future low-frequency gravitational wave experiments \citep[e.g.,][]{Colpi_Dotti_2011}. Given their importance for understanding the co-evolution of galaxies and SMBHs, and for testing fundamental physics, BBHs have been of both theoretical and observational interest since the 1970s \citep[e.g.,][]{Thorne_Braginskii_1976,Begelman_etal_1980,Roos_1981}. 

Theoretical studies on BBHs in the past two decades have laid down the basic framework for the dynamical evolution of BBHs within a merged galaxy \citep[for a review, see, e.g.,][]{Merritt_Milosavljevic_2005}. The two BHs sink to the center of the merged host via dynamical friction, and form a BBH. The BBH orbit continues to decay by scattering stars in the galactic nucleus until it depletes the loss cone. In the absence of other mechanisms to further extract energy from the binary, or to refill the loss cone, the BBH will stall at $\sim$ pc scale, establishing the ``final parsec barrier'' \citep[][]{Milosavljevic_Merritt_2001}. A number of mechanisms have been suggested to circumvent this barrier, though in most cases their effectiveness is quite uncertain: non-spherical stellar potentials \citep[e.g.,][]{Yu_2002,Merritt_Poon_2004,Khan_Holley-Bockelmann_2013}, interactions with gas \citep[e.g.,][]{Armitage_Natarajan_2002,Escala_etal_2005,Mayer_etal_2007,Cuadra_etal_2009}, or hierarchical triple BH systems \citep[e.g.,][]{Blaes_etal_2002} through the Kozai-Lidov mechanism \citep{Kozai_1962}. Once the BBH reaches orbital separations below $\sim 0.01-0.05$ pc (for a $\sim 10^8\,M_\odot$ BBH), gravitational radiation will dominate the successive decay of the binary orbit, and merge the two BHs within a Hubble time \citep[e.g.,][]{Peters_1964}. However there is considerable theoretical uncertainty in the BBH evolution from a few pc to the gravitational regime, and the time the BBH spends during this phase is not well constrained. 

The strongest indirect evidence for BBHs perhaps comes from the observed central light deficit in massive ellipticals \citep[e.g.,][]{Ferrarese_etal_1994,Faber_etal_1997}, where the scouring of the central stellar cusp is most easily explained by star interactions with the BBH \citep[e.g.,][]{Ebisuzaki_etal_1991,Milosavljevic_Merritt_2001}. Other indirect evidence includes unusual radio morphologies such as helical/winged/X-shaped jet or double-lobed morphologies, which are commonly explained by BBH models \citep[e.g.,][]{Roos_etal_1993,Romero_etal_2000,Merritt_Ekers_2002,Liu_etal_2003,Liu_2004,Liu_Chen_2007} -- although the BBH interpretation is not unique for these observations. 

Directly observing BBHs during different merger stages is challenging, given the stringent resolution requirement and the difficulty of identifying active galactic nuclei (AGNs). While close quasar pairs (binary quasars) on tens of kpc scales have been known for a while \citep[e.g.,][]{Djorgovski_1991,Kochanek_etal_1999,Hennawi_etal_2006,Hennawi_etal_2010}, clear examples of kpc-scale binary AGNs were substantially rarer until recently \citep[e.g.,][]{Owen_etal_1985,Komossa_etal_2003,Ballo_etal_2004,Bianchi_etal_2008}. 

Considerable progress has been made in the last decade on the identification of binary AGNs on kpc scales, mostly owing to the advent of modern, large-scale sky surveys. Below a few kpc, a growing number of confirmed or candidate binary AGNs have been selected from multiwavelength surveys \citep[e.g.,][]{Zhou_etal_2004,Comerford_etal_2009a,Comerford_etal_2009b,Wang_etal_2009a,Xu_Komossa_2009,Liu_etal_2010a,Liu_etal_2010b,Liu_etal_2011, Liu_etal_2012, Liu_etal_2013, Smith_etal_2010,Shen_etal_2011b,Rosario_etal_2011, Koss_etal_2011,Koss_etal_2012,McGurk_etal_2011,Fu_etal_2011,Fu_etal_2012,Comerford_etal_2012,Barrows_etal_2012,Ge_etal_2012}, increasing the inventory of confirmed kpc-scale binary AGNs already by several folds. Below $\sim 100$ pc, there is only one clear example, 0402+379, where a pair of BHs with a projected separation of $\sim 7$ pc were detected in the radio \citep[e.g.,][]{Rodriguez_etal_2006} with the very long baseline array (VLBA); in addition, there is one claimed binary AGN with a projected separation of $\sim 150$ pc in the nearby Seyfert galaxy NGC 3393 \citep[][]{Fabbiano_etal_2011}. 

So far there is no confirmed case of a BBH with sub-pc separation. This scale is unresolvable with current instruments except with long baseline radio interferometry. Thus the observational searches for sub-pc BBHs almost exclusively rely on spatially unresolved signatures. 

The first potential indirect signature is a periodic photometric light curve, where the periodicity may indicate the presence of a BBH. The best known candidate is OJ 287, a $z=0.306$ blazar with its light curve showing a quasi-period of $\sim 9$ yr in restframe, which is interpreted as a sub-pc BBH with a mass ratio $q\lesssim 0.01$ in the latest models \citep[e.g.,][]{Valtonen_etal_2008}. Other examples have been claimed in the literature but these are not as robust as OJ 287 \citep[see][and references therein]{Merritt_Milosavljevic_2005}.

A second signature commonly utilized to search for sub-pc BBHs is the velocity offset between the broad emission line centroid and the systemic velocity in broad-line quasars \citep[e.g.,][]{Gaskell_1983,Peterson_etal_1987,Gaskell_1996,Popovic_etal_2000,Shen_Loeb_2010}. Since the broad line region (BLR) is gravitationally confined to the BH on sub-pc scales \citep[e.g.,][and see below]{Peterson_1997}, any velocity offset may reflect the orbital motion of the BH in a hypothetical  binary. Two classes of broad line velocity offset have attracted attention in the literature as indicators of candidate BBHs: (1) a double-peaked broad line profile, which is interpreted to mean that both BHs are active and co-rotating along with their own distinct BLRs; (2) a single-peaked broad line profile, with the broad line centroid offset from the systemic velocity, in which case only one BH is active, or both BHs are active but their BLR emission cannot be separated in velocity space. The second feature is also often used to argue for recoiling BHs after the coalescence of the BBH \citep[e.g.,][]{Loeb_2007}. 

It is straightforward to select such candidate sub-pc BBHs based on these broad line diagnostics, especially in the era of massive spectroscopic surveys. Examples are presented in either large samples \citep[e.g.,][]{Eracleous_Halpern_1994,Strateva_etal_2003,Shen_etal_2011, Tsalmantza_etal_2011}, or individual discoveries \citep[e.g.,][]{Komossa_etal_2008,Boroson_Lauer_2009,Shields_etal_2009}. However, it is impossible to confirm these cases with single-epoch spectroscopy. In fact, most sub-pc BBH candidates proposed this way early on have been rejected as long-term monitoring did not observe the expected radial acceleration of the broad line from binary motion \citep[e.g.,][]{Halpern_Filippenko_1988,Eracleous_etal_1997,Shapovalova_etal_2001}. The main difficulty here involves the poorly understood BLR geometry and kinematics even for single BHs, which may mimic a BBH. For instance, the rival model for the double-peaked or offset broad lines is a disk emitter model \citep[e.g.,][]{Chen_Halpern_1989,Eracleous_Halpern_1994,Eracleous_etal_1995}, which does not require a BBH to explain the peculiar broad line profile. Thus extraordinary caution is required when interpreting these unusual broad line profiles as BBHs \citep[e.g.,][]{Gaskell_2010,Shen_Loeb_2010,Eracleous_etal_2012,Popovic_2012}. In addition, some theoretical arguments suggest that the double-peaked broad line profile is a very ineffective indicator for BBHs, given the requirement that both BLRs are largely dynamically distinct while their relative motion must be large enough to separate the two peaks in the spectrum -- such conditions are difficult to fulfill in reality \citep[e.g.,][]{Shen_Loeb_2010}. 

The most effective, and the most definitive, way to improve the power of the broad line diagnostics to identify BBHs is multi-epoch spectroscopy. Long-term spectroscopic monitoring of the broad lines can confirm or rule out the BBH hypothesis based on changes in the broad line radial velocity (acceleration), given that the time baseline is long enough to see the expected binary motion and the spectral quality is adequate. Earlier spectral monitoring of samples of double-peaked disk emitters did not focus on testing the BBH scenario \citep[e.g.,][]{Gezari_etal_2007,Lewis_etal_2010}. Spectroscopic monitoring programs that are specifically designed to find BBHs in quasars are just starting \citep[e.g.,][]{Eracleous_etal_2012,Decarli_etal_2013}.

We are conducting systematic searches for sub-pc BBHs in broad-line quasars using multi-epoch spectroscopy of large statistical quasar samples. In contrast to previous work \citep[e.g.,][]{Eracleous_etal_2012,Decarli_etal_2013} which only focused on objects with offset or double-peaked broad lines, we are also targeting normal quasars with no broad line velocity offset\footnote{During the preparation of this work, we became aware of an independent study by \citet{Ju_etal_2013} that also focuses on the normal quasar population. However, there are several key differences in our approach compared with theirs, which will be discussed in detail in \S\ref{sec:comp}.}. There are two reasons for targeting normal quasars. First, quasars with double-peaked or significantly offset broad lines are only a minority of the entire quasar population \citep[$<10\%$, e.g.,][]{Shen_etal_2011}. While they are more likely to be BBHs by selection, they do not represent the general quasar population where a BBH can be easily hidden in the spectrum (since we only observe the line-of-sight velocity). Second, the acceleration in the radial velocity from the BBH is the largest at zero velocity offset (see \S\ref{sec:prelim}), therefore on average it is easier to detect acceleration for BBHs with non-offset broad lines than those with offset broad lines.  

\begin{figure}
 \centering
 \includegraphics[width=0.48\textwidth]{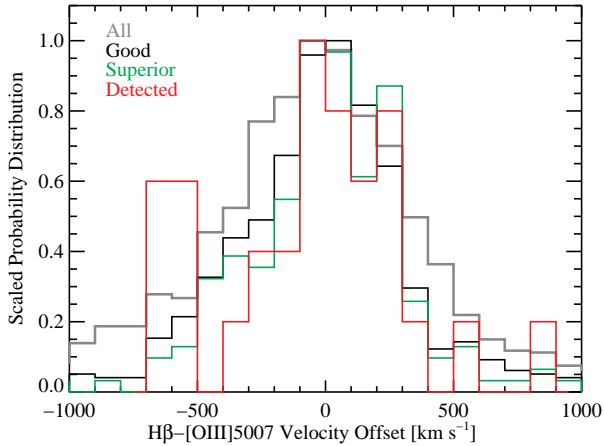}
 \caption{Distributions of the broad \hbeta\ velocity offset relative to the \OIIIb\ line for different samples in this study, which are all consistent with the bulk quasar population \citep[e.g.,][]{Shen_etal_2011}. Definitions of the ``good'' and ``superior'' samples can be found in \S\ref{sec:vel_shift} and \S\ref{sec:general_pop}, respectively. The ``detected'' sample consists of individual quasars in which a significant velocity shift has been detected (see 
\S\ref{sec:binary_candidate}). The individual detections do not seem to have a different distribution (see \S\ref{sec:result}). }
 \label{fig:Hb_OIII_shift}
\end{figure}

In this paper we present results for the general quasar population, i.e., without pre-selection of a subsample with double-peaked or offset broad lines, based on multi-epoch spectroscopy separated by up to several years. As shown in Fig.\ \ref{fig:Hb_OIII_shift}, the vast majority of our quasars have broad line velocity offset within $1000\ {\rm km\,s^{-1}}$ of the systemic velocity (based on \OIIIb). Thus these quasars will not have been included in the samples of \citet{Eracleous_etal_2012} and \citet{Decarli_etal_2013}. Our sample is thus substantially larger than the other two samples, allowing better statistical constraints. In the second paper of the series (hereafter Paper II), we will focus on a subset of quasars selected with offset broad lines, and present our own multi-epoch spectroscopic data and analysis in the context of the BBH scenario. 

The organization of the paper is as follows. In \S\ref{sec:prelim}, we provide preliminaries on the broad line diagnostics in the BBH scenario. The data and our approach to measure radial velocity acceleration with multi-epoch spectra are described in \S\ref{sec:data}. We pay particular attention to a proper error analysis, which is key to this kind of study. In \S\ref{sec:result} we present candidate detections of acceleration (which implies a possible BBH), and the statistical constraints on the general BBH population in quasars using (mostly) non-detections. In \S\ref{sec:disc} we discuss the caveats of our approach and compare to previous studies, and we summarize our conclusions in \S\ref{sec:con}. By default all time separations are in the restframe of the quasar, and $L$ refers to the quasar bolometric luminosity unless otherwise specified. We use ``offset'' to refer to the velocity difference between two lines in single-epoch spectra, and ``shift'' to refer to the changes in the line velocity between two epochs. For simplicity we assume circular BBH orbits, but the general methodology can be applied to a BBH population with an eccentricity distribution as well. 

\section{Preliminaries}\label{sec:prelim}

Since essentially all quasars in our sample have single-peaked broad lines, we initially assume that the broad line emission is dominated by one active BH. We will return to the possibility that the broad line emission comes from both BHs in \S\ref{sec:caveat}.

Consider a postulated SMBH binary on a circular orbit, where only BH 1 is active and powering its BLR. The orbital period,
line-of-sight (LOS) velocity and acceleration (relative to the systemic frame) of the active BH are:
\begin{eqnarray}\label{eqn:v_a}
P&=&2\pi d^{3/2}(GM_{\rm tot})^{-1/2}= 9.4d_{0.01}^{3/2}M_{8,\rm tot}^{-1/2}\quad {\rm yr}\ ,\nonumber\\
V_1&=&\frac{M_2}{M_{\rm tot}}\left(\frac{GM_{\rm tot}}{d}\right)^{1/2}\sin I\sin\phi\nonumber \\
&= & 6560\left(\frac{M_2}{M_{\rm tot}}\right)M_{8,\rm tot}^{1/2}d_{0.01}^{-1/2}\sin I\sin\phi \quad {\rm km\,s^{-1}}\ ,\nonumber\\
a_1 &=& \frac{GM_2}{d^2}\sin I\cos\phi \nonumber \\
&=& 4400\left(\frac{M_2}{10^8\,M_\odot}\right)d_{0.01}^{-2}\sin I \cos\phi\quad  {\rm km\,s^{-1}yr^{-1}}\ ,
\end{eqnarray}
where subscripts 1 and 2 refer to BH 1 and 2, $M_{\rm tot}=M_1+M_2$, $I$ is the inclination of the
orbit, $d$ is the binary separation, and $\phi=\phi_0+2\pi t/P$ is the orbit phase. We use the conventions $M_{\rm 8,tot}=(M_{\rm tot}/10^8\,M_\odot)$ and $d_{0.01}=(d/0.01\,$pc). For zero
velocity offset relative to the systemic velocity, $\phi=0$, and the acceleration is at its maximum. 
%, and $d_{0.1}=d/0.1$ pc.
%\nonumber \\
% &\sim& 44\left(\frac{M_2}{10^8\,M_\odot}\right)d_{0.1}^{-2}\sin I \cos\phi\quad {\rm km\,s^{-1}yr^{-1}}

The typical size of the BLR (for \hbeta) around a single BH with mass $M_1$ is\footnote{This BLR size is approximate only, assuming a fixed slope $\alpha=0.5$ in the BLR $R-L$ relation, and so slightly differs from directly observed relations \citep[e.g.,][]{Kaspi_etal_2005, Bentz_etal_2009}. It also changed slightly from \citet{Shen_Loeb_2010} to reflect the most recent measurements of the 
$R-L$ relation. In our analyses in \S\ref{sec:data} and \S\ref{sec:result} we estimate the BLR size using the best-fit relation reported in \citet{Bentz_etal_2009} instead of the approximation in Eqn.\ (\ref{eqn:R_L}). } (Shen \& Loeb 2010)
\begin{eqnarray}\label{eqn:R_L}
R_{\rm BLR}&\sim& R_0(L/L_0)^{1/2}\sim 2.7\times 10^{-2}\left(\frac{L}{10^{45}\ {\rm erg\,s^{-1}}}\right)^{1/2}\ {\rm pc} \nonumber \\
 &\sim& 3\times 10^{-2}\left(\frac{\lambda_{\rm Edd}}{0.1}\right)^{1/2}\left(\frac{M_1}{10^8\,M_\odot}\right)^{1/2}\ {\rm pc}\ ,
\end{eqnarray}
following the observed $R-L$ relation for the reverberation mapping AGN sample \citep[e.g.,][]{Kaspi_etal_2005, Bentz_etal_2009}. Here $R_0$ and $L_0$ are constants, and $\lambda_{\rm Edd}\equiv L/L_{\rm Edd}$ is the luminosity Eddington ratio with $L_{\rm Edd}=1.26\times 10^{38}(M_1/M_\odot)\ {\rm erg\,s^{-1}}$ the Eddington luminosity. This observed relation is tight, with only $<40\%$ scatter in the predicted BLR size \citep[e.g.,][]{Bentz_etal_2009}. The dynamical time of the BLR is
\begin{equation}\label{eqn:tdyn} 
t_{\rm dyn}\sim R_{\rm BLR}/{\rm FWHM}\sim 24(R_{\rm BLR,0.1}/{\rm FWHM}_{4000})\ {\rm yr}\ ,
\end{equation}
where ${\rm FWHM}_{4000}$ is the broad line FWHM in units of $4000\ {\rm km\,s^{-1}}$, and $R_{\rm BLR,0.1}$ is the BLR size in units of 0.1 pc.

%We should work in our ``comfort'' zone such that the BLR around one of the
%two BHs is dynamically dominated by its own BH, i.e., $R_{\rm BLR}<
%(M_1/3M_2)^{1/3}d$. Below this separation, BLRs can no longer be distinct,
%and becomes circumbinary; under such circumstances the interpretation of the
%peak velocity drift is ambiguous, but we will try to compare our constraints
%with some numerical simulation predictions. One should also imagine that this
%BLR contributes significant flux to the total broad line flux such that its
%motion can be reflected in the peak shift (tentatively I call such BLR
%component the dominant component).

Now we can derive some crude criteria for the detectability of the binary motion with two-epoch spectroscopy separated by restframe time $\Delta t$. For simplicity, we ignore the detailed shape of the line profile and consider only the width (FWHM) of the broad line. Suppose the sensitivity of our velocity shift measurement is a fraction of the FWHM of the broad line, $f_d$FWHM. Then we require the following condition to be met for the detection: 
\begin{equation}\label{eqn:fom_condition}
|a_1|\Delta t=f_a\frac{GM_2}{d^2}\Delta t > f_d{\rm FWHM}\ ,
\end{equation}
where $f_a = |\sin I\cos\phi|$. $f_d$ is typically {\em inversely} proportional to the S/N of the line \citep[e.g.,][and see \S\ref{sec:mc_sim}]{Lindegren_1978}. Obviously, detection of binary motion is easier if the time separation $\Delta t$ is large (so long as $\Delta t\ll P$, which is generally true), the binary separation $d$ is small, and the FWHM is small. 

The radius of the BLR must be smaller than the separation of the BHs, so $R_{\rm BLR}\le f_rd$, where $f_r$ is less than unity. Otherwise the BLR becomes circumbinary, the dynamics of the BLR becomes more complicated, and there may be no coherent velocity drift in the broad line emission \citep[e.g.,][]{Shen_Loeb_2010}. $f_rd$ is then the maximum size of the BLR before it is dynamically affected by the companion dormant BH in the system. We can define $f_r$ as the average radius of the Roche lobe in a circular binary system \citep[e.g.,][]{Paczynski_1971}:
\begin{eqnarray}\label{eqn:roche}
f_r & = & 0.38 - 0.2\log q\ ,\quad 0.05<q<1.88\nonumber \\
     & = & 0.46224(1+q)^{-1/3}\ ,\quad q>1.88
\end{eqnarray}
where $q\equiv M_2/M_1$ is the binary mass ratio. 

Now we consider this maximum BLR size, i.e., we set $R_{\rm BLR}= f_rd$, and derive some crude constraints on the detectability based on the properties of the active BH 1.  Since the BLR around BH 1 is dominated by the gravity of BH 1, we have 
\begin{equation}\label{eqn:mvir}
M_1=f_m{\rm FWHM}^2R_{\rm BLR}/G\ ,
\end{equation}
where $f_m$ is the virial coefficient, determined by the geometry of the BLR. An empirical estimate is $f_m\approx 1.4$ \citep[e.g.,][]{Onken_etal_2004}. Substituting Eqn.\ (\ref{eqn:mvir}) to  Eqn.\ (\ref{eqn:fom_condition}) we obtain
\begin{equation}
\frac{f_mf_af_r^2}{f_d}\frac{M_2}{M_1}\frac{{\rm FWHM}\Delta t}{R_{\rm BLR}}>1\ .
\end{equation}

Using the observed $R-L$ relation we can recast the above equation into:
\begin{equation}\label{eqn:fom_condition2}
\frac{f_mf_af_r^2}{f_d}\frac{M_2}{M_1}\frac{{\rm FWHM}\Delta t}{R_0(L/L_0)^{1/2}}>1\  ,
\end{equation}
from which we can define a {\rm figure of merit} (FoM) of the detection:
\begin{equation}\label{eqn:fom1}
{\rm FoM} = {\rm FWHM}/L^{1/2}\ .
\end{equation}
This FoM is based on two observables, line width and luminosity, and the larger it is, the more likely we will detect the velocity shift due to binary motion in a hypothetical BBH, assuming everything else is equal in Eqn.\ (\ref{eqn:fom_condition2}). 

If we substitute FWHM in Eqn.\ (\ref{eqn:fom_condition2}) with Eqn.\ (\ref{eqn:mvir}) and use the $R-L$ relation (\ref{eqn:R_L}) we obtain
\begin{equation}\label{eqn:fom_condition3}
C\frac{f_m^{1/2}f_af_r^2}{f_d}M_2\Delta t\frac{(L/L_{\rm Edd})^{1/2}}{L^{5/4}}>1\ ,
\end{equation}
where all the other known quantities are absorbed in $C$.  We can therefore define an alternative FoM based on the luminosity and Eddington ratio of the active BH 1:
\begin{equation}\label{eqn:fom2}
{\rm FoM}^\prime = \frac{(L/L_{\rm Edd})^{1/2}}{L^{5/4}}\ .
\end{equation}
From this we see that if all quasars have the same Eddington ratio and other conditions are the same, including the companion mass $M_2$, lower luminosity
(or equivalently, lower $M_1$) quasars are more likely to have detectable velocity shifts due to binary motion (if instead the mass ratio $M_2/M_1$ were constant, FoM$^\prime$ would vary as $L^{-1/4}$ at fixed Eddington ratio, so the conclusion would be similar but weaker). One caveat here for both figure-of-merit definitions is that $f_d$ in conditions (\ref{eqn:fom_condition2}) and (\ref{eqn:fom_condition3}) scales inversely with S/N, while generally lower luminosity objects have lower S/N at fixed redshift in a given survey, which will reduce the detectability.

Note that in our postulated binary system, there is only one active BH. This is different from the scenario studied in \citet{Shen_Loeb_2010}, where both BHs are active. Therefore we do not require the additional condition that the difference in LOS velocities of the two BLRs must exceed the FWHM of the line to make a double-peaked profile \citep[e.g., the upper limit of $d$ in eqn.\ 6 of][]{Shen_Loeb_2010}. 

\begin{deluxetable}{lcccccc}
%%\tabletypesize{\tiny}
\tablecaption{Sample statistics \label{table:sample}} \tablehead{
 & \multicolumn{2}{c}{All} & \multicolumn{2}{c}{Good} & \multicolumn{2}{c}{Superior}\\
 & $N_{\rm pair}$ & $N_{\rm qso}$ & $N_{\rm pair}$ & $N_{\rm qso}$ & $N_{\rm pair}$ & $N_{\rm qso}$}\startdata
    all & 1910 & 1347 &  688 & 521 &  193 & 163\\
$\Delta t<30$ days   & 757 & 613 &   235 & 207 &  0  & 0 \\
$\Delta t>100$ days & 834 & 674 &   316 & 268 &  193 & 163 \\
%\quad$\sigma_{a,{\rm meas}}<500\,{\rm [km\,s^{-1}\,yr^{-1}]}$ & 437 &  367\\
%\quad$\sigma_{a,{\rm meas}}<200\,{\rm [km\,s^{-1}\,yr^{-1}]}$ & 224 &  192\\
%\quad$\sigma_{a,{\rm meas}}<100\,{\rm [km\,s^{-1}\,yr^{-1}]}$ & 119 &  97\\
%\quad$\sigma_{a,{\rm meas}}<50\,{\rm [km\,s^{-1}\,yr^{-1}]}$ & 54 &  45\\
%\hline\\
$\Delta t>1$ yrs & 251 & 223 &   104 &  94 &  98  & 90 \\
$\Delta t>2$ yrs & 117 & 106 &    53   &  49 &   53  & 49 \\
$\Delta t>3$ yrs & 69 & 65 &    27    &  26 &   27  & 26 \\
$\Delta t>5$ yrs & 6 & 6 &    4      & 4  &  4      & 4  
\enddata
\tablecomments{All times are in the restframe of the quasar. The ``good'' sample is defined in \S\ref{sec:vel_shift}, in which we search for binary candidates. The ``superior'' sample is a subset of the good sample, defined in \S\ref{sec:general_pop}, which we use to place statistical constraints on the binary population. }
\end{deluxetable}

\section{Data}\label{sec:data}

We start from the DR7 compilation of the spectroscopic
quasar catalog \citep{Schneider_etal_2010} from SDSS.
This sample contains 105,783 quasars brighter than $M_{i}=-22.0$ that have at
least one broad emission line with FWHM larger than 1000 ${\rm
 km\ s^{-1}}$ or have interesting/complex absorption features. The reduced and calibrated one-dimensional (1D) spectral data used in this study are available through the SDSS Data Archive
Server\footnote{http://das.sdss.org/spectro/} (DAS). The spectral resolution is
$R\sim 1850-2200$, and the 1D spectra are stored in vacuum
wavelength, with a pixel scale of $10^{-4}$ in log-wavelength, which
corresponds to $69\ {\rm kms^{-1}}$. All spectra are wavelength calibrated to the heliocentric reference, with an accuracy of better than $5\ {\rm km\,s^{-1}}$. Throughout the paper, we refer to the signal-to-noise ratio per pixel as S/N. 

Several thousand of these DR7 quasars have multiple spectra taken at different epochs by SDSS. Most of them have two epochs, and a small fraction of them have multiple epochs. The default spectrum reported in the DR7 quasar catalog is usually the one with the highest S/N, and we always take that default spectrum as Epoch 1 regardless of the relative time between the two epochs. Epoch 2 is then the other available spectra in SDSS (could be more than one). We call each Epoch 1/2 combination a pair of observations. These quasars with repeated spectroscopy in SDSS form the sample in which we search for velocity shift of the broad line. We will
focus on the \hbeta\ line for this systematic study, which restricts us to
$z\lesssim 0.9$, and reduces the number of available quasars to below two thousand. 

We choose the \hbeta\ line rather than other broad lines for the following reasons: (1) it has
robust systemic redshift from the \OIII\ lines (which are shifted out of SDSS spectrum for $z\gtrsim 0.8$); (2) the presence of \OIII\ line in
the SDSS spectrum provides an empirical way to subtract the narrow \hbeta\
component; (3) the restframe time separation $\Delta t$ between two epochs is longer for the \hbeta\
sample  
than for higher-redshift quasars; (4) the broad Balmer lines are better studied and better
understood than \MgII\ or \CIV\ lines, especially in terms of reverberation
mapping studies, therefore their BLR size and virial BH mass estimates are the most reliable \citep[e.g.,][and references therein]{Shen_2013}; (5) \MgII\ is difficult because of the complication of the strong \FeII\ emission
underneath \MgII; \CIV\ is difficult because this line is more asymmetric and
may arise from a disk-wind component; in addition, for both \MgII\ and \CIV\
the narrow line subtraction will be ambiguous; (6) finally, \halpha\ is a line complex enclosing more than one narrow line (\NIIab, narrow \halpha), and the broad line component isolation is seemingly difficult. 

Nevertheless, if we detect any \hbeta\ velocity shift, we will investigate the \MgII\ and \halpha\ lines at the two epochs, if covered in the SDSS spectrum.

The basic sample statistics are summarized in Table 1. For restframe $\Delta
t>100\,$days, there are $834$ pairs of observations over 674 unique quasars;
most of them have a restframe time span of less than a year. We do not remove duplicated objects, and treat multiple pairs for the same object as independent measurements in our statistical analysis in \S\ref{sec:result}. 

\subsection{Measurements of spectral properties}\label{sec:fit}

\begin{figure}
 \centering
 \includegraphics[width=0.5\textwidth]{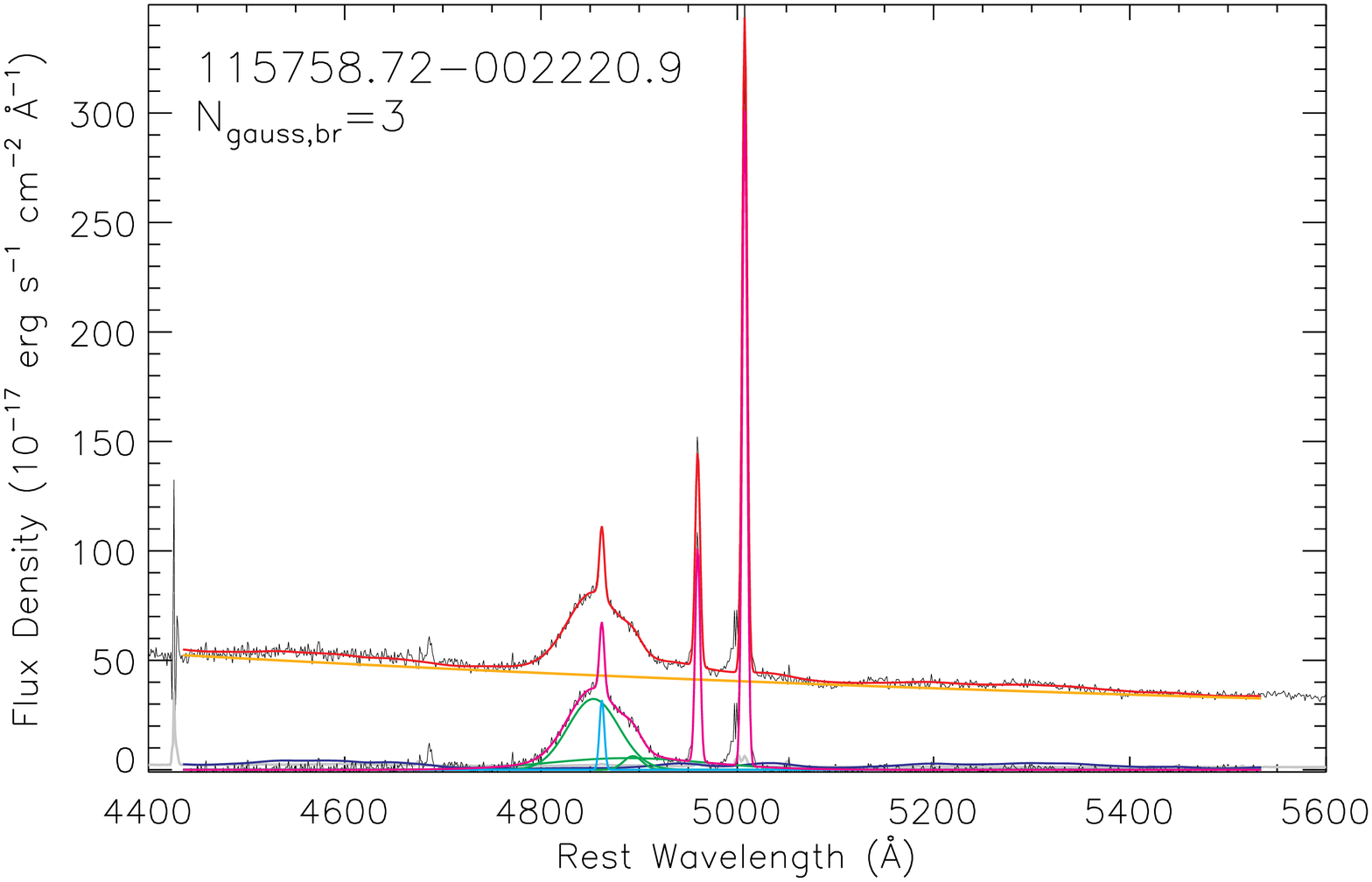}
 \includegraphics[width=0.5\textwidth]{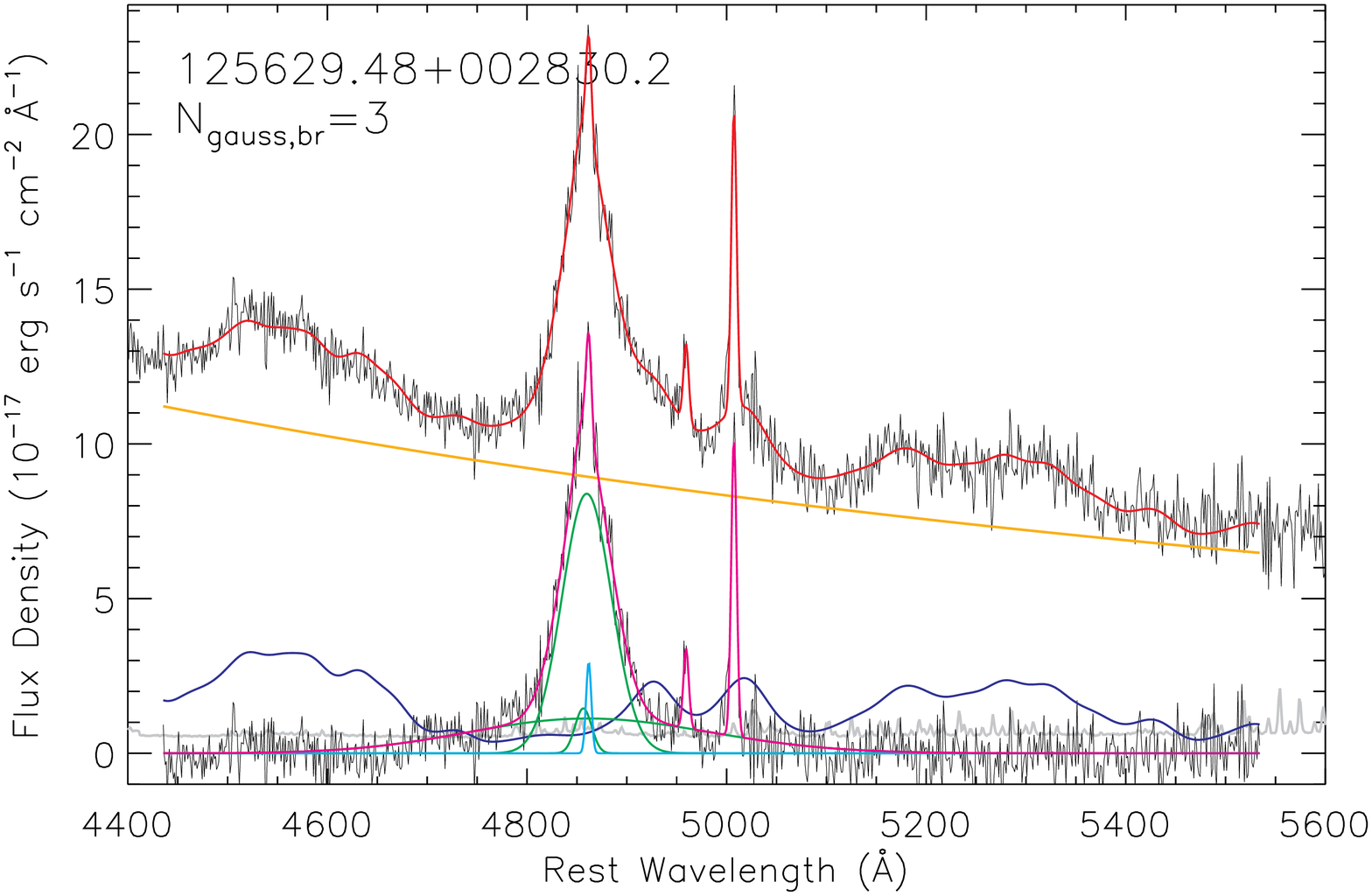}
 \caption{Examples of our model fits. In both panels the upper spectrum is the full spectrum while the lower spectrum is the pseudo-continuum subtracted spectrum.
 Yellow is the power-law continuum model, blue is the \FeII\ emission model, cyan is the narrow \hbeta\ component and green is the broad \hbeta\ components. The SDSS designation and the number of Gaussians for the broad \hbeta\ line are marked in the top left corner of each panel. These two examples represent different broad line widths, \OIII\ and \FeII\ strengths.  
 }
 \label{fig:fit_examp}
\end{figure}

In order to characterize the properties of quasars in our sample, we measure the continuum and emission line properties using
a combination of model fitting and direct measurements on the spectrum. This process also produces the broad-line spectra (i.e., continuum and narrow line removed) on which we will perform the cross-correlation to measure velocity shifts in \S\ref{sec:vel_shift}.

The basic spectral fitting approach follows our earlier work in
\citet[]{Shen_etal_2008,Shen_etal_2011}. In short, an \FeII\ template plus a
power-law continuum is fit to several line-less windows around \hbeta\ to
form a pseudo-continuum; this pseudo-continuum model is then subtracted from
the spectrum, leaving the emission lines. The narrow emission lines are
modeled by Gaussians (in logarithmic wavelength) with the same width and their
relative velocity offsets fixed to those based on their restframe wavelengths. Whenever necessary we fit two Gaussians to the \OIII\
lines to account for the core and the wing components often seen in \OIII;
the narrow \hbeta\ is tied to the core \OIII\ component in such cases. The
broad \hbeta\ profile is fit with a mixture of Gaussians (up to 5, but 3 are
enough in most cases). These Gaussian functions are merely used to reproduce
the line profile and we assign no physical meaning to individual Gaussian
components. We visually inspect all fits to make sure they are
reasonable in the sense that the line is well reproduced by the model, and we require a
reduced $\chi^2<5$ for the fits. Fig.\ \ref{fig:fit_examp} shows two examples of the decomposition of
different spectral components. The \OIIIb/narrow \hbeta\ flux ratio is tied to be the same at both epochs. This is important for cases where the narrow \hbeta\ contributes
a substantial amount to the \hbeta\ profile, and incorrect subtraction of the
narrow \hbeta\ emission will likely impact the cross-correlation analysis in
\S\ref{sec:vel_shift}. Due to aperture effects and seeing/guiding variations between the two epochs, as well as the extended nature of the narrow line regions, the absolute narrow line flux could change, and so we do not fix the normalization of the narrow line flux at the two epochs. Visual inspections were performed to ensure there is no artificial line variation near the systemic velocity between the two epochs due to the narrow line removal. 

Once we have decomposed the spectrum around the \hbeta\ region, we measure a variety of broad line properties both using the model fit and using the raw (broad-line only) spectrum:

\begin{itemize}

\item[1.] Line peak, FWHM, line flux, and restframe equivalent width (EW) of the
    broad line from the model fit;
    
\item[2.] Line centroid or first moment of the line:
    $\bracket\lambda\equiv \sum\lambda_if_i/\sum f_i$, where $\lambda_i$ and $f_i$ are the
    pixel wavelength and flux density. To compute the centroid and higher
    order moments we use the raw spectrum, and we only use pixels that
    enclose $2\times {\rm FWHM}$ of the line peak. This is to reduce
    the adverse effects of the noisy wings in determining the line
    moments.

%\item[4.] Line mode: ($3\times$median - $2\times$mean) of all the pixels
%    within $2\times {\rm FWHM}$ of the peak, computed in logarithmic
%    wavelength. This often agrees with the peak value, but could
%    substantially deviate from the peak value in some objects.

\end{itemize}

We also compute the second moment, skewness and kurtosis of the broad line,
again using all the pixels enclosing $2\times$FWHM of the line peak. The $n$-th moment of the line is defined as $\mu_n\equiv\sum (\lambda_i-\bracket\lambda)^nf_i/\sum f_i$, the skewness is defined as $s=\mu_3/\mu_2^{3/2}$, and the kurtosis is defined as $k=\mu_4/\mu_2^2-3$. We compared the distributions of these broad line profile characteristics to other studies \citep[e.g.,][]{Zamfir_etal_2010,Eracleous_etal_2012} whenever available, and found good agreement. 

\begin{figure}
 \centering
 \includegraphics[width=0.5\textwidth]{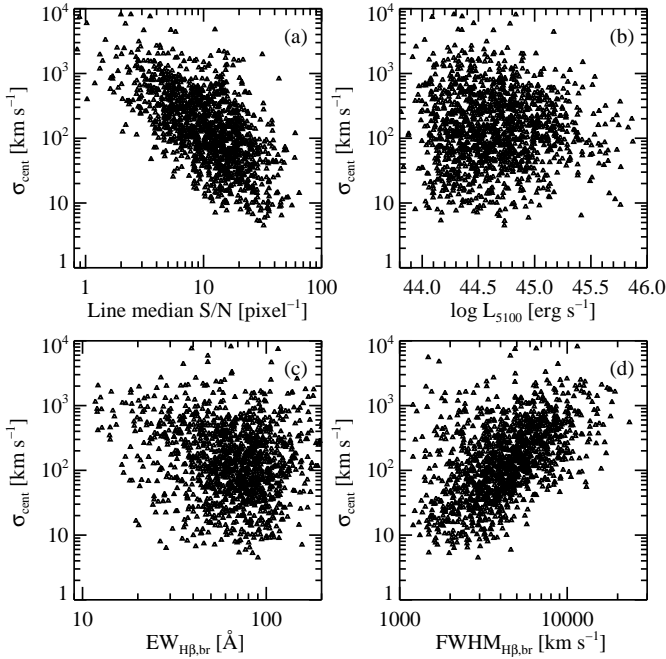}
 \caption{Measurement error in line centroid as functions of various broad line properties: (a) median S/N per pixel for the broad
 line flux; (b) Continuum luminosity at 5100 \AA; (c) Restframe equivalent width of broad \hbeta; (d) FWHM of the
 broad \hbeta. The measurement error decreases as S/N increases, and increases as line width increases, as expected. In addition, stronger lines (larger EW) have on average smaller measurement errors in line centroid, but the trend is not as strong as those with S/N and FWHM, and is likely a secondary effect of S/N. }
 \label{fig:err_study_cent}
\end{figure}

We use a Monte Carlo approach to estimate the measurement errors in the above various quantities
\citep[e.g.,][]{Shen_etal_2008,Shen_etal_2011}. In practice we
add artificial noise (using the reported flux errors at each pixel) to the model spectrum to create a mock spectrum, repeat our fitting and record the results, for 50 trials. We then use the semi-quantile of the $68\%$ range of these mock results as the nominal measurement uncertainty of each
measured quantity. 

%All the measurements and errors are compiled in Table 2 for our \hbeta\ sample. 

%\textbf{We assume these measurement errors are Gaussian,
%although for a few peculiar profiles this is not necessarily the case; but
%for the overall population it should be OK. }

%We show some basic statistical properties of our measurements in Fig.\
%\ref{fig:vel_diff} and Fig.\ \ref{fig:err_study_peak}. This gives you an idea
%on what typical time span our sample covers, and what are the typical
%measurement errors. Some further discussions are included in the captions.
%
%There are some general similarities between different line center
%definitions, but for specific objects they could differ a lot. We will be
%primarily using line peak for our investigation, with the following caveat
%kept in mind: if the associated BLRs moving with one or two BHs do not
%dominate the broad line emission, their motion may be hidden in the wings of
%the broad line rather than in the peak. To test such a scenario we will use
%line centroid, or variations in the line width, or skewness, after we discuss
%results on line peak. Line mode generally is more noisy than the other two
%line center definitions, and so it is probably the least useful.

Fig.\ \ref{fig:err_study_cent} shows how the measurement error in the broad line centroid, $\sigma_{\rm cent}$, changes with
various quantities. The most significant trends are the correlations of $\sigma_{\rm cent}$ with S/N and FWHM: $\sigma_{\rm cent}$ decreases as S/N increases, and increases as FWHM increases. This is expected, as the uncertainty in measuring the line centroid is proportional to line width, and is inversely proportional to S/N, i.e., $\sigma_{\rm cent}\propto {\rm FWHM/SN}$. However, the trends shown in Fig.\ \ref{fig:err_study_cent} somewhat deviate from the expected dependence on FWHM and S/N, which is likely caused by the following: a) there is a slight anti-correlation between FWHM and S/N for our objects; b) the line shape also changes with line width, which may increase the difficulty of measuring line centroid for large line widths. Finally, there is a weak dependence of measurement errors on the strength of the line, such that stronger lines on average have smaller errors in measuring their centroid. This weak trend with line strength is likely a secondary effect of S/N, i.e., given fixed continuum level and photon noise, weaker lines have smaller S/N. 

Line centroid (or line peak) is not a robust indicator for the velocity shift we are trying to measure, because it is only one characteristic of the broad line: slight changes in the line shape may produce artificial centroid shifts or hide a real shift. In the next section, we will measure the broad line velocity shift with a cross-correlation technique, which is more robust in measuring the broad line velocity shift between the two epochs.

The measured broad \hbeta\ shape parameters will be used in our analyses in \S\ref{sec:result}.

\subsection{$\chi^2$ cross-correlation between two epochs}\label{sec:vel_shift}

In order to constrain the velocity shift between two epochs and hence the
orbital acceleration in the hypothetical binary, we deploy a
$\chi^2$ cross-correlation technique (``ccf'' for short) similar to that used in
\citet{Eracleous_etal_2012}. We describe this approach in detail below.

We consider the pseudo-continuum and narrow-line subtracted spectra from \S\ref{sec:fit}, i.e.,
broad line spectra only. This is different from the approach in
\citet{Eracleous_etal_2012}, which used several broad line windows that avoid
the narrow \hbeta\ line and did not subtract the underlying pseudo-continuum
in the cross-correlation. We prefer to use the broad line spectra in the
cross-correlation, because possible changes in the continuum slope will
potentially bias the cross-correlation result if the full spectrum is used.
We also want to use a large portion of the broad line profile to increase
the statistical power in the cross-correlation, which requires removal of the central
narrow \hbeta\ component. As discussed in \S\ref{sec:fit}, we carefully
controlled the systematics from narrow \hbeta\ subtraction in our fitting, and visually inspected all fits to ensure that the narrow line subtraction does not introduce artificial changes between the two epochs near the systemic velocity. 

We shift the second epoch broad line spectrum by an increasing number of
pixels, and compute the $\chi^2$ as a function of the shift:
\begin{equation}\label{eqn:chi2}
\chi^2 = \sum_i\frac{(f_{1,i}-f_{2,i}^\prime)^{2}}{\sigma_{1,i}^2 + \sigma_{2,i}^{\prime,2}}\ ,
\end{equation}
where $f_{1,i}$ and $\sigma_{1,i}$ are the flux density and flux density
errors of the $i$th pixel in the Epoch 1 spectrum, and $f^{\prime}_{2,i}$
and $\sigma^{\prime}_{2,i}$ are the corresponding quantities of
the $i$th pixel in the shifted Epoch 2 spectrum. We only use pixels
within [4800,4940]\AA\ to compute the $\chi^2$, which encloses most of the
broad line flux and excludes the noisy wings. During the $\chi^2$
calculation, we also scale the second-epoch spectrum by the ratio of the
integrated broad line flux within [4800,4940]\AA\ of the two epochs. The
latter step is necessary because the broad line flux could vary between the
two epochs, which will reduce the statistical power of the cross-correlation.

One great advantage of using the SDSS data is that both spectra were reduced
using the same software and binned on the same logarithmic
wavelength grid, i.e., 1 pixel shift corresponds to $69\,{\rm km\,s^{-1}}$
shift. The step size in the grid of shifts that we use to compute the cross-correlation is
1 pixel, i.e., we do not use sub-pixel shifts and rebinning of the second-epoch
spectra as in \citet{Eracleous_etal_2012}. We chose to do so because the S/N per pixel is typically several, and so by interpolating on a sub-pixel grid and rebinning we would likely introduce systematics, which would complicate the resulting error estimation of the measured velocity shift.

Once we have the $\chi^2$ values, we determine the minimal $\chi^2$ and its
location as follows: we fit the 20 points enclosing the minimal $\chi^2$
point with a 6th-order B-spline function. The minimal $\chi^2$ value is then
determined from the spline, which allows us to measure sub-pixel shifts, as
well as reduce the impact of noisy $\chi^2$ curve. We estimate the confidence
range by finding the intercepts of the B-spline at $\Delta\chi^2=\chi^2_{\rm
min}+6.63$, which corresponds to $99\%$ confidence range ($\sim 2.5\sigma$). In this way we measure the velocity shift $V_{\rm ccf}$ and the associated uncertainty between the two epochs. 

\begin{figure}
 \centering
 \includegraphics[width=0.48\textwidth]{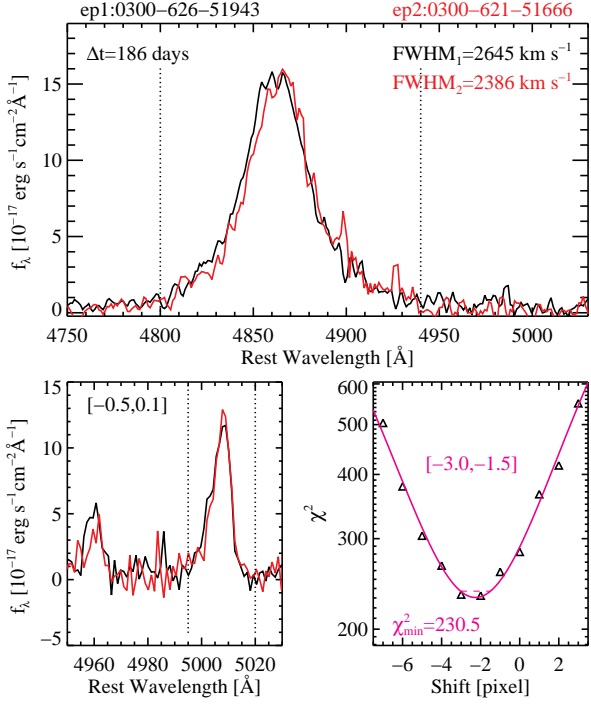}
 \caption{An example of the cross-correlation procedure to measure the broad line velocity shift between
 two epochs. The top panel shows the broad line spectra at the two epochs, and the bottom left panel shows that for the narrow \OIII\ region. All spectra have been boxcar smoothed by 3 pixels for better display. The two dotted vertical lines mark the range where
 we perform the $\chi^2$ cross-correlation.  The Epoch 2 spectrum has been scaled to the Epoch 1 spectrum using the integrated line flux within the cross-correlation wavelength range. The bottom right
 panel shows the $\chi^2$ curve for \hbeta\ as a function of pixel shift (recall that 1 pixel corresponds to 69 ${\rm km\,s^{-1}}$) in open triangles, where the magenta line is the 6th-order B-spline on the
 20 points centered on the point with the minimal $\chi^2$. The dashed horizontal segment indicates
 the $\Delta\chi^2=6.63$ (2.5$\sigma$) range in pixel shift, also indicated in the magenta square bracket. In this example a velocity shift is detected at $>2.5\sigma$ significance. The $\chi^2$ cross-correlation test on the \OIIIb\ line yields a 2.5$\sigma$ velocity shift range of $[-0.5,0.1]$ pixels, not enough to account for the velocity shift measured for \hbeta. }
 \label{fig:examp_ccf}
\end{figure}

\begin{figure}
 \centering
 \includegraphics[width=0.48\textwidth]{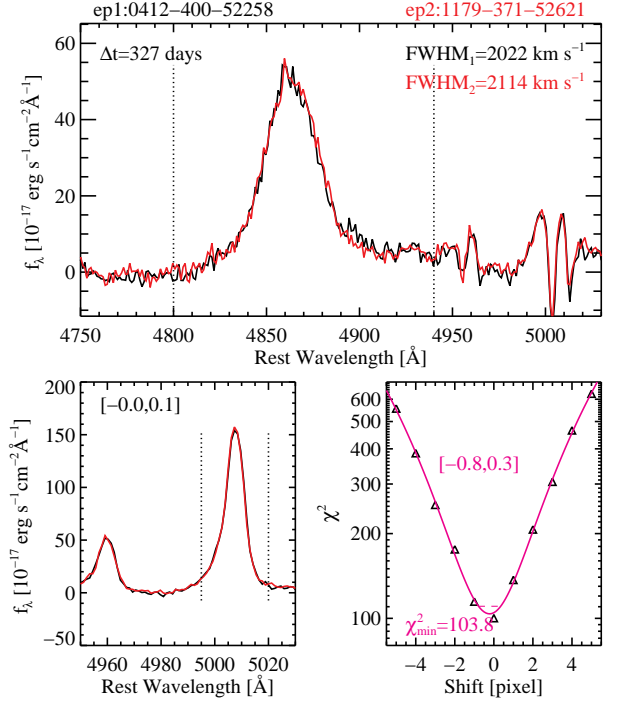}
 \caption{Same as Fig.\ \ref{fig:examp_ccf}, but for a case where no significant velocity shift is detected over a restframe time span of $\sim$ a year.  }
 \label{fig:examp_ccf2}
\end{figure}

\begin{figure}
 \centering
 \includegraphics[width=0.48\textwidth]{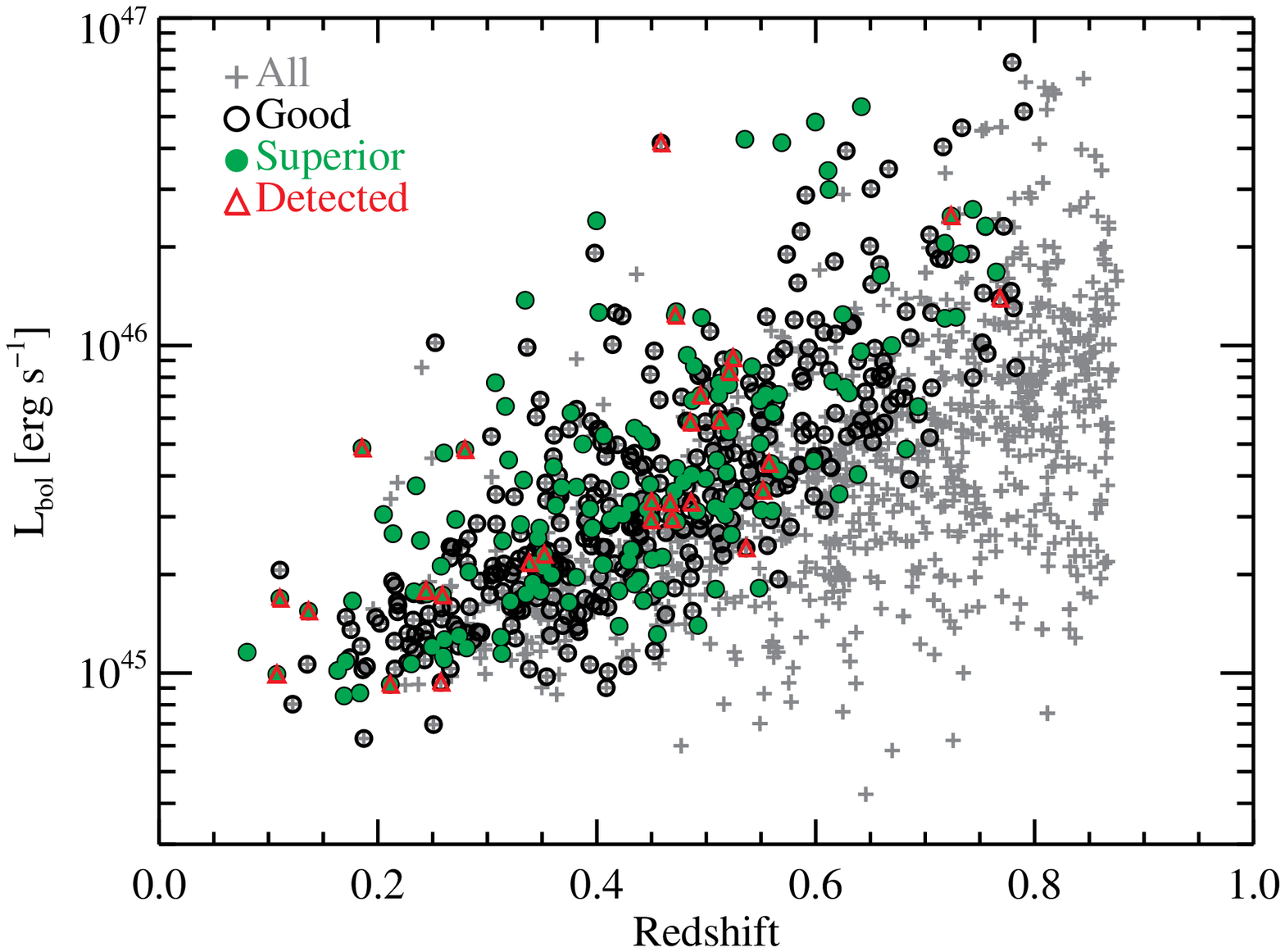}
 \includegraphics[width=0.48\textwidth]{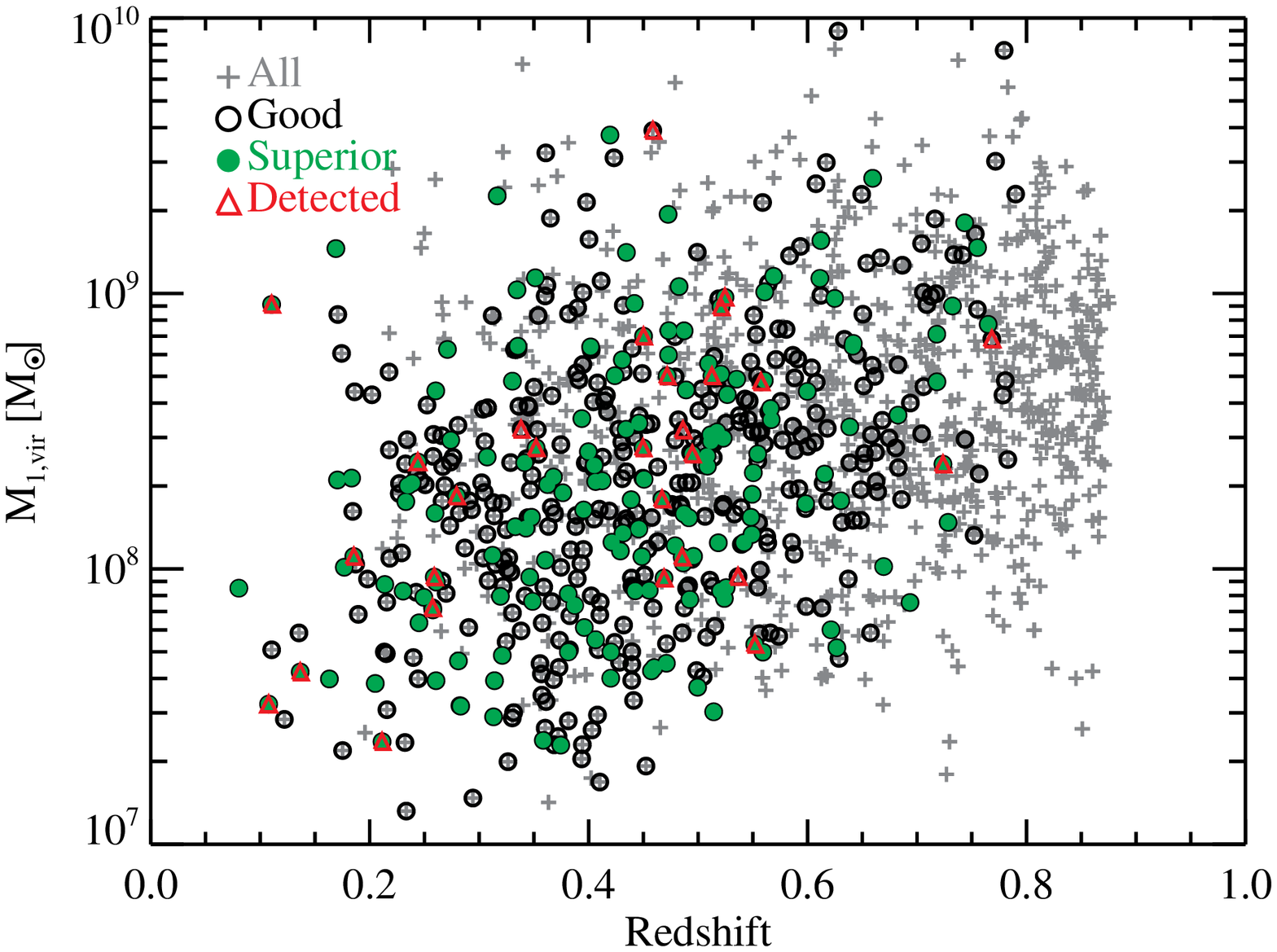}
 \includegraphics[width=0.48\textwidth]{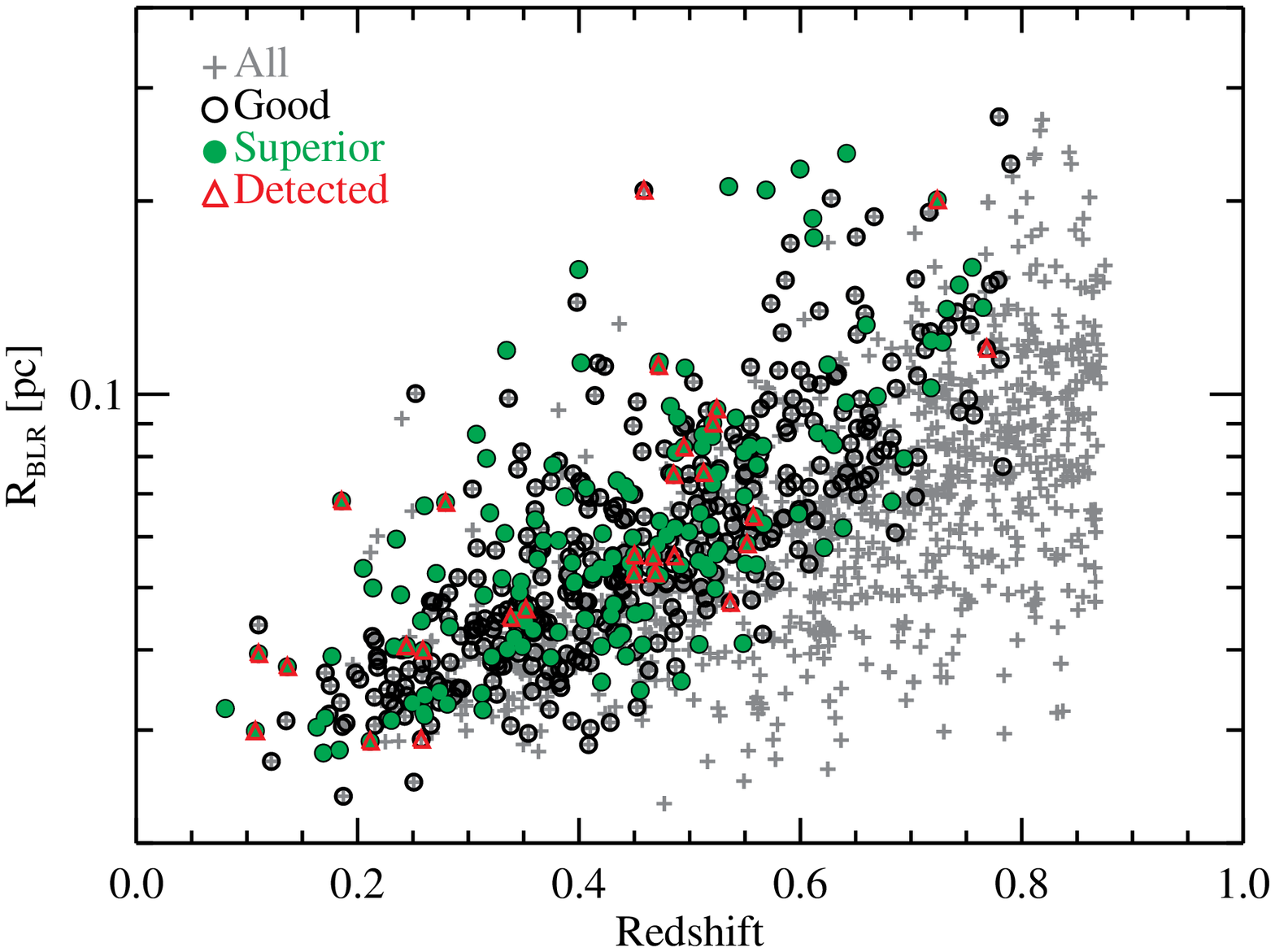}
 \caption{Distributions of quasar properties for different samples. From top to bottom, bolometric luminosity, virial BH mass estimates, and \hbeta\ BLR size versus redshift. The bolometric luminosity and virial BH mass estimates were taken from the \citet{Shen_etal_2011} catalog. The BLR sizes were estimated from the $5100$ \AA\ continuum luminosity using the best-fit $R-L_{5100}$ relation in \citet{Bentz_etal_2009}, therefore they have explicit dependence on luminosity. The gray points are for our parent sample, and the black points are the ``good'' subsample defined in \S\ref{sec:vel_shift} in which we search for velocity shifts and statistical constraints on the general quasar population. The green points are the ``superior'' sample defined in \S\ref{sec:general_pop}, which is a subset of the ``good'' sample. Finally the red points are individual quasars in which a significant velocity shift has been detected (see \S\ref{sec:binary_candidate}). Other than removing objects with low-quality spectra at high redshift and low luminosity, the good sample is statistically indistinguishable from the parent sample. In addition, the ``superior'' sample is statistically indistinguishable from the ``good'' sample, as confirmed by standard KS tests. 
 }
 \label{fig:Lbol_z}
\end{figure}

\begin{figure}
 \centering
 \includegraphics[width=0.48\textwidth]{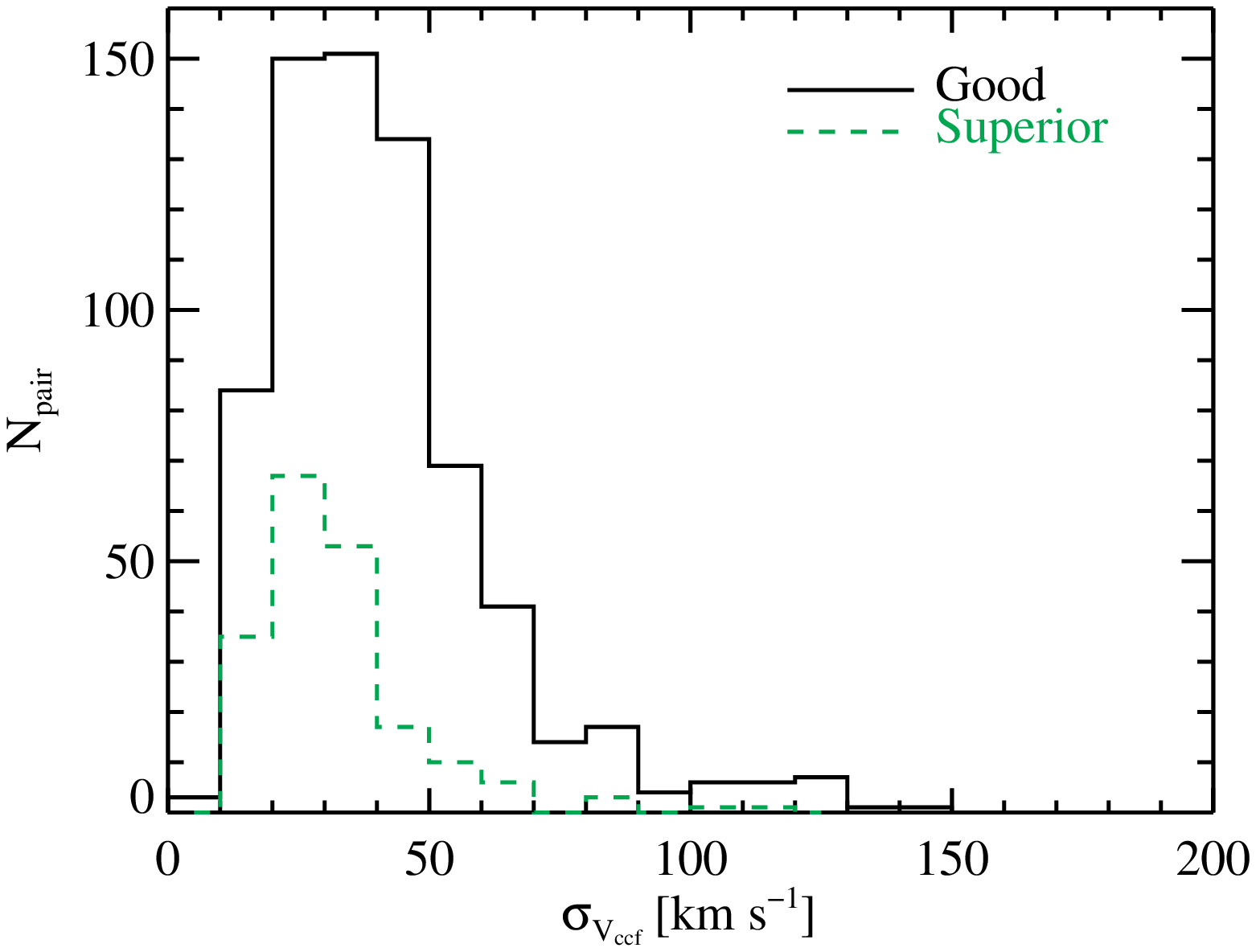}
 \includegraphics[width=0.48\textwidth]{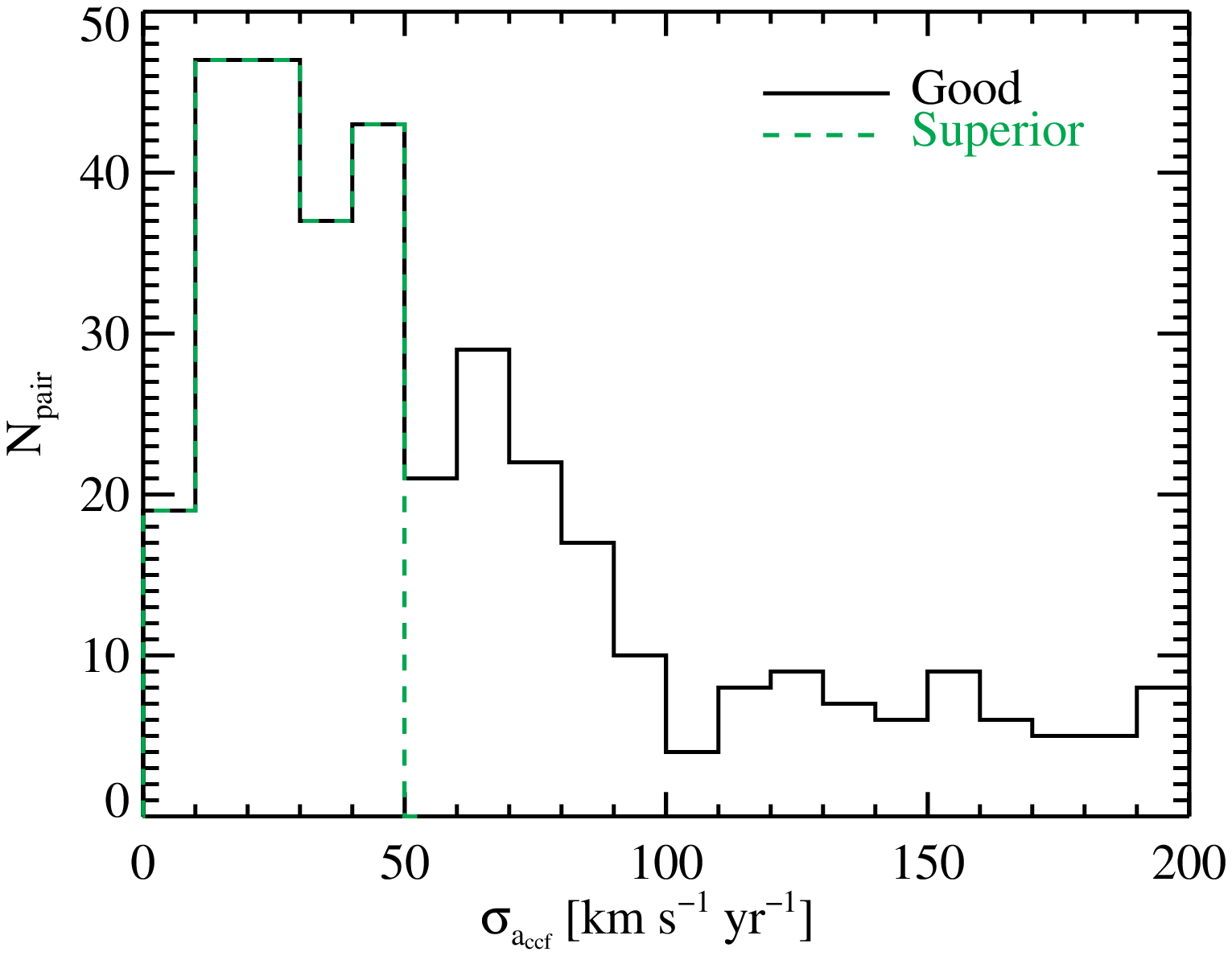}
 \caption{{\em Top}: Distributions of the measurement errors in the velocity shift $V_{\rm ccf}$ between two epochs using the $\chi^2$ ccf method. {\em Bottom}: distributions of the measurement errors in the LOS acceleration measured from two epochs, i.e., $a_{\rm ccf}=V_{\rm ccf}/\Delta t$. The black lines are for the ``good'' sample (\S\ref{sec:vel_shift}), while the green dashed lines are for the ``superior'' sample (\S\ref{sec:general_pop}). 
 }
 \label{fig:sig_v}
\end{figure}

We also apply this cross-correlation method to the \OIIIb\ line, using pixels
within [4995,5020]\AA. This allows us to check possible issues of bad
wavelength calibration associated with either epoch. In most cases the shift
of the \OIIIb\ line is constrained to be $< 20\,{\rm
km\,s^{-1}}$, which is expected as the SDSS wavelength calibration is better than $5\ {\rm km\,s^{-1}}$. We will use the uncertainty in the \OIII\ shift as an additional constraint
to select individual detections of broad \hbeta\ velocity shift in \S\ref{sec:binary_candidate}.

Fig.\ \ref{fig:examp_ccf} shows an example of our $\chi^2$ cross-correlation procedure, where a velocity shift is detected at the 2.5$\sigma$ significance level, and the restframe time separation is $\sim 6$ months. Fig. \ref{fig:examp_ccf2} shows another example, where no significant velocity shift is detected over a restframe timescale of about a year.  In both examples the shape of the broad \hbeta\ remains more or less the same, as indicated by the FWHM. 

Many SDSS spectra are of low quality and often the resulting $\chi^2$ curve
is too noisy to measure a well-defined minimum and confidence ranges. To eliminate these, we
first filter our sample using a simple S/N cut, keeping only objects where both
epochs have detected broad \hbeta\ line at $>6\sigma$ significance (where line
fluxes and errors are taken from our spectral fits in \S\ref{sec:fit}). We
then visually inspect the B-spline fits for all objects, and discard cases
where the $\chi^2$ curve is noisy, such that either a meaningful spline fit
is not possible, or that the formal confidence range from the spline fit is
questionable. For such cases
the two epochs cannot constrain the velocity shift well using this
cross-correlation method. These objects usually have noisy spectra, or very
broad lines (FWHM$\gtrsim 10,000\,{\rm km\,s^{-1}}$). For the remaining
objects, the $\chi^2$ curve shows a well-determined minimum and the B-spline
provides reasonably good measurements of its minimum location and associated
confidence levels.

\begin{figure}
 \centering
 \includegraphics[width=0.48\textwidth]{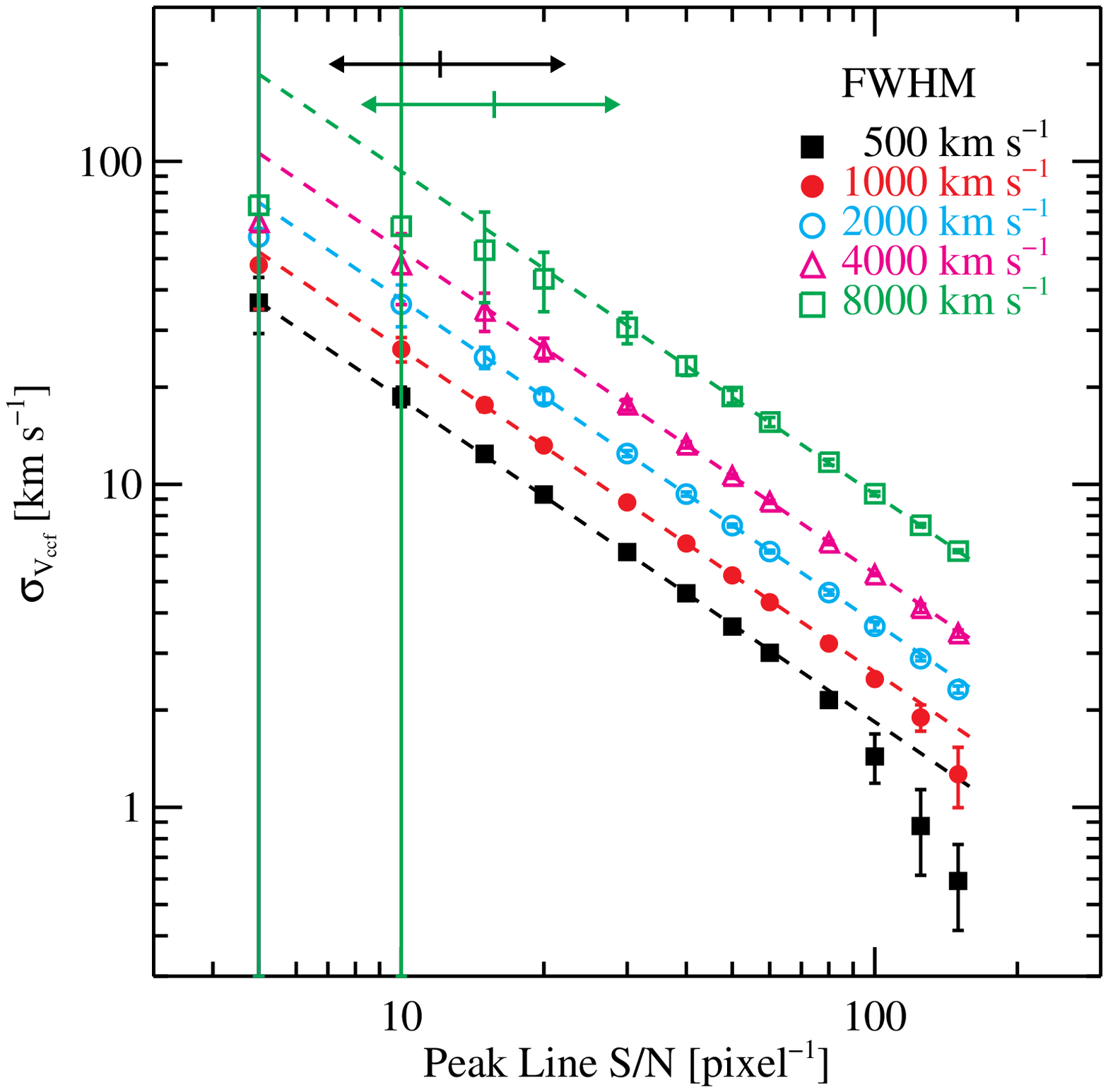}
 \caption{Tests on the measurement errors in the velocity shift using the $\chi^2$ ccf method, with mock spectra. Shown here are the measurement errors as functions of peak line S/N (per pixel) and FWHM. The points and error bars are the median and standard deviations from 5000 trials in our Monte Carlo mock tests. The dashed lines are not a fit to the points, but straight lines with a power-law slope of $-1$. This test indicates that the measurement error decreases as S/N increases and FWHM decreases, as expected. The two-sided arrows at the top indicate the $68\%$ quantile range of actual peak line S/N for the ``good'' sample (black) and the ``superior'' sample (green). Our quasars have typical broad \hbeta\ FWHM$\sim 3500\ {\rm km\,s^{-1}}$. Thus the expected typical measurement error is $\sim 40\ {\rm km\,s^{-1}}$ (1$\sigma$), fully consistent with the actual values shown in Fig.\ \ref{fig:sig_v}. }
 \label{fig:mc_sigV}
\end{figure}

\begin{figure}
 \centering
 \includegraphics[width=0.48\textwidth]{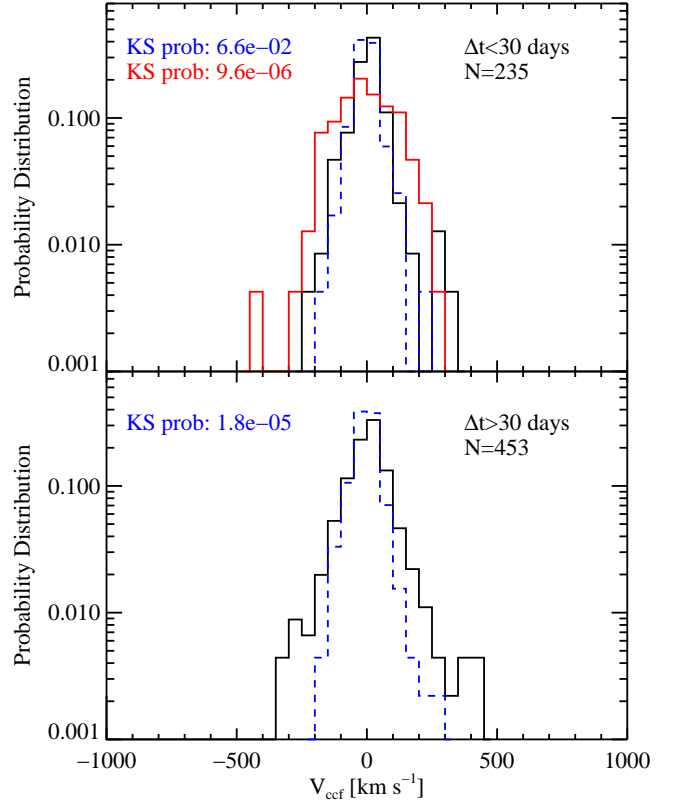}
 \caption{Tests on the measurement errors from the $\chi^2$ ccf approach in the ``good'' sample defined in \S\ref{sec:vel_shift}. {\em Top:} results for a subsample of observations with time span less than
 30 days (in restframe). The dispersion of velocity shift within such short period is due to a combination of measurement errors and potential systematic errors due to data reduction and short-term variability. The black histogram is the observed distribution of velocity shift between two epochs. The blue dashed line is the simulated
 distribution using a sample of zero velocity shifts, shuffled with the measurement error distribution. 
 The red line is the same as the blue line, but shuffled by an additional Gaussian error of $100\ {\rm km\,s^{-1}}$. The KS test probabilities of both simulated distributions being drawn from the same distribution as the observed distribution (black) are marked on the upper left corner. This test suggests that any additional systematic error terms are small, and our
 measurement error distribution is reasonable. {\em Bottom:} same as the top panel, but for observations with a time span longer than 30 days.  }
 \label{fig:test_meas_errors}
\end{figure}

\begin{figure}
 \centering
 \includegraphics[width=0.48\textwidth]{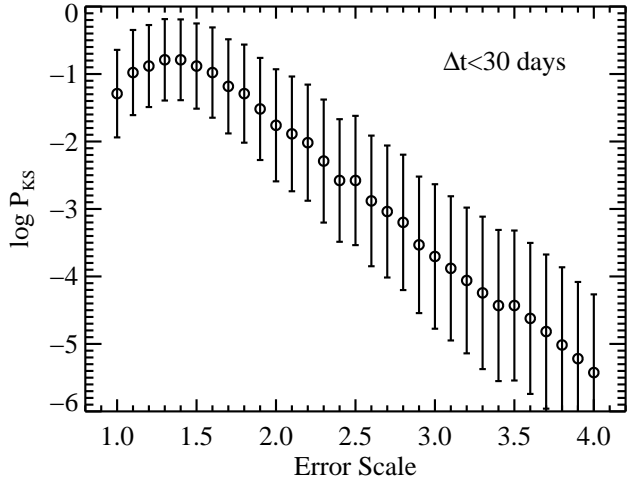}
 \caption{The KS probability between the observed velocity shift distribution for objects in the ``good'' sample with $\Delta t<30$ days and a simulated distribution of zero velocity shifts broadened by the measurement errors scaled by a constant. Under the assumption that the intrinsic distribution for these objects should be consistent with zero, this test suggests that our formal estimates of the 1$\sigma$ measurement error in velocity shift slightly underestimate the true error by $\sim 30\%$. }
 \label{fig:test_meas_errors_KS}
\end{figure}

\begin{figure}
 \centering
 \includegraphics[width=0.5\textwidth]{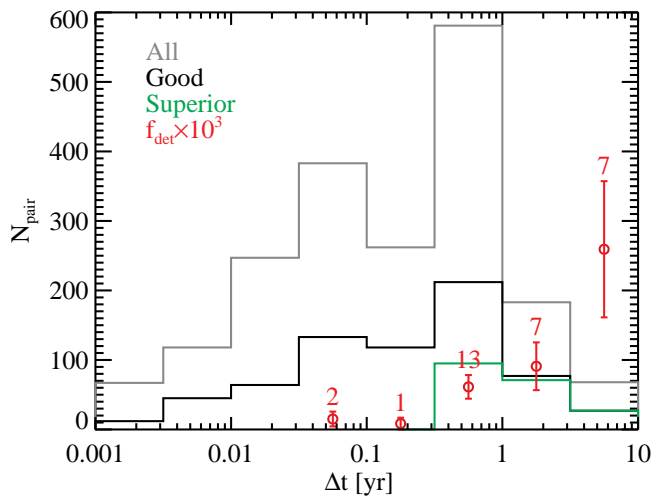}
 \caption{Distributions of restframe time separation for different samples. The gray line is our parent (``all'') sample, and the black line is the ``good'' sample defined in \S\ref{sec:vel_shift}, for which we have measured a reliable velocity shift between the two epochs with the $\chi^2$ cross-correlation method. The reduction from the parent sample to the good sample is mostly due to the quality of the spectra, and does not change the distribution of $\Delta t$ significantly. The green line is the ``superior'' sample defined in \S\ref{sec:general_pop}, which is a subset of the ``good'' sample with small measurement errors on the measured radial acceleration. As a result, the ``superior'' sample requires on average longer time separations than the ``good'' sample. Finally, the red points show the fraction of detections (multiplied by $10^3$) in the good sample, in each time separation bin, and the error bars are Poisson. The numbers above the red points are the actual number of detections. The detection fraction generally increases as a function of time separation. 
 }
 \label{fig:hbeta_hist}
\end{figure}

\begin{figure}
 \centering
 \includegraphics[width=0.5\textwidth]{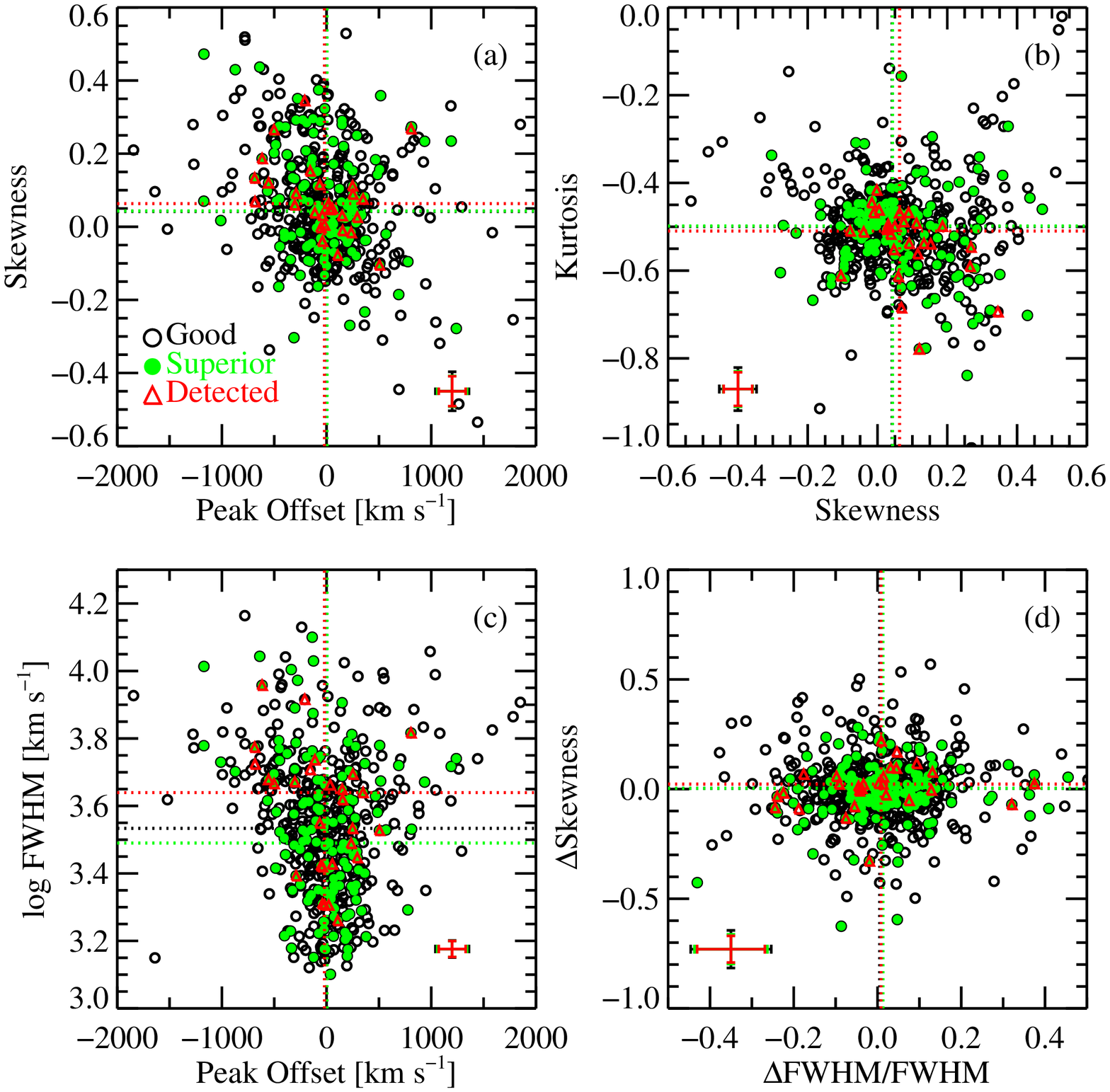}
 \caption{Broad \hbeta\ line shape properties for various samples. Typical measurement errors are shown in each panel. The dotted lines mark the median values for each sample, in their corresponding colors. Panel (a) shows an anti-correlation between the broad line peak offset (relative to \OIIIb) and the skewness of the line, which is consistent with fig.\ 5 in \citet{Eracleous_etal_2012}. Note that the definition of skewness in the latter work has the opposite sign to ours. Panel (b) plots the skewness against the kurtosis. Panel (c) plots FWHM against broad line peak offset. Panel (d) plots the relative changes in FWHM against the absolute changes in skewness between the two epochs. Significant shape changes between the two epochs are not common. These four plots indicate that objects with detected broad \hbeta\ velocity shifts between the two epochs do not stand out as a distinct population. 
 }
 \label{fig:sample_line_prop}
\end{figure}

While our rejection based on visual inspection of the $\chi^2$ curves is not rigorous, it is
necessary to reject false positives and obtain a clean sample. As long as  the remaining objects probe the same parameter space as the parent sample, we are not biasing the sample selection by this visual screening (see below). There are 688
objects kept in our final sample, which we call the ``good'' sample. We will
focus on this sample in the following analysis.

Fig. \ref{fig:Lbol_z} shows the statistical quasar properties of the good sample and the parent
sample. Objects at high redshift and low luminosities are preferentially removed from our good sample because they have low quality spectra. Other than that, the good sample probes the same parameter space of the general SDSS quasar population, thus there is no obvious bias
from our sample refining. The resulting error distributions of the measured velocity shift and acceleration of the good sample are shown in Fig.\ \ref{fig:sig_v}. 

The virial BH masses probed by our sample have a median value of $1.8\times 10^8\ M_\odot$ with a dispersion of $\sim 0.4$ dex. We fully appreciate the uncertainty in these mass estimates \citep[for a comprehensive review of the caveats of virial BH masses, see, e.g.,][]{Shen_2013}. The $\sim 0.4$ dex dispersion in these virial BH masses is comparable to the typical virial mass uncertainty, and so it is reasonable to assume that our sample only probes a limited dynamic range in BH mass. The BLR size of our quasars is typically below $0.1$ pc, with a median value of $\sim 0.06$ pc. The uncertainty of these BLR size estimates based on $L_{5100}$ is $<40\%$ \citep[e.g.,][]{Bentz_etal_2009}, much better than that for the BH mass estimates, and so we will use these BLR size estimates as true values. These physical properties will be used in our later analyses in \S\ref{sec:result}.

\subsection{Tests of the $\chi^2$ cross-correlation method}\label{sec:mc_sim}

To use the cross-correlation results in our statistical analyses, we must make sure that our error estimation of the velocity shift is reliable. We first test this using simple Monte Carlo experiments on mock data. 

We generate mock broad line spectra with different FWHM and S/N, sampled on the same wavelength grid as the SDSS spectra. For simplicity we assume a single-Gaussian intrinsic profile for the broad line, and we use the peak S/N (per pixel) as the indicator for the spectral S/N. We then run exactly the same $\chi^2$ cross-correlation procedure on the mock data, and derive the errors of the measured velocity shift between two epochs. 

Fig.\  \ref{fig:mc_sigV} summarizes the results. The data points show the median and standard deviations from 5000 trials. For given FWHM, the measurement uncertainty decreases with S/N roughly as $\sigma_{V_{\rm ccf}}\propto ({\rm S/N})^{-1}$. This trend breaks down at both the large FWHM/low-S/N end and the small FWHM/high-S/N end. In the former regime it is very difficult to measure the velocity shift of a noisy and broad line profile, as indicated by the large scatter in the error estimations. The latter is likely inherent to our $\chi^2$ error estimation procedure, such as the finite bin size in the CCF and the B-spline, but is not a concern since it is beyond the regime of interest in terms of S/N and line width. On the other hand, at fixed S/N, the measurement uncertainty increases as FWHM increases. This trend with FWHM is slightly slower than $\sigma_{V_{\rm ccf}}\propto {\rm FWHM}$ because we are using the peak S/N instead of the S/N integrated over the line profile. 

Overall these trends are fully consistent with the expectations of the precision in measuring the profile center with the cross-correlation method \citep[e.g.,][]{Lindegren_1978}, suggesting that our error estimation is reasonable. 

The second test is to study the distributions of measured velocity shifts for the good sample.  Fig.\ \ref{fig:test_meas_errors} shows the distributions of measured velocity shifts for pairs separated by $\Delta t<30$ and $\Delta t>30$ days in restframe (black histograms). For $\Delta t<30$ days, if we assume there is no intrinsic change in the broad line, given that this time is much shorter than the BLR dynamical timescale or possible binary period, then the expected velocity shift distribution should be consistent with zero, broadened by the error distribution. The blue dashed line in the upper panel shows the expected zero velocity shift shuffled by our measurement errors for these objects, and it is not too much different from the observed distribution. If we add an additional error component of $100\ {\rm km\,s^{-1}}$, the resulting velocity shift distribution would be the red histogram, which is inconsistent with what we observe. Therefore we are not {\em underestimating} our errors in the velocity shift by much. To quantify this statement we scale the errors up by a constant factor, and perform KS tests between the observed distribution and a simulated distribution of zero velocity shifts shuffled by the scaled-up errors. The results are shown in Fig.\ \ref{fig:test_meas_errors_KS}, where we plot the median and standard deviations from $5000$ random realizations at each error scale. We achieve the best agreement with the observed distribution if our nominal errors were scaled up by only $30\%$. 

We also verify that the measurement error in the velocity shift, $\sigma_{V_{\rm ccf}}$, does not vary with the time separation between two epochs, as also indicated by the similar distributions of the blue dashed lines in the two panels of Fig.\ \ref{fig:test_meas_errors}. Thus the observed velocity shift distribution for $\Delta t>30$ days shows excess variance compared with zeros broadened by errors. However, the errors in the measured velocity shifts are still quite large for many objects, prohibiting tight constraints on the intrinsic velocity shift distribution. Therefore in \S\ref{sec:general_pop} we use a subset of the good sample with small measurement errors to place more stringent constraints on the intrinsic distribution. 

The tests performed in this section suggest that our error estimation of the measured velocity shift is reliable, and hence we will proceed to statistical constraints and individual detections in the next section. 

\section{Results}\label{sec:result}

For the 688 pairs of observations that define our good sample, we have derived meaningful
constraints on the velocity shift between two epochs from the
cross-correlation analysis. As shown in Fig.\ \ref{fig:sig_v}, the
typical uncertainty in the measured velocity shift is $\sigma_{V_{\rm ccf}}\sim 40\ {\rm km\,s^{-1}}$. Although our spectra are generally of lower quality, this typical uncertainty
is comparable to or better than that achieved in \citet{Eracleous_etal_2012}, mainly due to
the fact that we are using a larger portion of the broad line in the
cross-correlation than in \citet{Eracleous_etal_2012}.

We have 30 detections ($\sim 2.5\sigma$, 28 unique objects) among the 688 pairs in the good sample, 2 of which
are detections with a restframe time difference of less than one month.  Fig.\ \ref{fig:hbeta_hist} shows the detection fraction with Poisson errors for objects in the
good sample, as a function of time separation. There is an apparent increase of the detection fraction as the
time baseline increases. The detection fraction is $\lesssim$ a few percent at
$\Delta t<1\,$yr, but increases to $\sim 20\%$ at $\Delta t>3\,$yr. Since
there is no dependence of the spectral quality or line width (which determine the precision of our measurements) on restframe time separation,
this increased detection rate with time is real. It indicates either that the broad lines are
more variable on longer timescales, in agreement with the general expectation of
broad line variability (as the dynamical time of the BLR is on the order of a few years, see \S\ref{sec:prelim}), or that we are detecting binary BH orbital acceleration, which leads to a velocity shift that grows linearly with time. The detection rate with a restframe time
baseline longer than a few years is consistent with the results in
\citet{Eracleous_etal_2012} and Paper II, but we caution that the statistics are poor and can still hide large differences.  

The detection rate at $\Delta t<30$ days can be used to estimate the
false-positive rate due to statistics and systematics. We only detected
2 out of $\sim 230$ pairs with such short time intervals, or $\sim 1\%$ -- this is consistent with our detection threshold of $\sim 2.5\sigma$. This should be considered as an upper limit on the false-positive rate as both of the detections could be real with large accelerations (see further discussion of these two objects in \S\ref{sec:class2}). Thus the majority of detections with longer time baselines should be genuine.

Fig.\ \ref{fig:sample_line_prop} shows the broad \hbeta\ line shape properties of the ``good'', ``superior'' (defined in \S\ref{sec:general_pop}) and the detected samples, which suggest that objects with detected broad \hbeta\ velocity shifts are not a distinct population from the parent sample. In addition, significant profile changes (in terms of FWHM and skewness) are not common for the timescales probed by our sample. 

Below we will first derive constraints on the general population with the full distribution of observed velocity shifts (mostly non-detections) in \S\ref{sec:general_pop}, and then discuss individual detections in \S\ref{sec:binary_candidate}.

\begin{figure}
 \centering
 \includegraphics[width=0.5\textwidth]{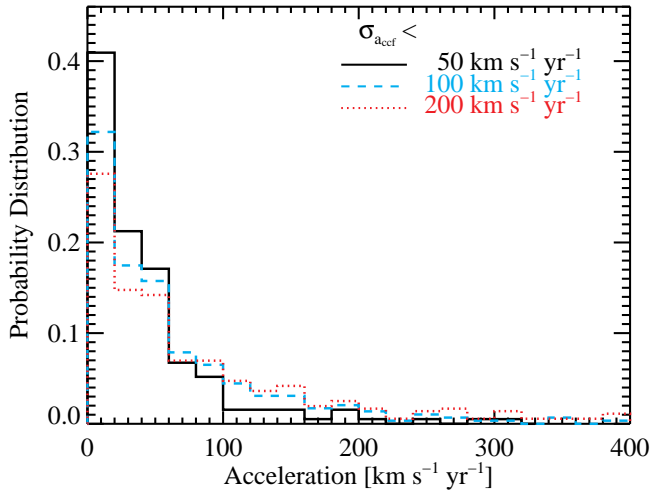}
 \caption{Distributions of the measured acceleration for subsets of the ``good'' sample with measurement errors smaller than different threshold values. The distribution gets broader when the threshold on measurement errors increases, as expected. Our ``superior'' sample defined in \S\ref{sec:general_pop} is the black histogram, with $\sigma_{a_{\rm ccf}} < 50\ {\rm km\,s^{-1}\,yr^{-1}}$. }
 \label{fig:accel_err_cut}
\end{figure}

\subsection{Constraints on the general population}\label{sec:general_pop}

%Given the constraints on the velocity shift between the two epochs, we can
%derive the limit on the acceleration of a hypnotized binary system. For
%simplicity, we use the $99\%$ CL to compute the limiting acceleration for
%objects in the good sample.
%
%\textbf{Actually, consider an ``intrinsic'' cumulative distribution of
%acceleration convolved with the error distribution, to compare with the
%observed distribution. Looks like an intrinsic Gaussian distribution with
%$\sigma=50\,{\rm km\,s^{-1}}$ convolved with errors fits the observed
%acceleration distribution better than $\sigma=0\,{\rm km\,s^{-1}}$ or
%$\sigma>100\,{\rm km\,s^{-1}}$, based on the KS probability distributions
%from random mock samples. }
%
%For objects with short time span between the two epochs, or with large
%uncertainties in the velocity shifts, the constraints on acceleration are too
%wide to be useful. Therefore, we limit ourselves to objects with stringent
%limits on acceleration. As long as these objects probe the same parameter
%space as the general population, the statistical limits we derive for the
%subsets are also applicable to the general population.
%
%We consider two  with $\Delta t>0.1\,$yr and $\sigma_v<200\,{\rm
%km\,s^{-1}}$. These criteria remove about half of the objects in the good
%sample, leaving 351 objects. The distributions of broad line properties of
%this subset, however, are indistinguishable from those for the good sample.

Most of our objects do not show detectable acceleration between the two epochs. We can use these non-detections to place constraints on a hypothetical binary population. Specifically we will compare the distribution of measured accelerations with predictions from a binary population, using the measurement error distribution in our $\chi^2$ ccf approach. In doing so, we have to assume that all the observed accelerations are either due to binary motion, or due to errors in the measurements, i.e., we neglect the complication that some apparent ``acceleration'' may be due to broad line variability. We also stick to the simplified scenario described in \S\ref{sec:prelim} that only one BH is active and dominates its BLR dynamics. We will come back to the caveats in our assumptions in \S\ref{sec:disc}. Throughout \S\ref{sec:general_pop} we adopt a constant mass for BH 1, $M_1=1.8\times 10^8\ M_\odot$, i.e., the median virial mass estimate for our sample. The reason for adopting a constant value rather using the full distribution of viral masses is because the latter is dominated by the large uncertainties associated with these virial BH masses (see discussion in \S\ref{sec:vel_shift}). But we will use the full distribution of the measured BLR sizes, since the uncertainties associated with these  estimates are substantially smaller. 

\begin{figure}
 \centering
 \includegraphics[width=0.5\textwidth]{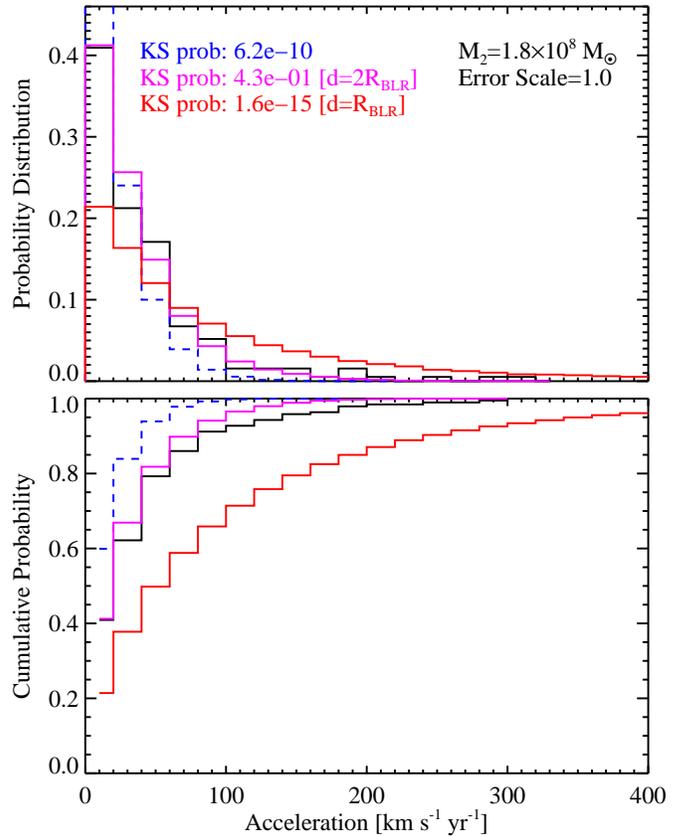}
 \caption{An example of constraining the binary population with the observed acceleration distribution of the ``superior'' sample defined in \S\ref{sec:general_pop}. Shown here is the case where the binary fraction is $100\%$ with a fixed companion BH mass $M_2=1.8\times 10^8\ M_\odot$, the median virial mass for the active BH. The
 black histogram shows the observed acceleration distribution. The blue dashed histogram shows a simulated sample of zero acceleration broadened by the measurement error distribution, which is narrower than the observed distribution, indicating non-negligible acceleration. We consider two simple cases: 1) the binary separation is twice the BLR size of the active BH; 2) the binary separation equals the BLR size. The simulated distributions (with measurement errors) of the two cases are shown in the magenta and red lines. The case with the smaller binary separation is ruled out at a high significance. {\em Top:} differential probability distributions. {\em Bottom:} cumulative probability distributions. }
 \label{fig:test_ccf_aerr}
\end{figure}

\begin{figure}
 \centering
 \includegraphics[width=0.5\textwidth]{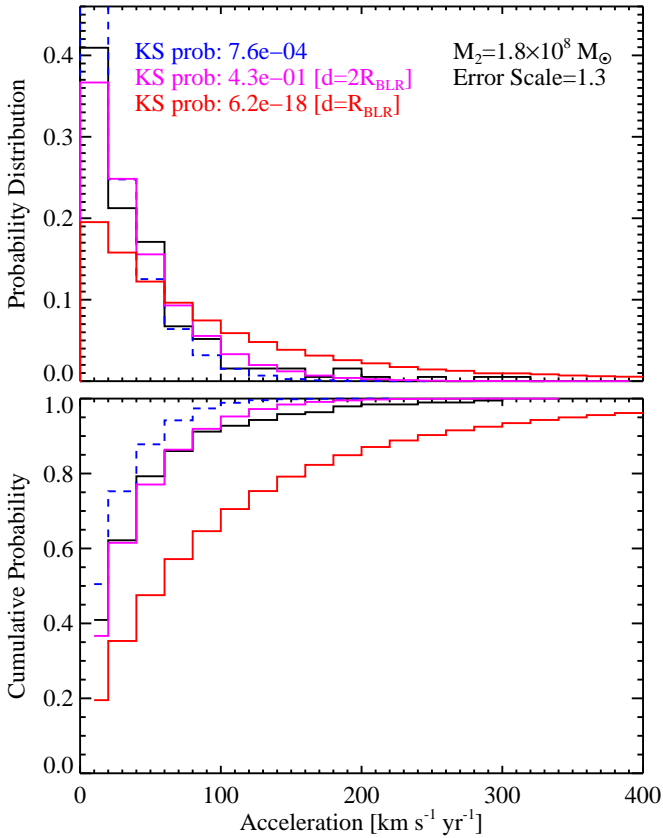}
 \caption{Same as Fig.\ \ref{fig:test_ccf_aerr}, but with the measurement errors scaled up by $30\%$. The pure error distribution is still narrower than the observed distribution, requiring an intrinsic acceleration distribution. }
 \label{fig:test_ccf_aerr2}
\end{figure}

To get the strongest constraints possible, we consider a further refinement of the ``good'' sample, with stringent limits on the precision of the measured acceleration between the two epochs, i.e., $\sigma_{a_{\rm ccf}}<50\ {\rm km\,s^{-1}\,yr^{-1}}$. Fig.\  \ref{fig:accel_err_cut} shows the acceleration distributions for subsets of the ``good'' sample with different cuts on $\sigma_{a_{\rm ccf}}$. As we increase the threshold error, the observed distribution broadens. Therefore to better constrain the intrinsic acceleration distribution, while still keeping a sufficient number of observations, we use the stringent $50\ {\rm km\,s^{-1}\,yr^{-1}}$ cut in refining the ``good'' sample. There are $193$ pairs that satisfy this criterion, and all pairs are separated by a restframe time span $>0.4$ yr. We refer to this sample as the ``superior'' sample. As shown in Fig.\ \ref{fig:Lbol_z}, the distribution of these objects in the quasar parameter space is statistically indistinguishable from the ``good'' sample, hence we are not biasing the sample by imposing this cut on $\sigma_{a_{\rm ccf}}$. We also verified that the error distribution in measured accelerations does not correlate with the quasar continuum luminosity (hence the broad line size via the $R-L$ relation) or virial BH masses, which simplifies our simulations below using the measured BLR size distribution. 

Fig.\ \ref{fig:test_ccf_aerr} demonstrates how we use the observed acceleration distribution to constrain the binary population. In this example, we fix the companion BH mass to be $M_2=1.8\times 10^8\ M_\odot$, the median virial BH mass of the active BHs in the ``good'' sample. We also consider a $100\%$ binary fraction, i.e., all the objects in our sample are in a binary system with only one BH active. To obtain the expected acceleration distribution for this hypothetical binary population, we must also assume the distribution of binary separations. We consider an ideal case where the binary separation is a fixed scaling of the BLR size of the active BH, $d=(1/f)R_{\rm BLR}$ with $f<1$. The intrinsic distribution of the radial acceleration can then be derived using Eqn.\ (\ref{eqn:v_a}) assuming random orientations and orbital phases of the binary, and using the distribution of BLR sizes measured from $L_{\rm 5100}$ using the best-fit $R-L$ relation in \citet{Bentz_etal_2009}. We then convolve the intrinsic distribution with the measurement error distribution to obtain the simulated acceleration distribution expected from a binary BH population.

\begin{figure*}
 \centering
 \includegraphics[width=0.45\textwidth]{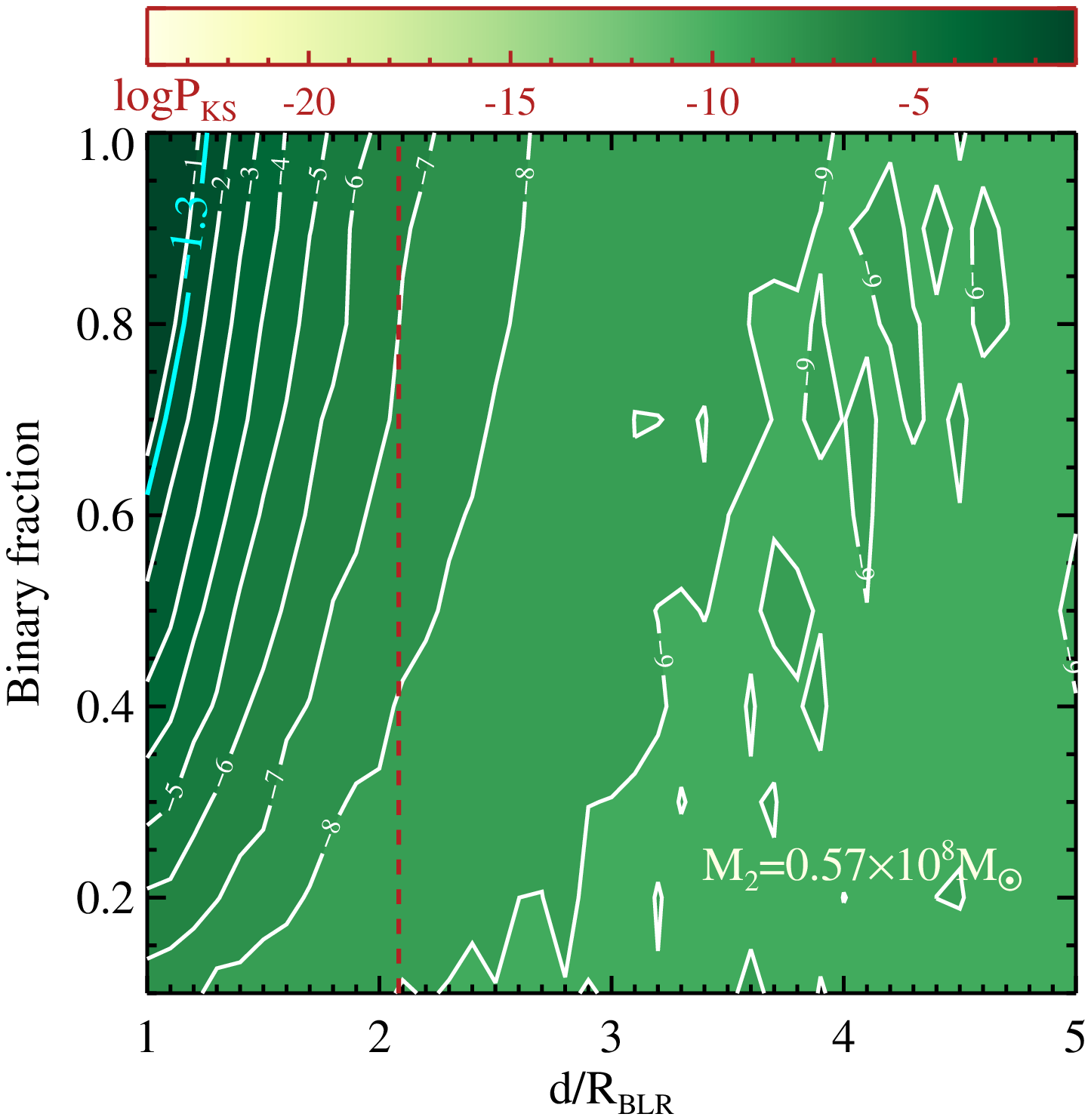}
 \includegraphics[width=0.45\textwidth]{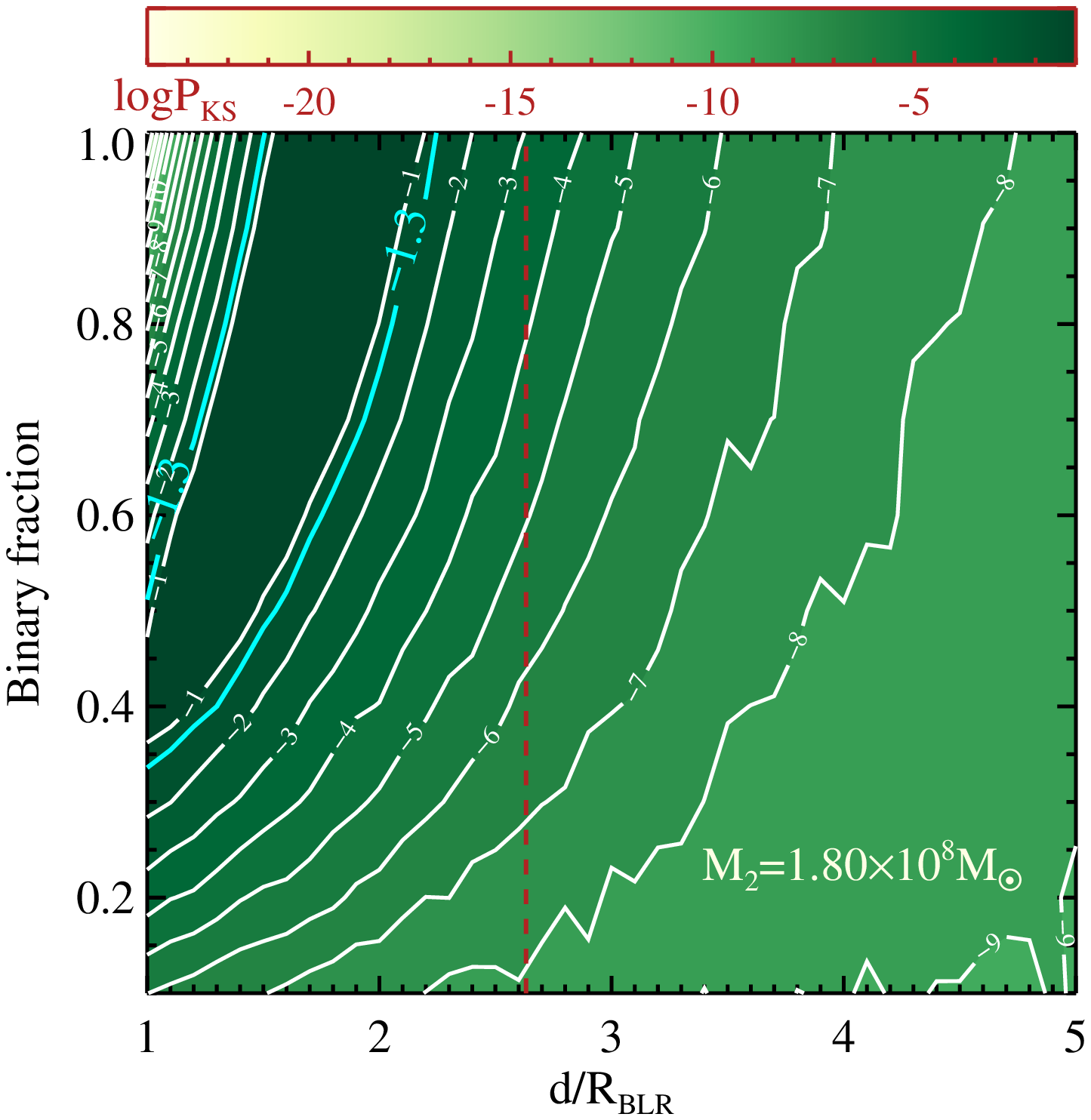}
 \includegraphics[width=0.45\textwidth]{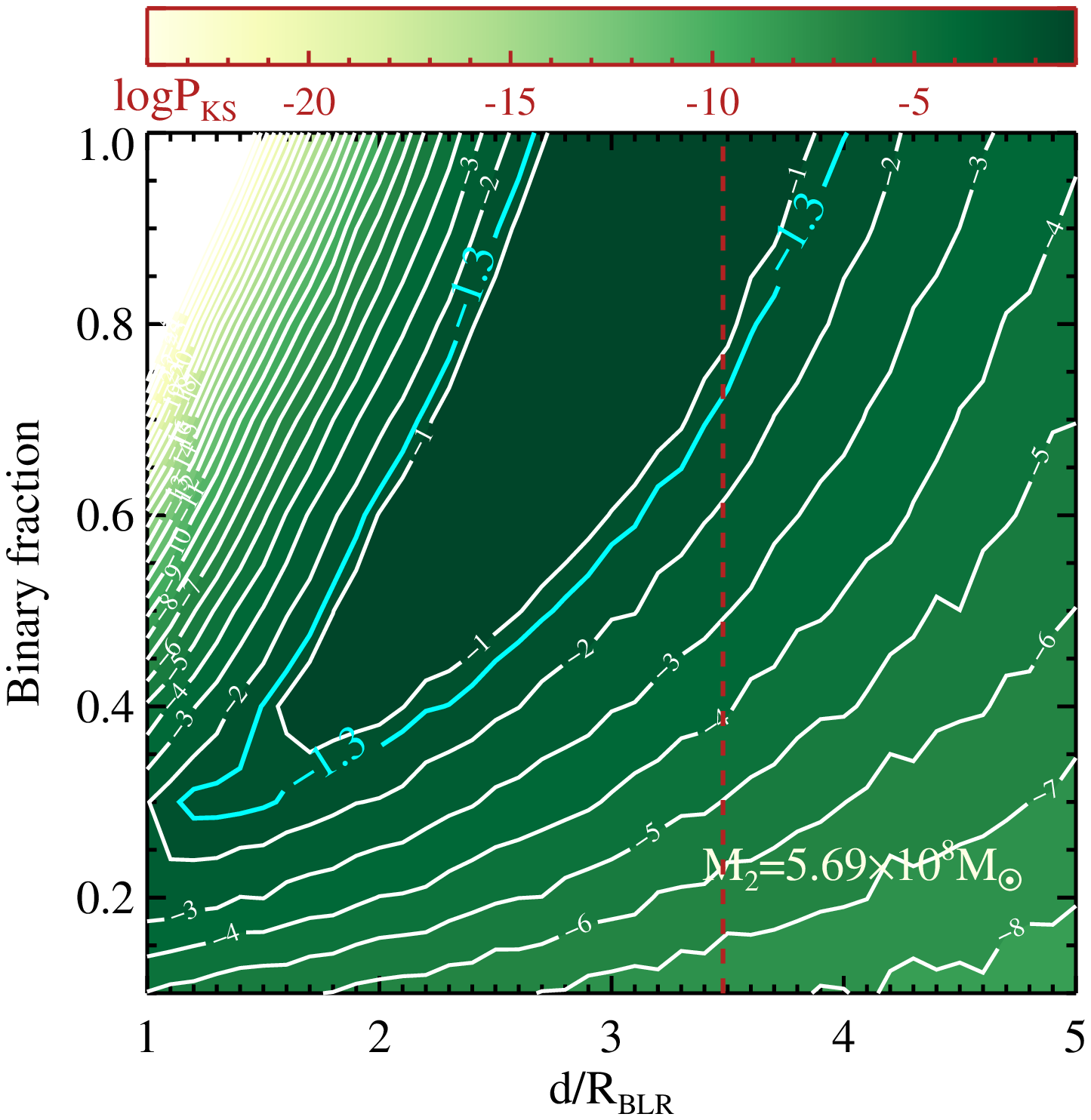}
 \includegraphics[width=0.45\textwidth]{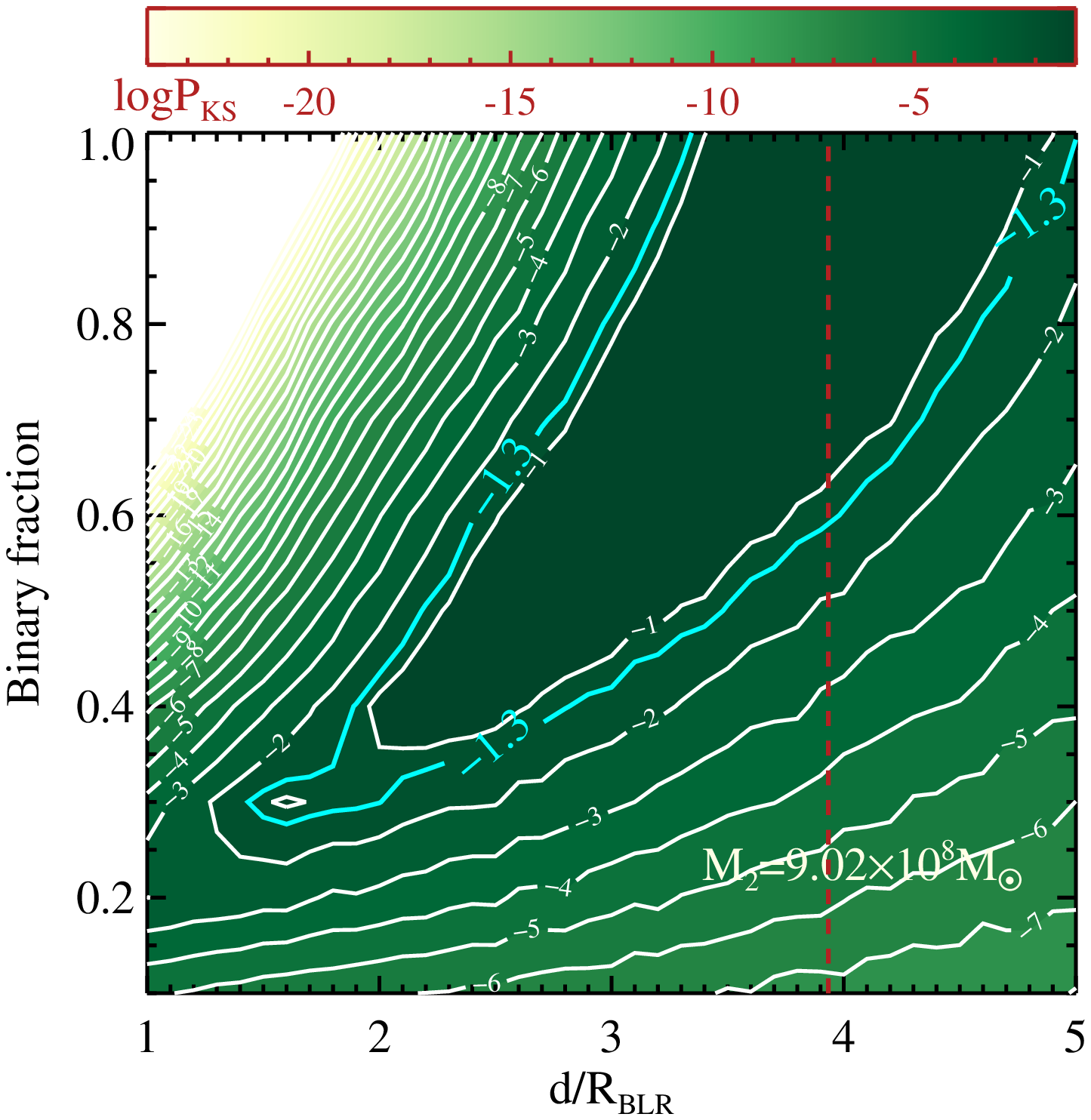}
 \caption{Parameter constraints on a hypothetical binary BH population, using the observed acceleration distribution. We have assumed a constant mass $M_1=1.8\times10^8\ M_\odot$ (see discussion in text). In each panel, the companion BH mass $M_2$ is fixed, and the contours show the logarithmic KS probability between the simulated acceleration distribution with the given binary fraction and the separation/BLR size ratio, and the observed distribution. The simulated distribution at each grid point has $10^5$ mock systems, and we have convolved the intrinsic distribution of accelerations in the hypothetical binary BH population with the measurement errors. The dashed vertical lines mark the threshold on binary separation, $d>f_r^{-1}R_{\rm BLR}$, where $f_r$ is given in Eqn.\ (\ref{eqn:roche}). The cyan lines indicate the $5\%$ probability contour. Darker regions are more probable. }
%More massive companion mass, smaller binary separation, and/or larger binary fraction are all less favored based on the observed distribution. 
\label{fig:ks_prob}
\end{figure*}

\begin{figure*}
 \centering
 \includegraphics[width=0.45\textwidth]{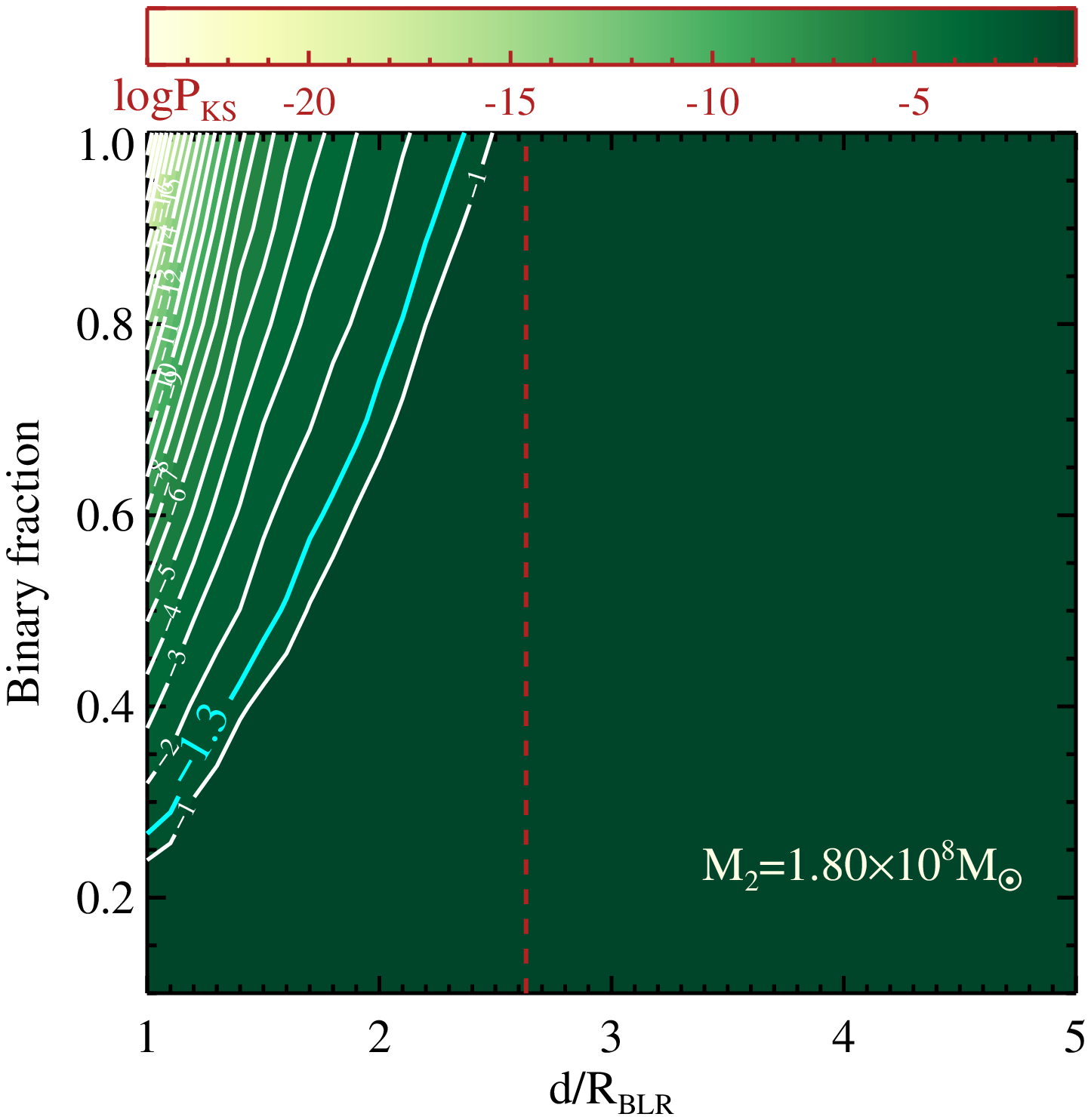}
 \includegraphics[width=0.45\textwidth]{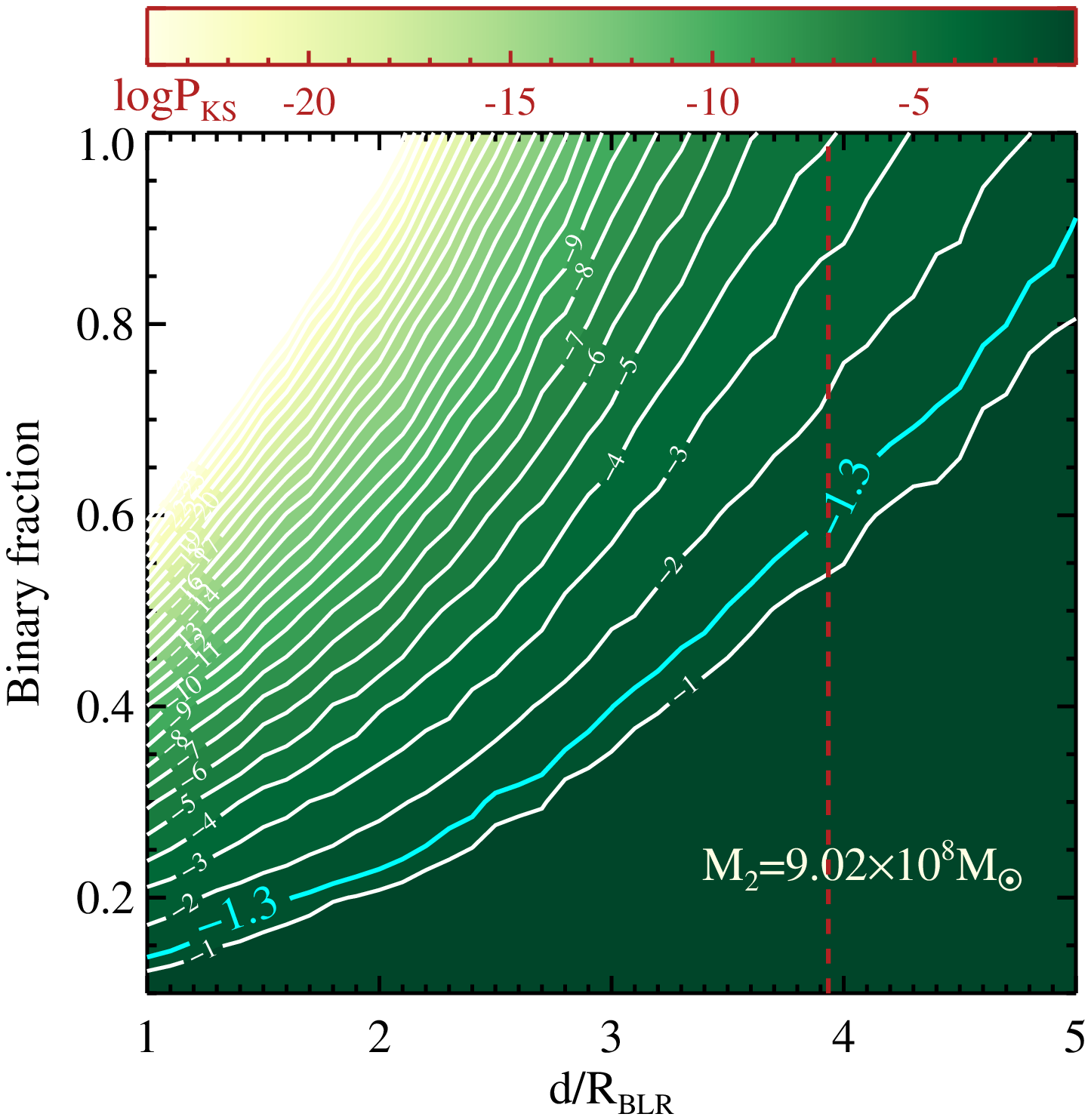}
 \caption{Same as Fig.\ \ref{fig:ks_prob}, but for the extreme case where all the variance in the observed acceleration distribution is due to broad line variability in single BHs and measurement errors. Two examples are shown here for $q\equiv M_2/M_1=1$ (left) and 5 (right), where $M_1=1.8\times 10^8\ M_\odot$ is the typical mass of the active BHs in our sample. Since all the variance has been accounted for, we no longer require a BBH population to contribute to the excess variance in the observed accelerations. As a result, there is essentially no constraint on the binary parameters when the companion BH is less massive than $M_1$ (left panel). On the other hand, for more massive companion masses, certain regions in the two-dimensional parameter space are ruled out, as the resulting accelerations will exceed the observed variance (right panel). The constraint on the exclusion of the BBH scenario becomes stronger for more massive companion BHs. }
 \label{fig:ks_prob2}
\end{figure*}

The black histogram in Fig.\ \ref{fig:test_ccf_aerr} is the observed acceleration distribution, and the blue dashed histogram is a simulated distribution of zero acceleration broadened by the error distribution. The simulated distribution of zero accelerations with errors is too narrow to account for the observed distribution, indicating non-negligible accelerations in the intrinsic distribution. The magenta and red histograms are the simulated distributions for a binary population with $d=2R_{\rm BLR}$ and $d=R_{\rm BLR}$, convolved with measurement errors. The case with $d=2R_{\rm BLR}$ produces consistent acceleration distribution with the observed one, but the case with $d=R_{\rm BLR}$ is strongly ruled out (again, assuming that {\em all} quasars are in equal-mass binaries with this separation).

As we have discussed in \S\ref{sec:mc_sim}, it is likely that our estimated measurement errors in velocity shifts are too small by $\sim 30\%$. We therefore scale our formal errors up by $30\%$ and repeat the above exercise in Fig.\ \ref{fig:test_ccf_aerr2}. Again, this test suggests there is non-negligible intrinsic acceleration, as the simulated distribution of zero accelerations is still narrower than the observed one. The $100\%$ binary scenario with $d=R_{\rm BLR}$ is even more strongly ruled out. 

%\begin{deluxetable*}{lccccccccccccccccc}
%\tablecaption{Summary of detections \label{table:detection}}
\begin{table*}
\caption{Summary of the detections}\label{table:detection}
\scalebox{0.9}{
\begin{tabular}{lccccccccccccccccc}
\hline\hline
%\tablehead{ SDSS designation & $z_{\rm sys}$ & $\log M_{\rm 1,vir}$ & $R_{\rm BLR}$ & \multicolumn{3}{c}{Epoch1} & \multicolumn{3}{c}{Epoch2} & $\Delta t$ & \multicolumn{3}{c}{$V_{\rm ccf}^*$} & \multicolumn{3}{c}{$a_{\rm ccf}^*$} & category \\
%hhmmss.ss$\pm$ddmmss.s & & $[M_\odot]$ & [pc] & plate & fiber & mjd & plate & fiber & mjd & [days] & \multicolumn{3}{c}{$[{\rm km\,s^{-1}}$]} & \multicolumn{3}{c}{$[{\rm km\,s^{-1}\,yr^{-1}}]$} & }\startdata
SDSS designation & $z_{\rm sys}$ & $\log M_{\rm 1,vir}$ & $R_{\rm BLR}$ & \multicolumn{3}{c}{Epoch 1} & \multicolumn{3}{c}{Epoch 2} & $\Delta t$ & \multicolumn{3}{c}{$V_{\rm ccf}^*$} & \multicolumn{3}{c}{$a_{\rm ccf}^*$} & category \\
hhmmss.ss$\pm$ddmmss.s & & $[M_\odot]$ & [pc] & plate & fiber & mjd & plate & fiber & mjd & [days] & \multicolumn{3}{c}{$[{\rm km\,s^{-1}}$]} & \multicolumn{3}{c}{$[{\rm km\,s^{-1}\,yr^{-1}}]$} & \\
\hline
012016.72$-$092028.8 &0.495 & 8.42 & 0.083 & 0660 &586 &52177 &2864 &621 &54467 &1532 &158 &102 &215 &37 &24 &51 &3 \\
015836.26$+$010632.0 &0.724 & 8.38 & 0.201 & 2045 &417 &53350 &2851 &417 &54459 &643 &-103 &-200 &-14 &-58 &-113 &-7 &3 \\
021225.56$+$010056.1 &0.513 & 8.70 & 0.075 & 0405 &380 &51816 &0703 &334 &52209 &259 &131 &58 &201 &184 &81 &282 &3 \\
030100.23$+$000429.3 &0.486 & 8.50 & 0.056 & 0411 &265 &51817 &0709 &477 &52205 &261 &420 &163 &545 &588 &229 &762 &1 \\
030639.57$+$000343.1 &0.108 & 7.51 & 0.030 & 0412 &400 &52258 &0412 &397 &51871 &349 &117 &43 &190 &122 &45 &199 &2 \\
\dotfill  & &  &  &   &  &  &0412 &394 &51931 &295 &89 &32 &151 &110 &40 &186 &  \\
032205.04$+$001201.4 &0.472 & 8.70 & 0.111 & 0413 &598 &51929 &0712 &431 &52199 &183 &62 &11 &117 &123 &21 &234 &3 \\
032213.89$+$005513.4 &0.185 & 8.04 & 0.068 & 0414 &341 &51901 &1181 &329 &53358 &1229 &48 &21 &78 &14 &6 &23 &1 \\
074700.19$+$285608.5 &0.257 & 7.86 & 0.029 & 1059 &327 &52618 &1059 &329 &52592 &20 &296 &151 &422 &5241 &2670 &7467 &2 \\
080811.00$+$070510.6 &0.467 & 8.25 & 0.056 & 1756 &475 &53080 &2076 &584 &53442 &246 &131 &55 &209 &194 &82 &309 &3 \\
083042.19$+$415142.8 &0.469 & 7.97 & 0.053 & 0761 &053 &52266 &0761 &079 &54524 &1537 &96 &40 &151 &22 &9 &36 &3 \\
093502.52$+$433110.6 &0.459 & 9.59 & 0.208 & 0870 &080 &52325 &0940 &321 &52670 &236 &172 &63 &287 &266 &98 &444 &3 \\
112010.15$+$184313.3 &0.768 & 8.83 & 0.118 & 2495 &144 &54175 &2872 &074 &54533 &202 &-144 &-276 &-18 &-261 &-498 &-32 &3 \\
122909.52$-$003530.0 &0.450 & 8.84 & 0.056 & 0289 &150 &51990 &2568 &032 &54153 &1491 &324 &197 &453 &79 &48 &111 &1 \\
135252.14$+$003758.6 &0.485 & 8.04 & 0.075 & 0300 &626 &51943 &0300 &621 &51666 &186 &-158 &-209 &-103 &-310 &-409 &-202 &3 \\
135829.58$+$010908.6 &0.244 & 8.39 & 0.041 & 0531 &207 &52028 &0301 &458 &51641 &311 &138 &59 &215 &162 &70 &252 &2 \\
\dotfill &  &   &  &   &  &  &0301 &457 &51942 &69 &82 &15 &148 &437 &80 &783 &  \\
140030.13$+$002208.5 &0.259 & 7.97 & 0.040 & 0301 &545 &51942 &0301 &550 &51641 &239 &-124 &-195 &-47 &-189 &-298 &-72 &3 \\
141020.57$+$364322.7 &0.450 & 8.44 & 0.053 & 1643 &359 &53143 &2931 &496 &54590 &998 &227 &142 &309 &83 &52 &113 &1 \\
153705.95$+$005522.8 &0.137 & 7.63 & 0.038 & 2955 &289 &54562 &0315 &419 &51663 &2550 &110 &43 &181 &15 &6 &26 &1 \\
154656.62$+$005719.6 &0.211 & 7.37 & 0.029 & 2955 &038 &54562 &0342 &454 &51691 &2370 &151 &47 &261 &23 &7 &40 &3 \\
155053.16$+$052112.1 &0.110 & 8.96 & 0.039 & 1822 &308 &53172 &2951 &326 &54592 &1278 &-1034 &-1183 &-882 &-295 &-338 &-251 &1 \\
165638.86$+$362121.0 &0.536 & 7.97 & 0.047 & 0819 &618 &52409 &0820 &490 &52438 &18 &-144 &-231 &-56 &-2803 &-4482 &-1100 &2 \\
224623.54$+$130335.9 &0.521 & 8.95 & 0.090 & 0740 &208 &52263 &1893 &018 &53239 &641 &-269 &-368 &-166 &-153 &-210 &-94 &3 \\
230946.14$+$000048.8 &0.352 & 8.44 & 0.046 & 0381 &485 &51811 &0678 &555 &52884 &793 &138 &40 &226 &63 &18 &104 &3 \\
232525.32$+$000352.1 &0.338 & 8.51 & 0.045 & 0383 &518 &51818 &0681 &355 &52199 &284 &-186 &-288 &-77 &-239 &-369 &-99 &3 \\
234018.85$-$011027.2 &0.552 & 7.73 & 0.058 & 0385 &207 &51877 &0682 &053 &52525 &417 &82 &15 &152 &72 &13 &133 &3 \\
234145.51$-$004640.5 &0.524 & 8.99 & 0.095 & 0385 &139 &51877 &0682 &020 &52525 &425 &248 &160 &326 &213 &138 &280 &3 \\
234932.77$-$003645.8 &0.279 & 8.26 & 0.068 & 0386 &215 &51788 &0684 &261 &52523 &574 &-68 &-118 &-17 &-43 &-75 &-11 &1 \\
235545.51$+$004923.1 &0.557 & 8.68 & 0.065 & 0387 &324 &51791 &0685 &358 &52203 &264 &-275 &-422 &-128 &-381 &-583 &-177 &3 \\
\hline\\
\end{tabular}
}
%\caption{$^*$Velocity shift and acceleration are calculated based on the broad \hbeta\ line, and the first column is the value corresponding to the minimum $\chi^2$ and the last two columns enclose the $2.5\sigma$ confidence range ($\Delta\chi^2=6.63$). We classify the detections into three categories (see \S\ref{sec:binary_candidate}): 1) binary candidates; 2) broad line variability; 3) ambiguous cases. }\label{table:detection}
\begin{tablenotes}
      \small
      \item NOTE. --- $^*$Velocity shift and acceleration are calculated based on the broad \hbeta\ line, and the first column is the value corresponding to the minimum $\chi^2$ and the last two columns enclose the $2.5\sigma$ confidence range ($\Delta\chi^2=6.63$). We classify the detections into three categories (see \S\ref{sec:binary_candidate}): 1) binary candidates; 2) broad line variability; 3) ambiguous cases. 
\end{tablenotes}
\end{table*}
%\enddata
%\tablecomments{$^*$Velocity shift and acceleration are calculated based on the broad \hbeta\ line, and the first column is the value corresponding to the minimum $\chi^2$ and the last two columns enclose the $2.5\sigma$ confidence range ($\Delta\chi^2=6.63$). We classify the detections into three categories (see \S\ref{sec:binary_candidate}): 1) binary candidates; 2) broad line variability; 3) ambiguous cases. }
%\end{deluxetable*}

\begin{figure*}
 \centering
 \includegraphics[width=0.45\textwidth]{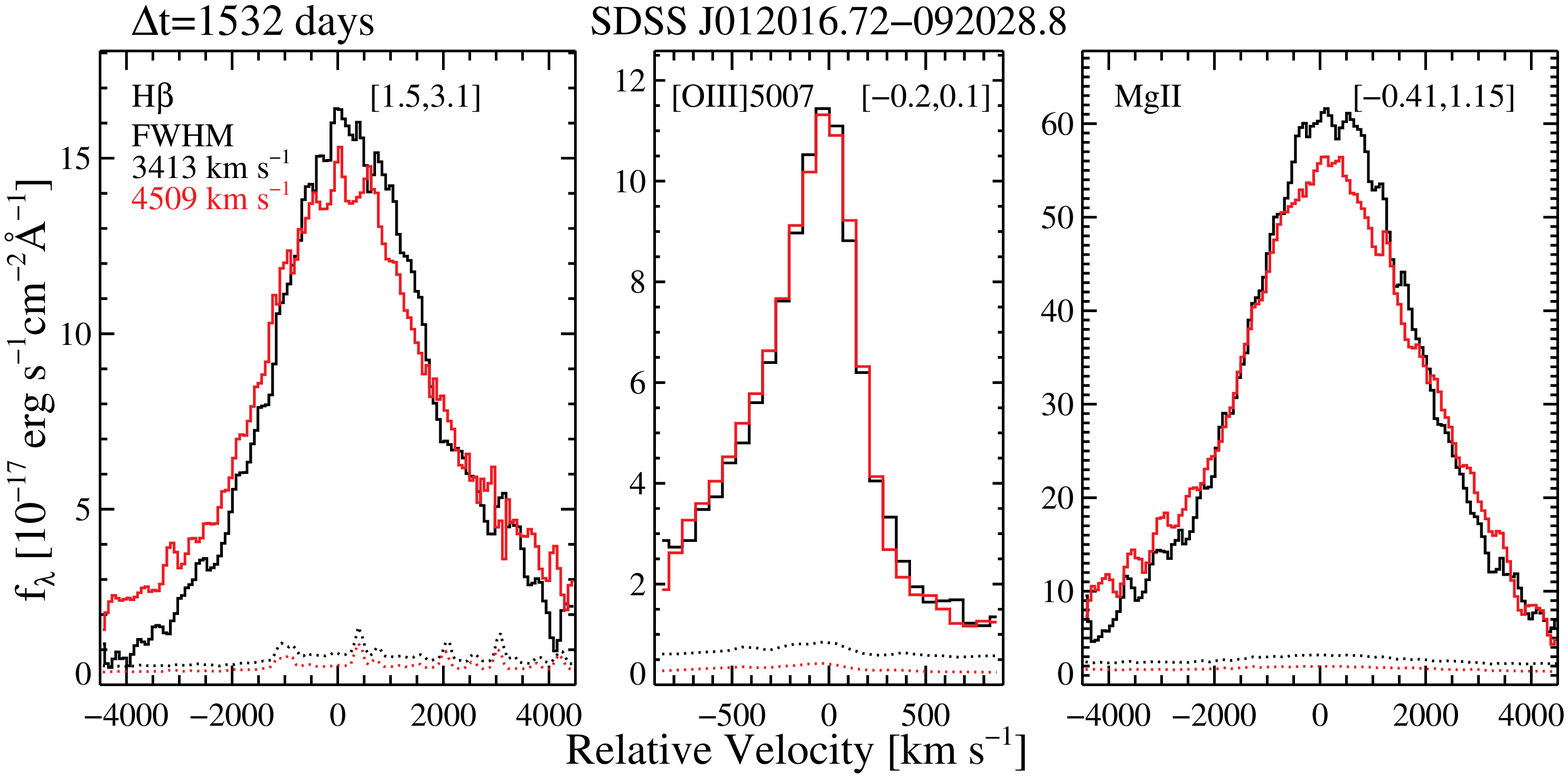}
 \includegraphics[width=0.45\textwidth]{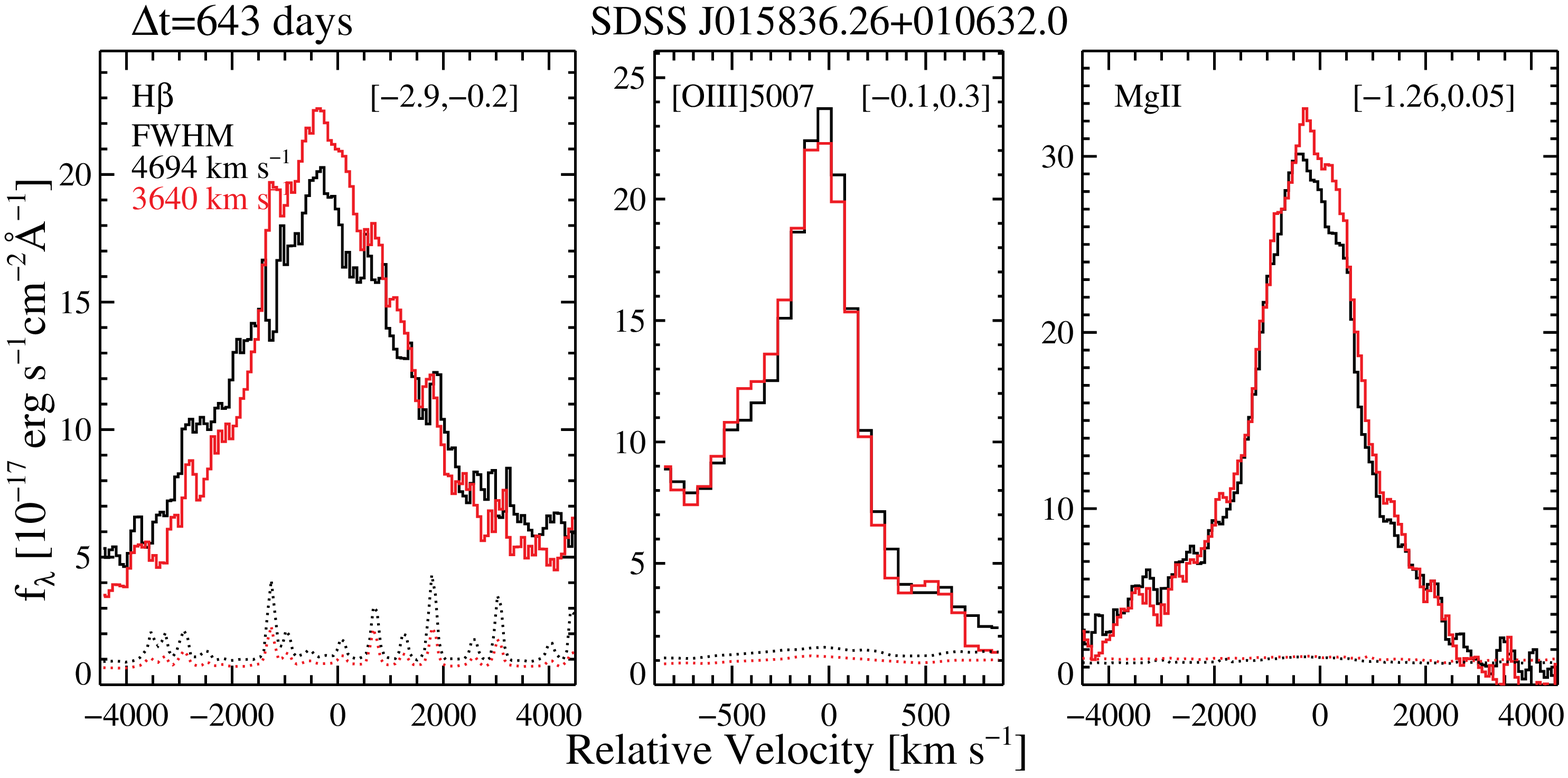}
 \includegraphics[width=0.45\textwidth]{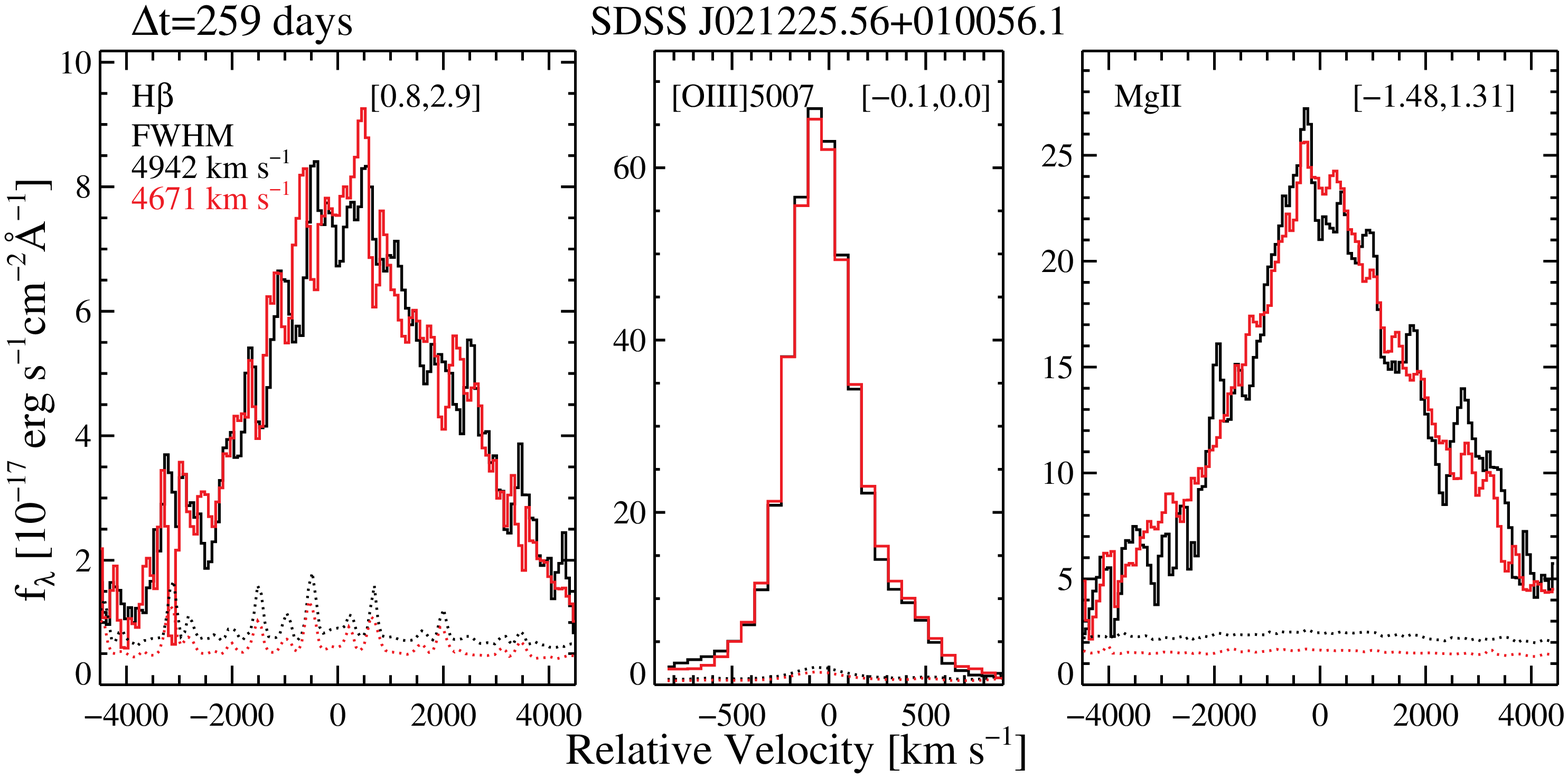}
 \includegraphics[width=0.45\textwidth]{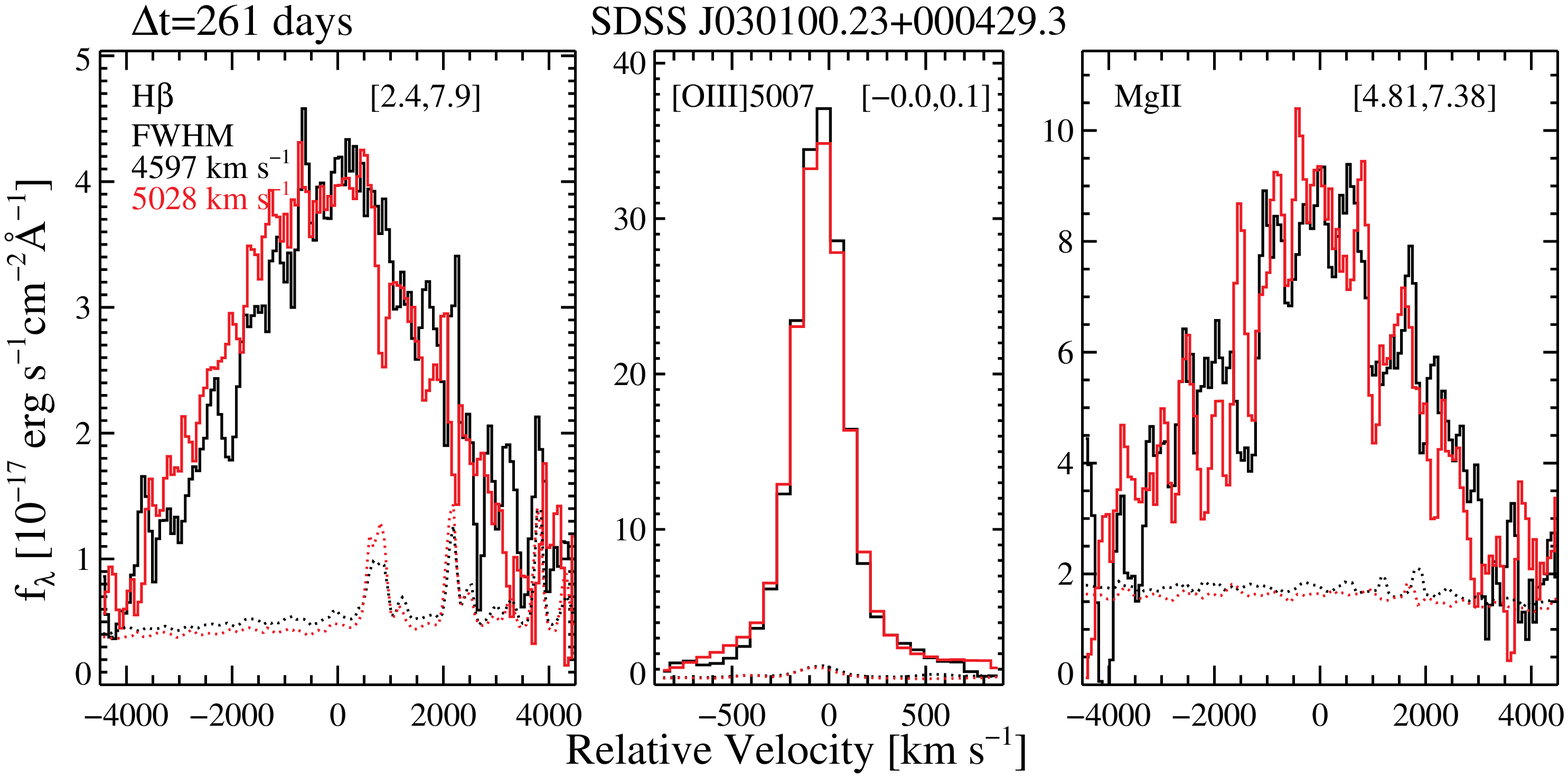}
 \includegraphics[width=0.45\textwidth]{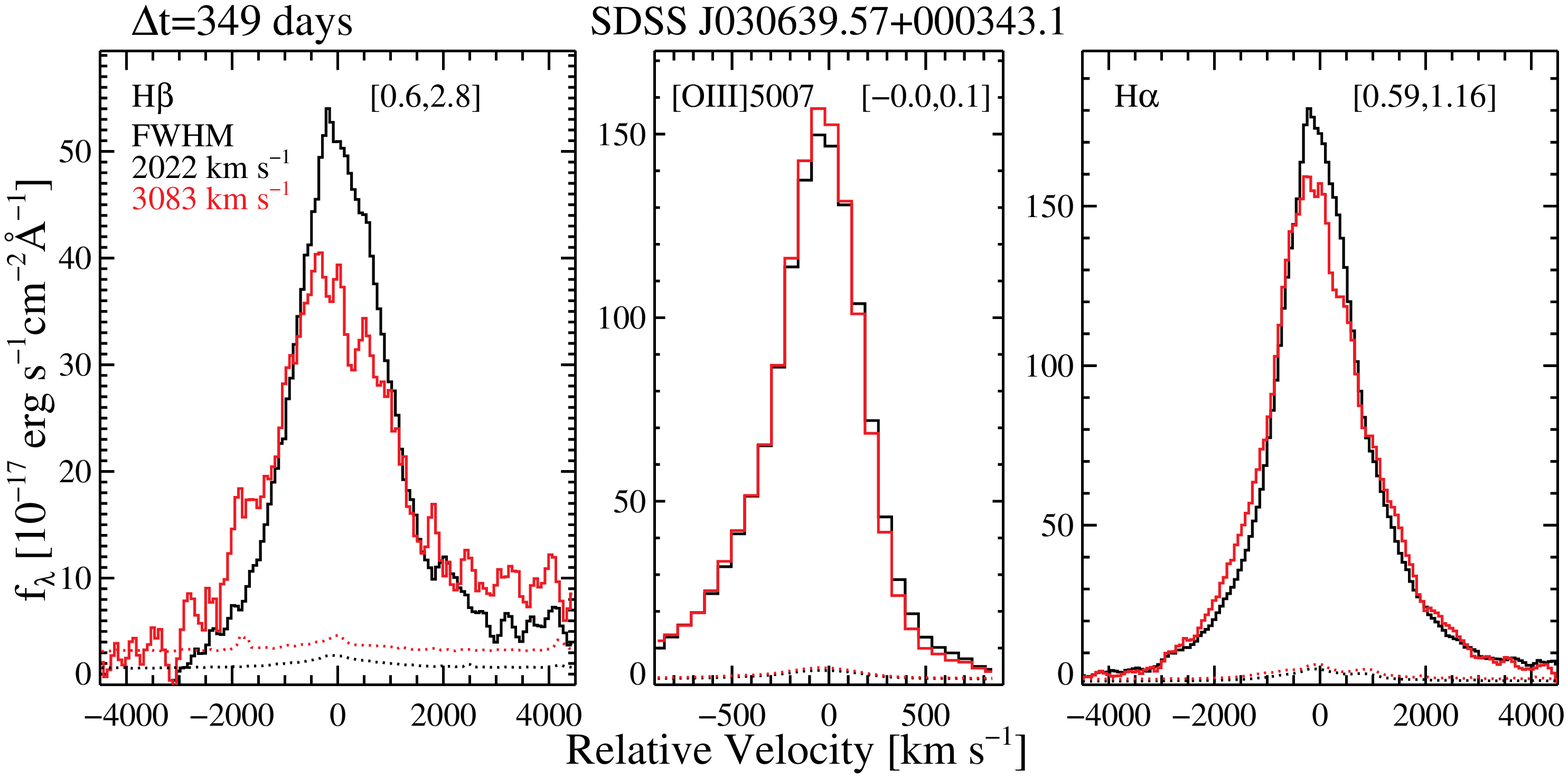}
 \includegraphics[width=0.45\textwidth]{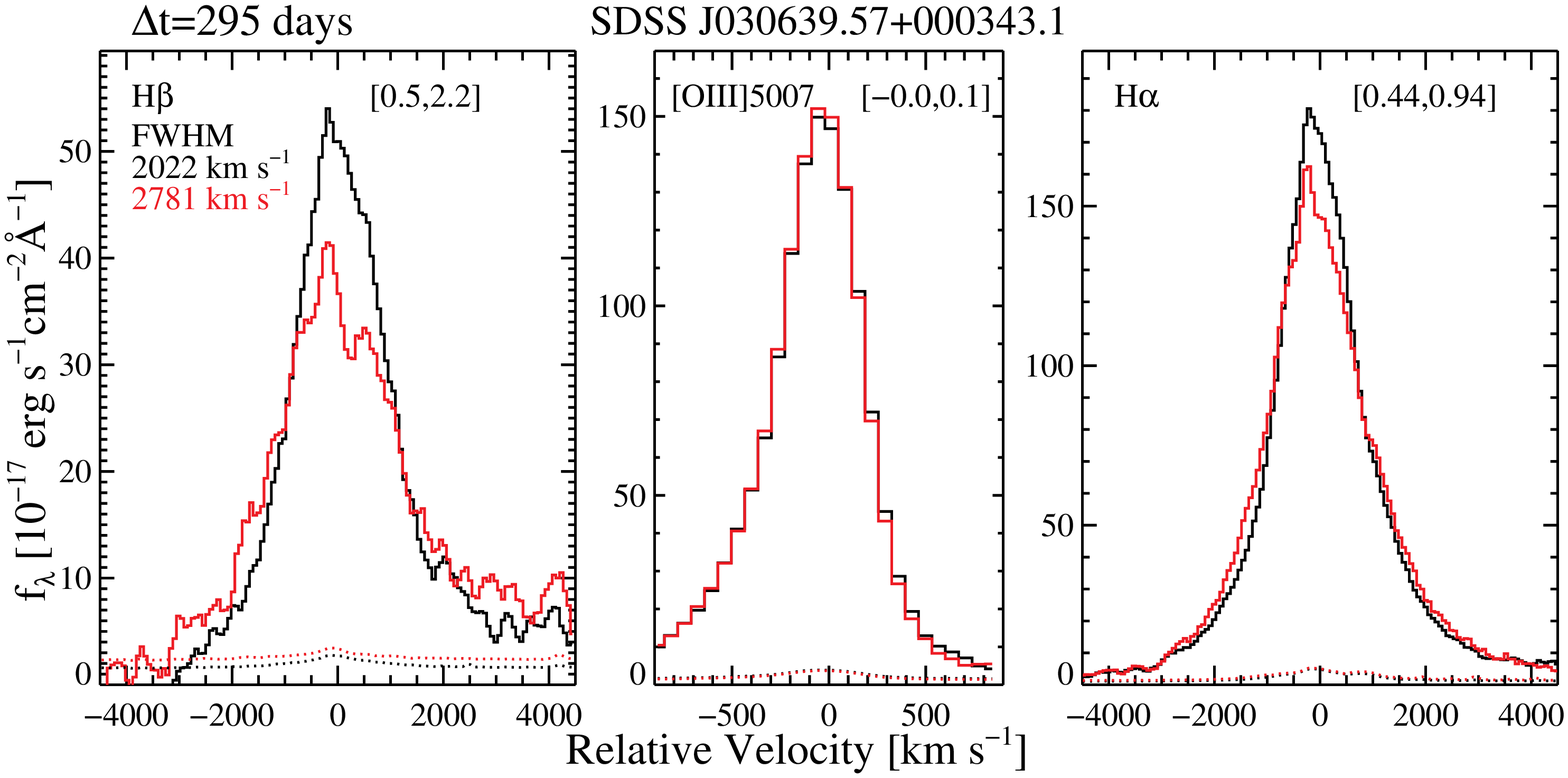}
 \includegraphics[width=0.45\textwidth]{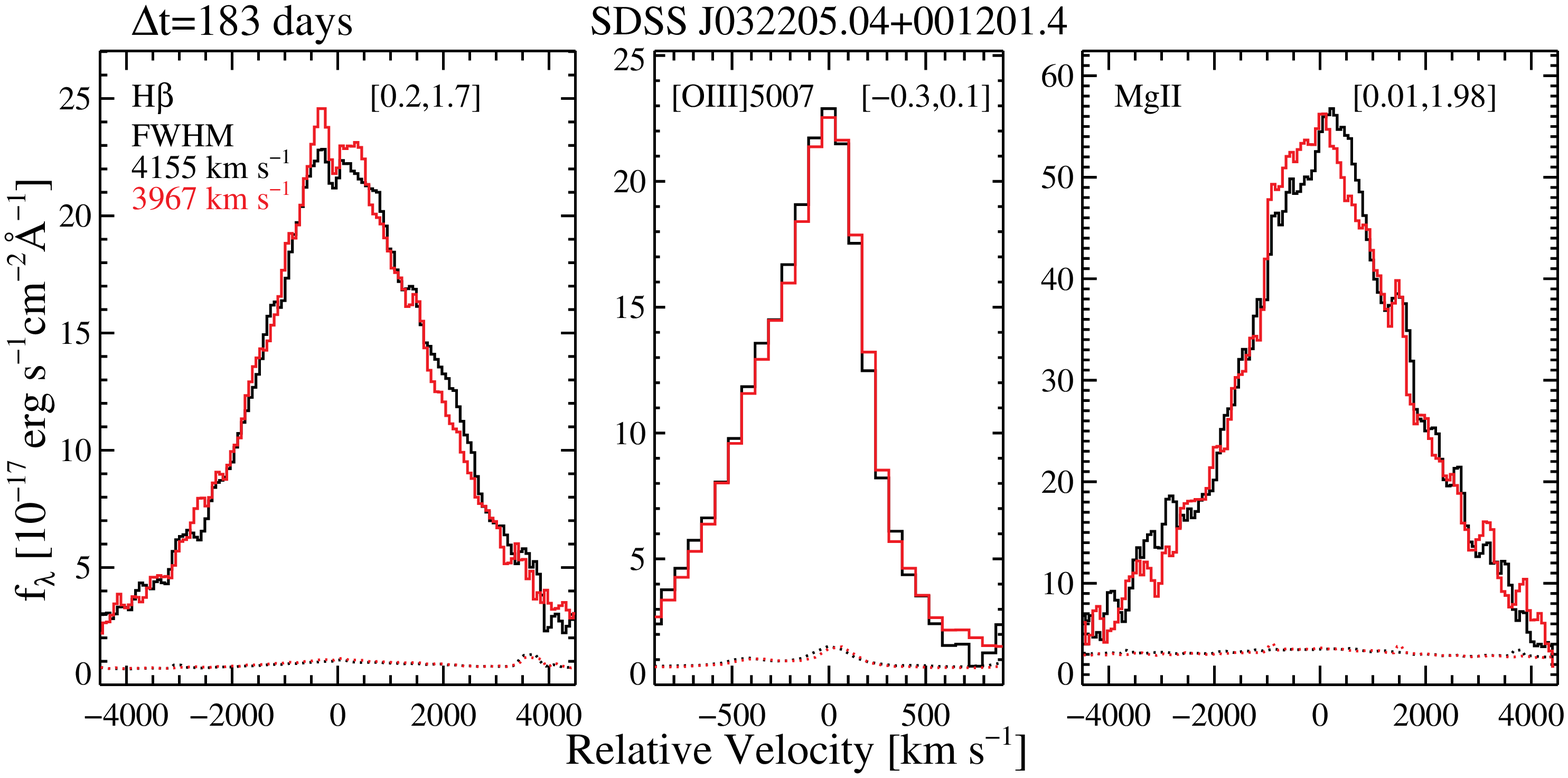}
 \includegraphics[width=0.45\textwidth]{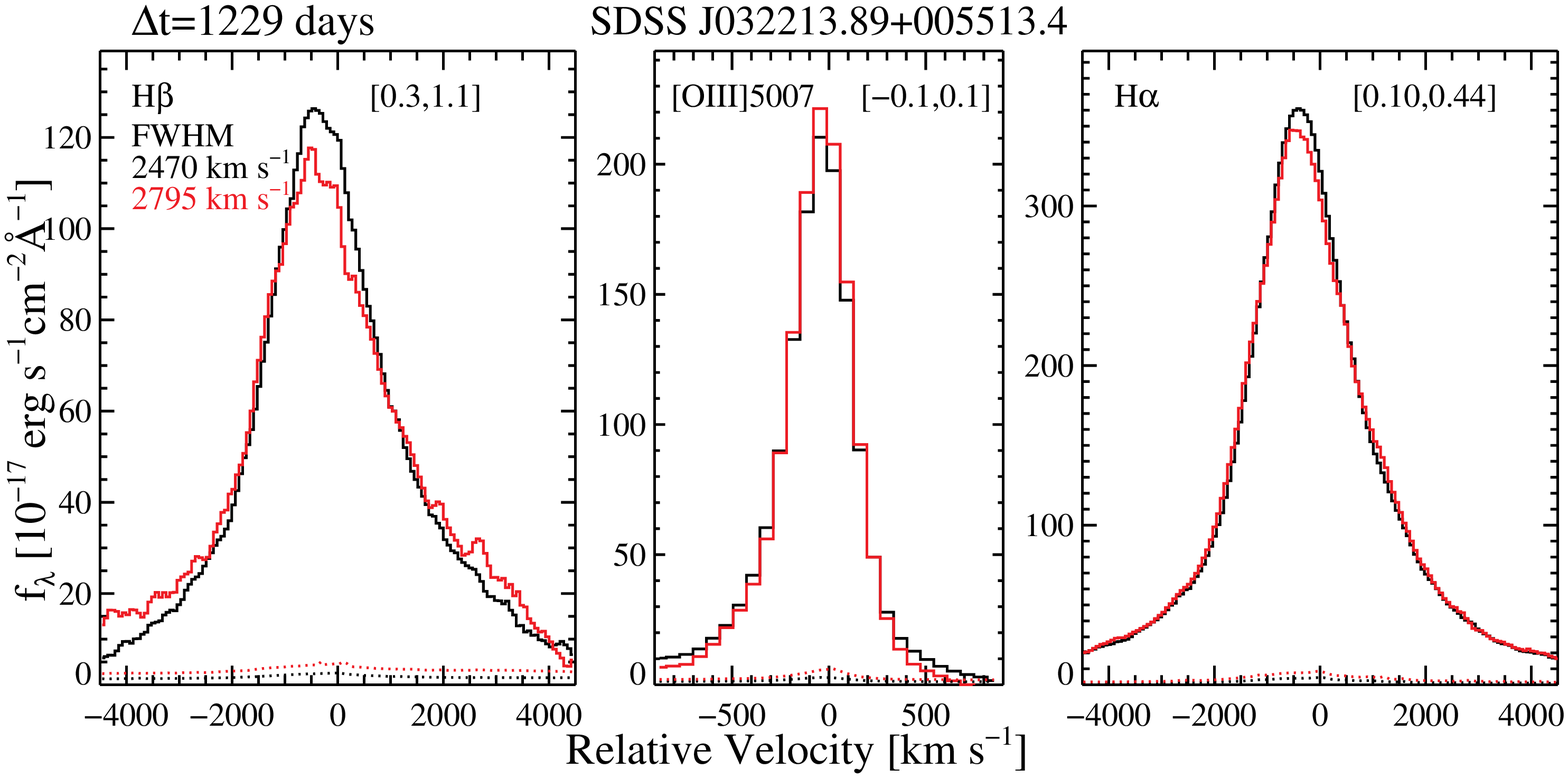}
 \includegraphics[width=0.45\textwidth]{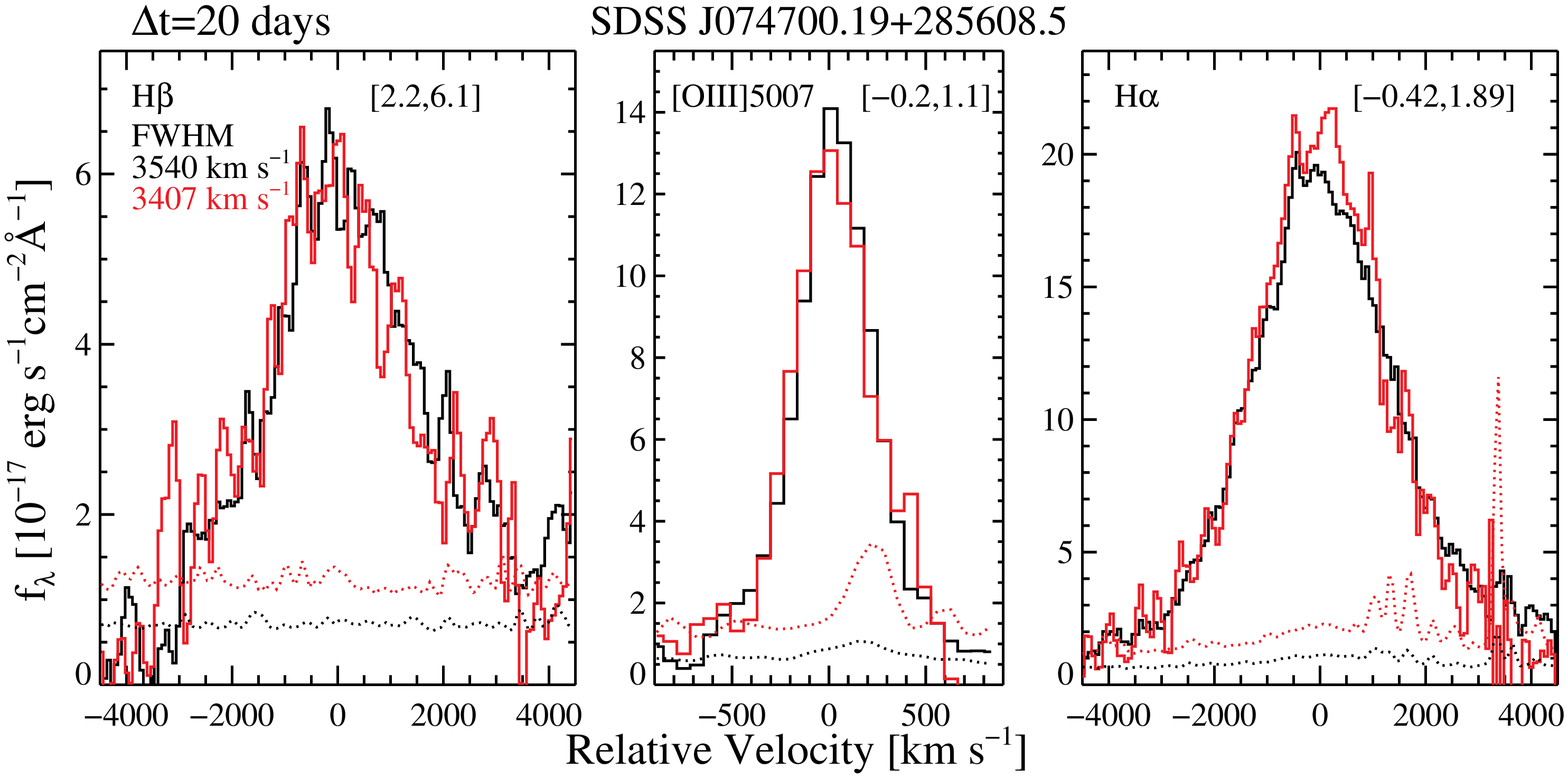}
 \includegraphics[width=0.45\textwidth]{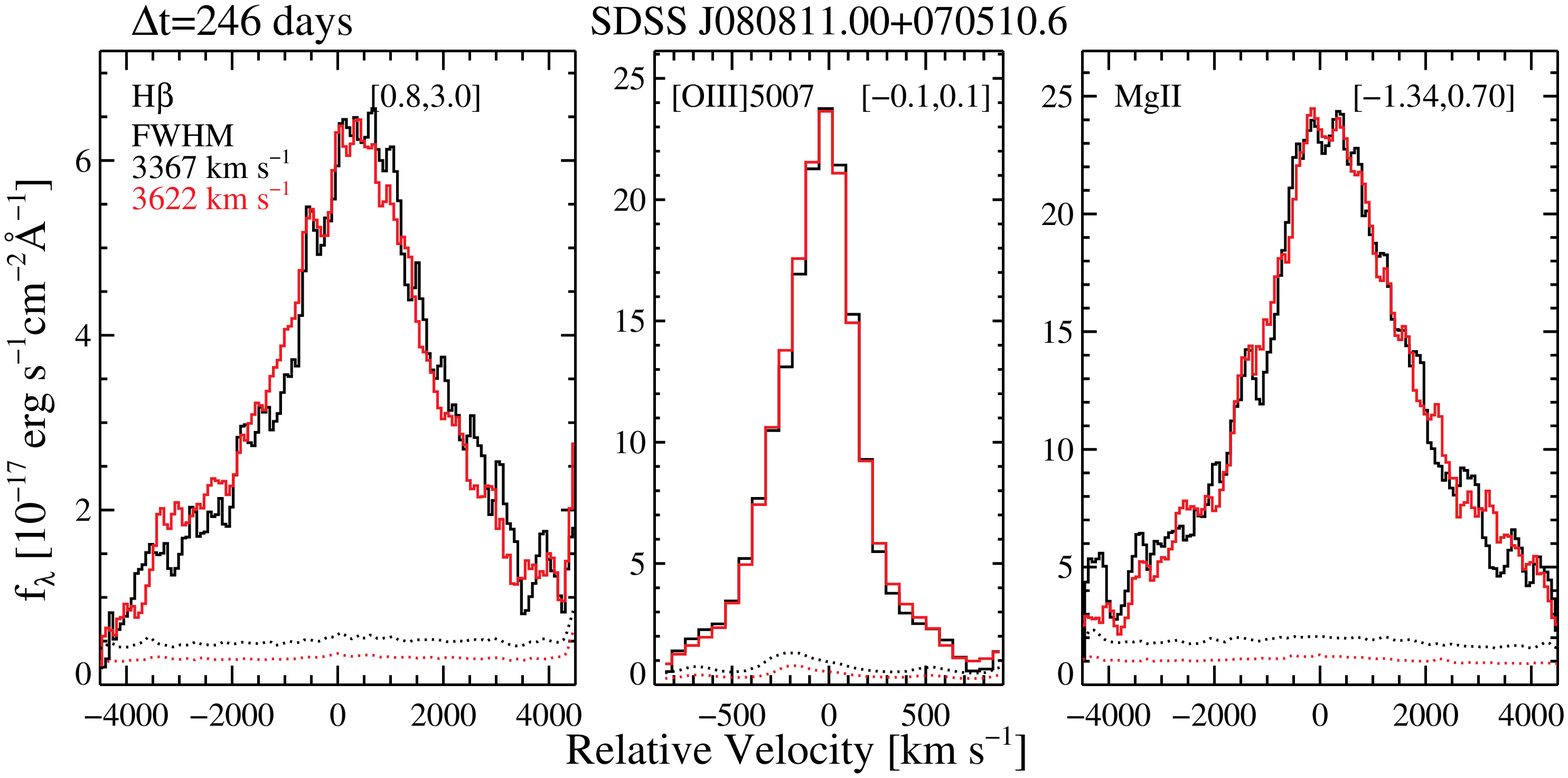}
 \caption{Line spectra at both epochs for the 30 detections in our good sample. Black is for Epoch 1 and red is for Epoch 2. The dotted lines are the corresponding flux density errors. We show the broad \hbeta\ and \OIIIb\ lines in velocity space relative to the systemic velocity. We also show the broad \MgII\ or \halpha\ line, if covered in the spectra and detected at $>6\sigma$ significance. The measured velocity shift in units of pixels (1 pixel$=69\,{\rm km\,s^{-1}}$) is marked in each sub-panel, which encloses the 2.5$\sigma$ confidence range in our $\chi^2$ cross-correlation approach. Discussions of individual detections are presented in \S\ref{sec:binary_candidate} and summarized in Table \ref{table:detection}.
 }
 \label{fig:det_obj1}
\end{figure*}

\begin{figure*}
 \centering
 \includegraphics[width=0.45\textwidth]{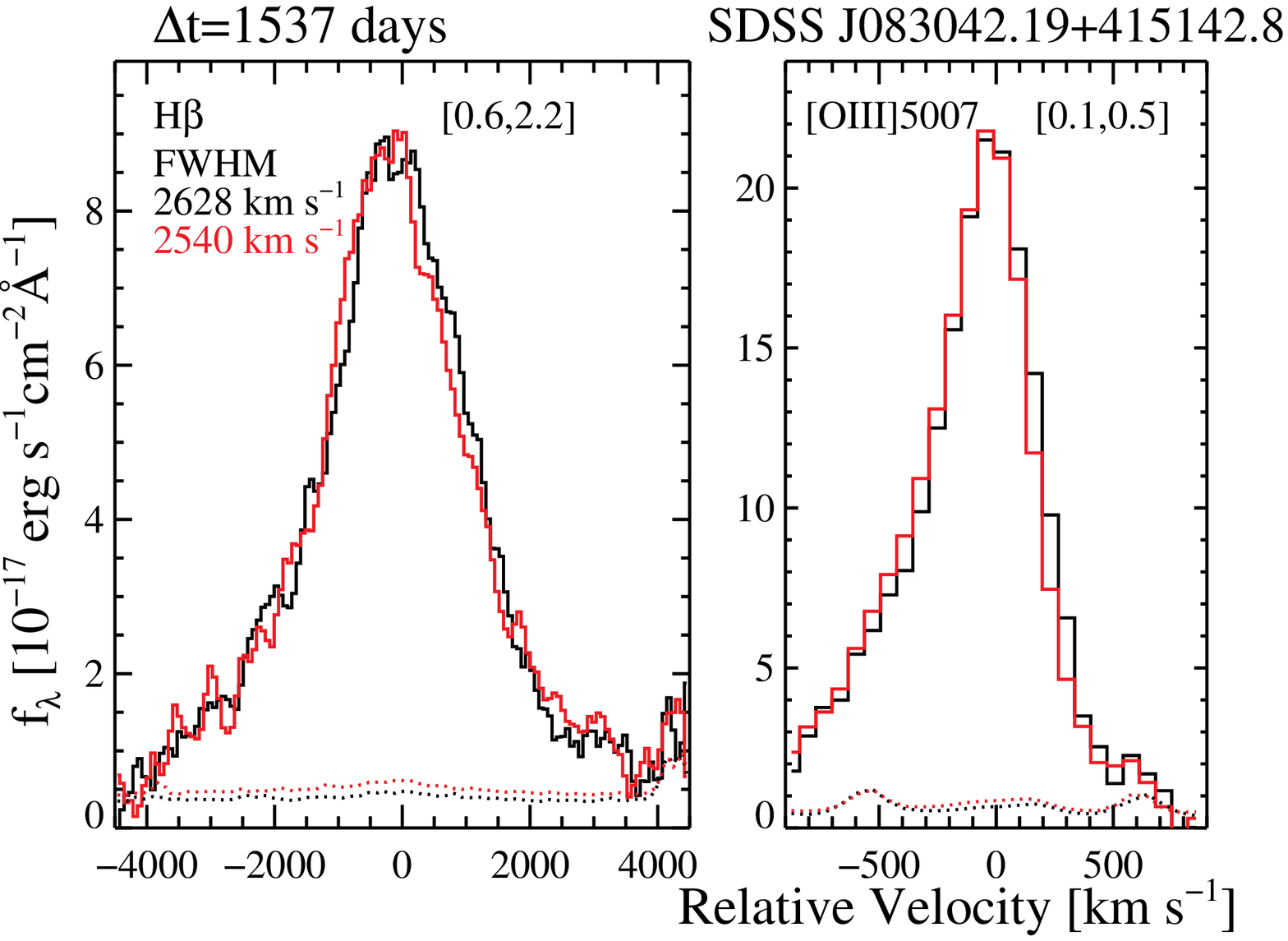}
 \includegraphics[width=0.45\textwidth]{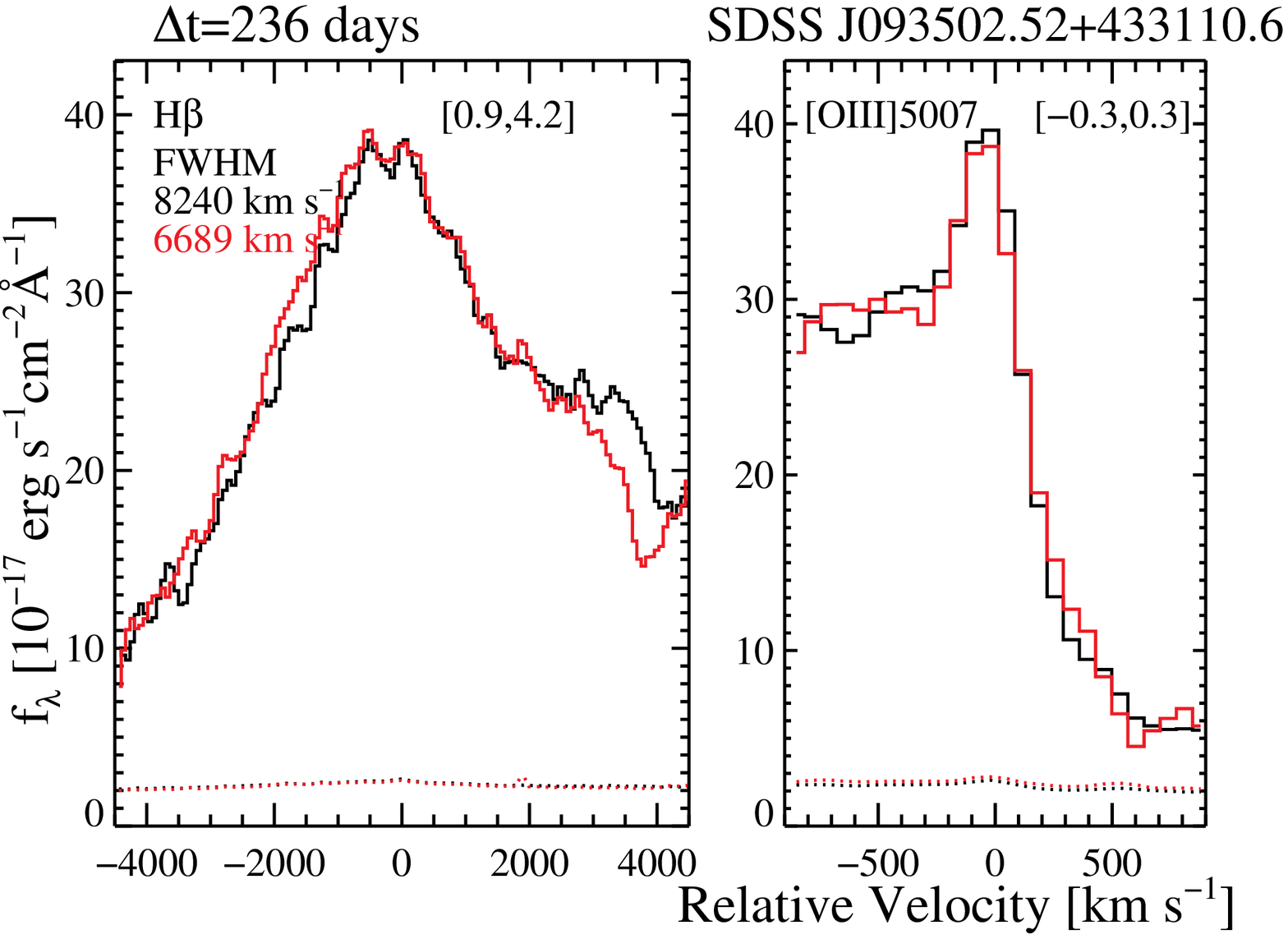}
 \includegraphics[width=0.45\textwidth]{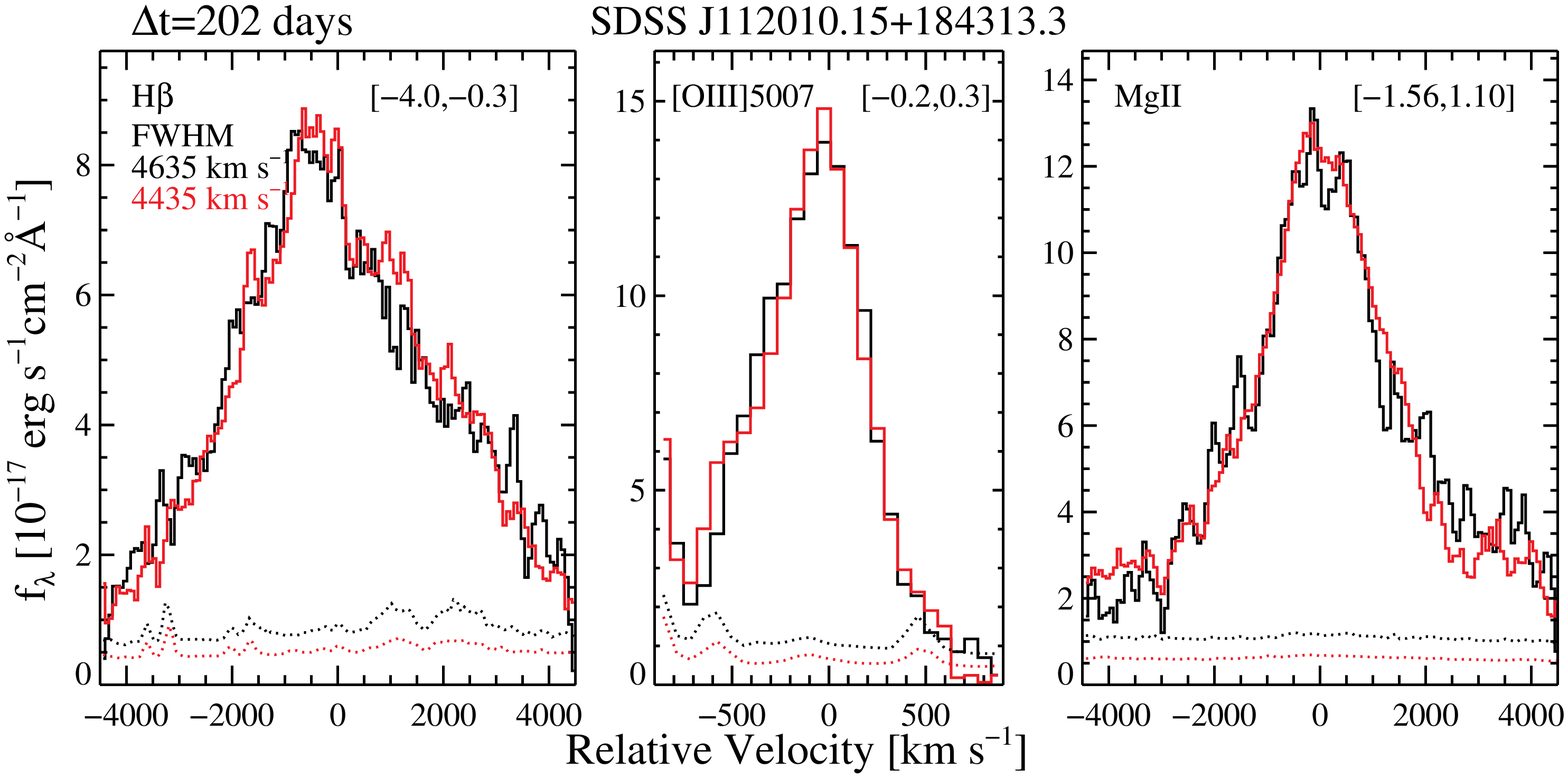}
 \includegraphics[width=0.45\textwidth]{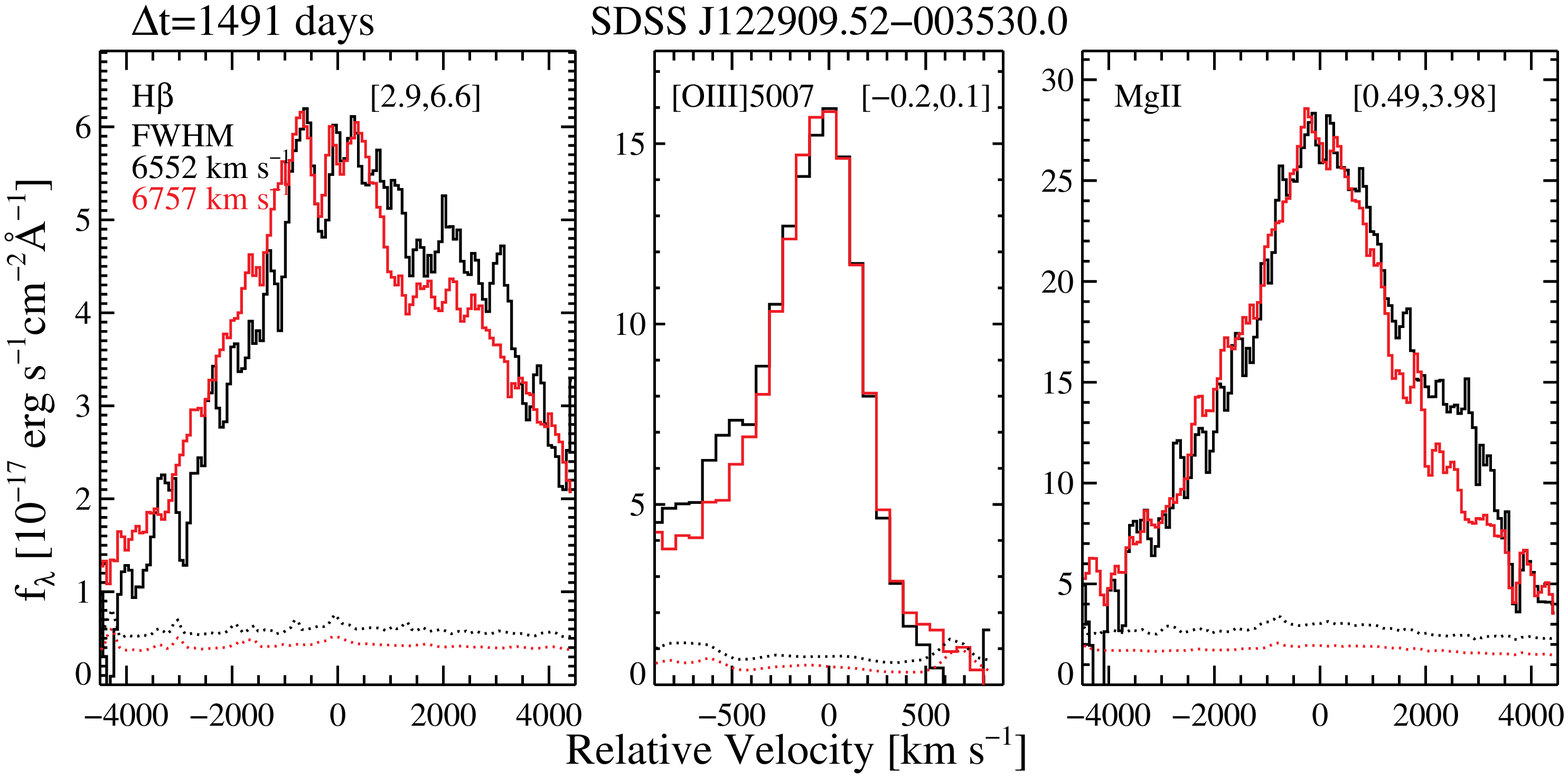}
 \includegraphics[width=0.45\textwidth]{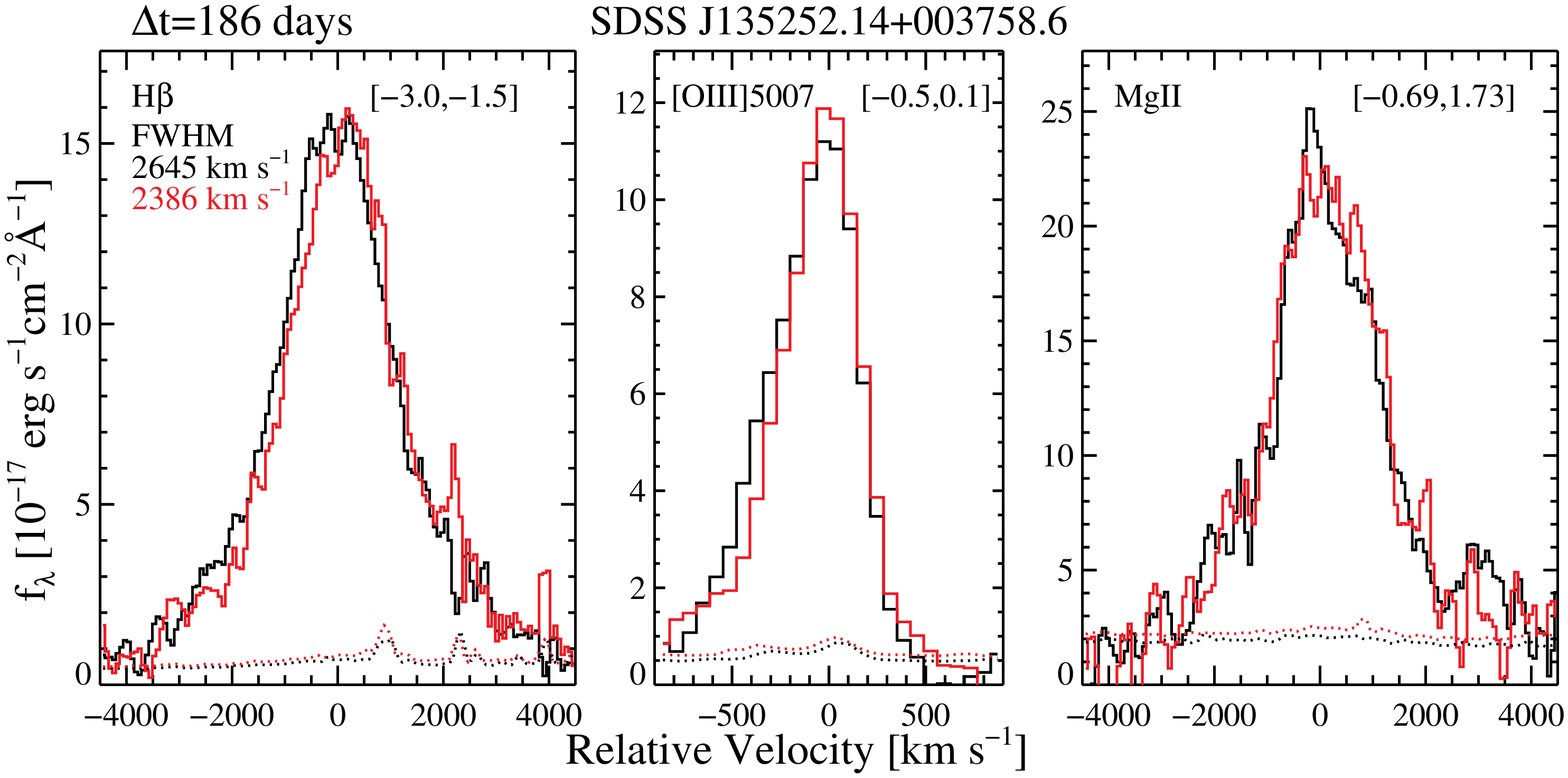}
 \includegraphics[width=0.45\textwidth]{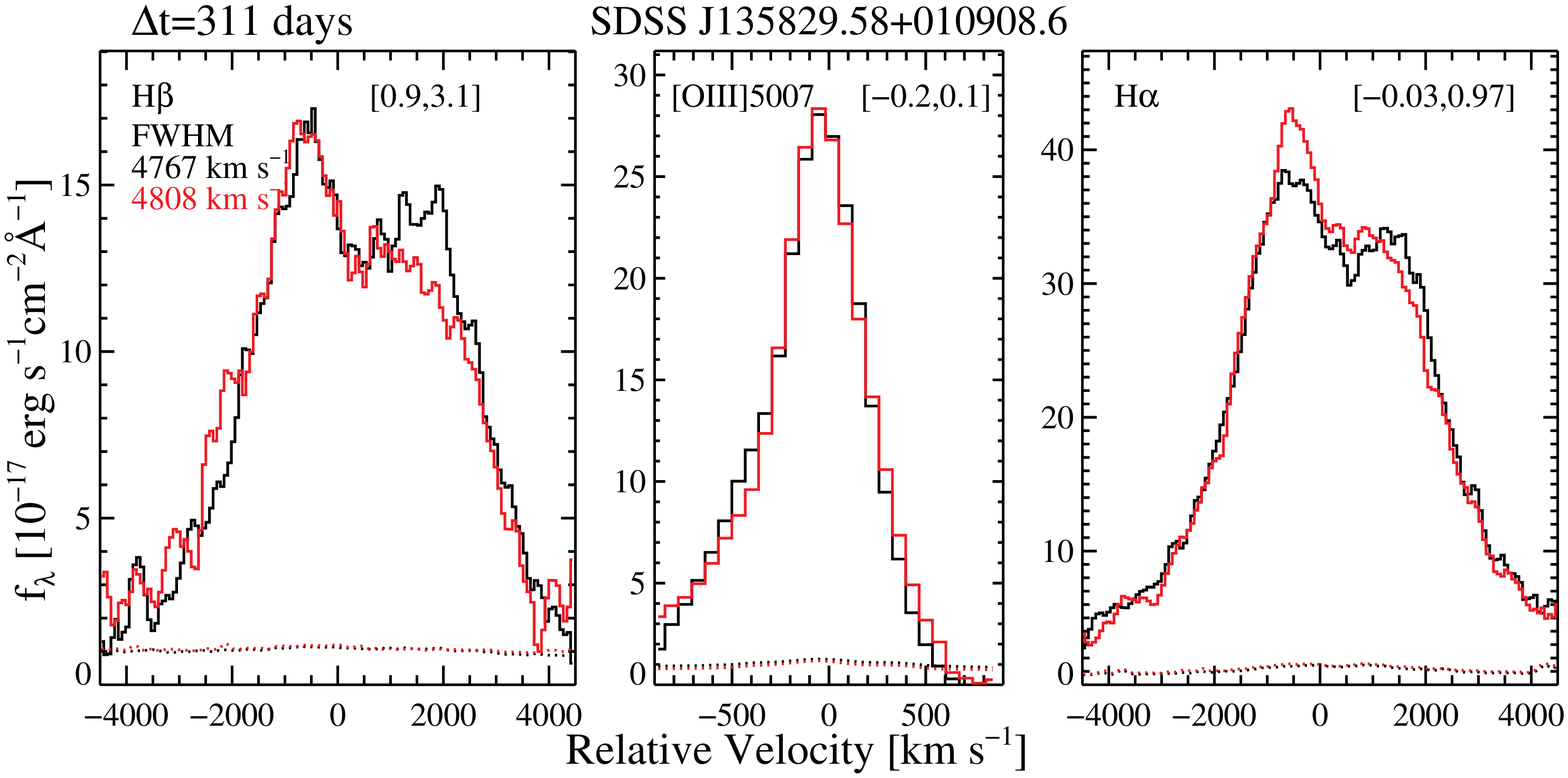}
 \includegraphics[width=0.45\textwidth]{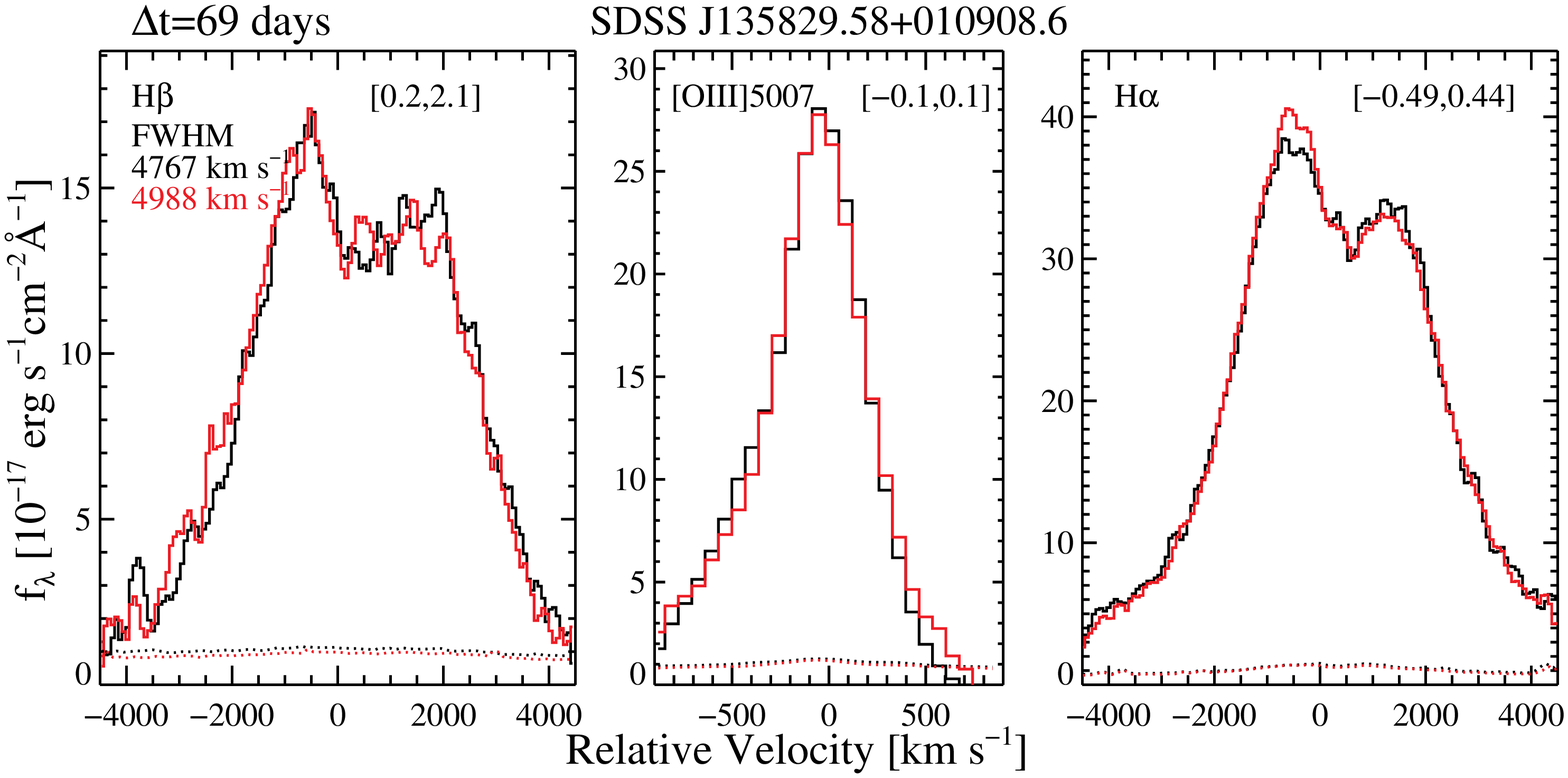}
 \includegraphics[width=0.45\textwidth]{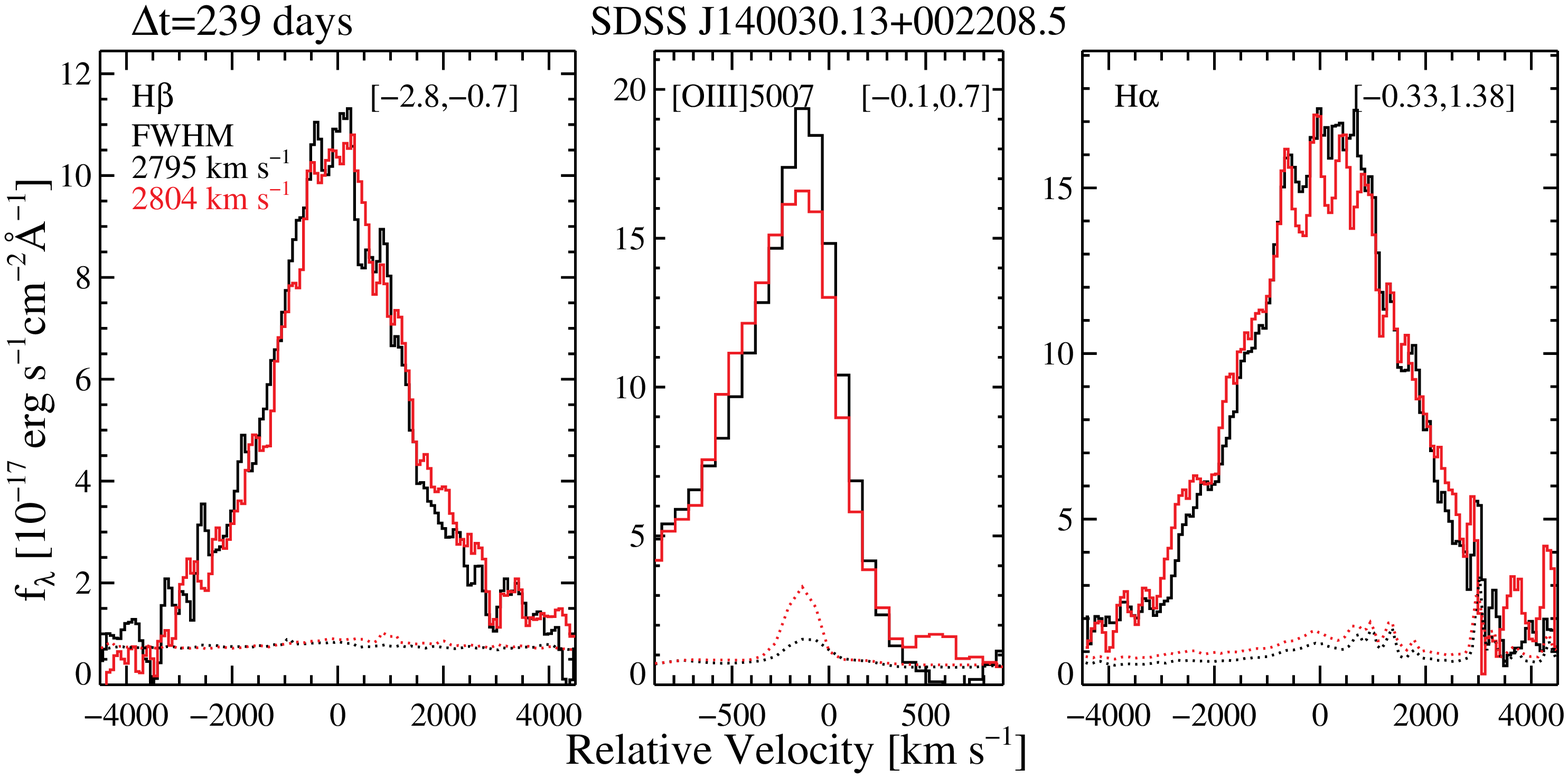}
 \includegraphics[width=0.45\textwidth]{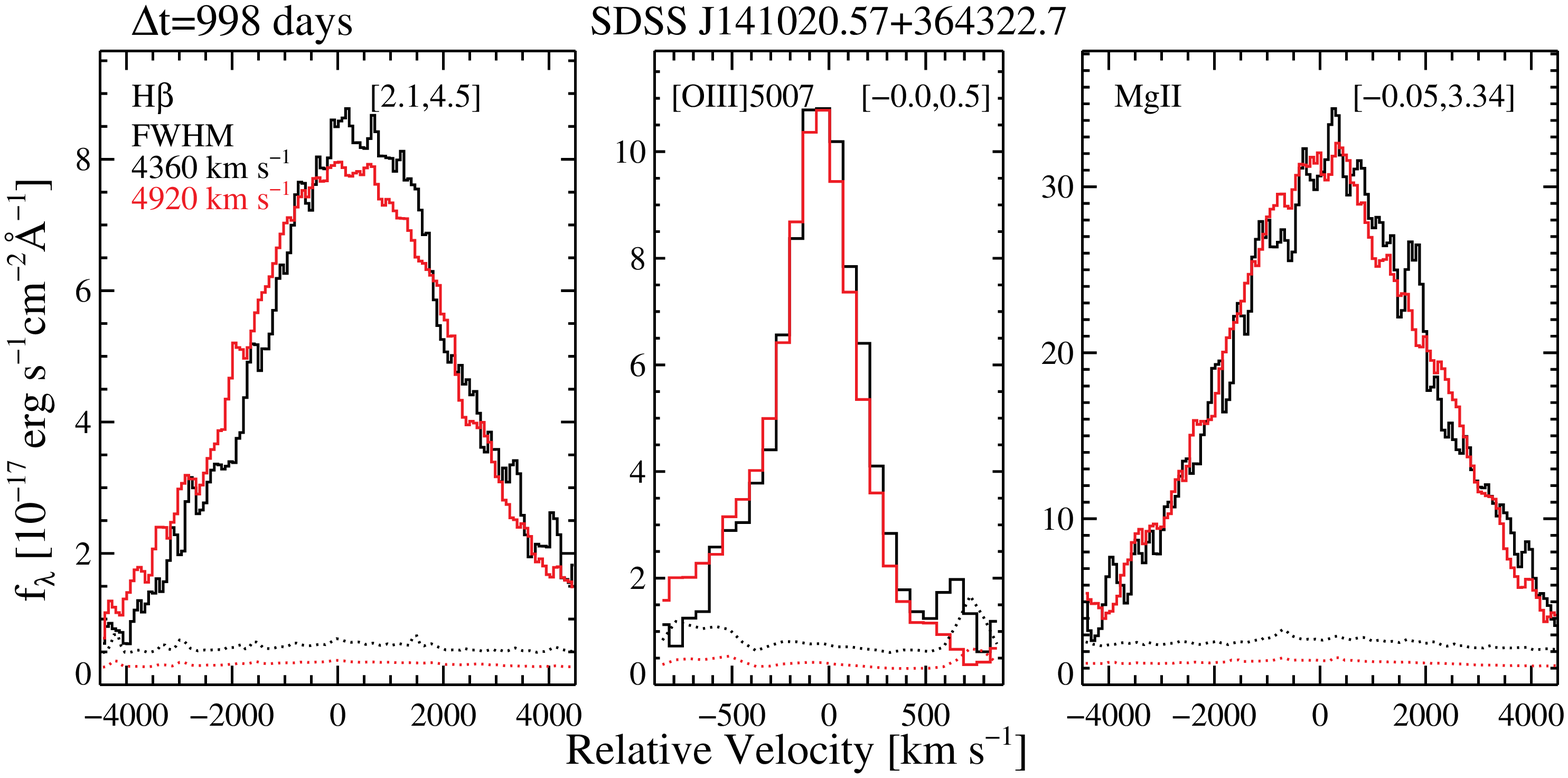}
 \includegraphics[width=0.45\textwidth]{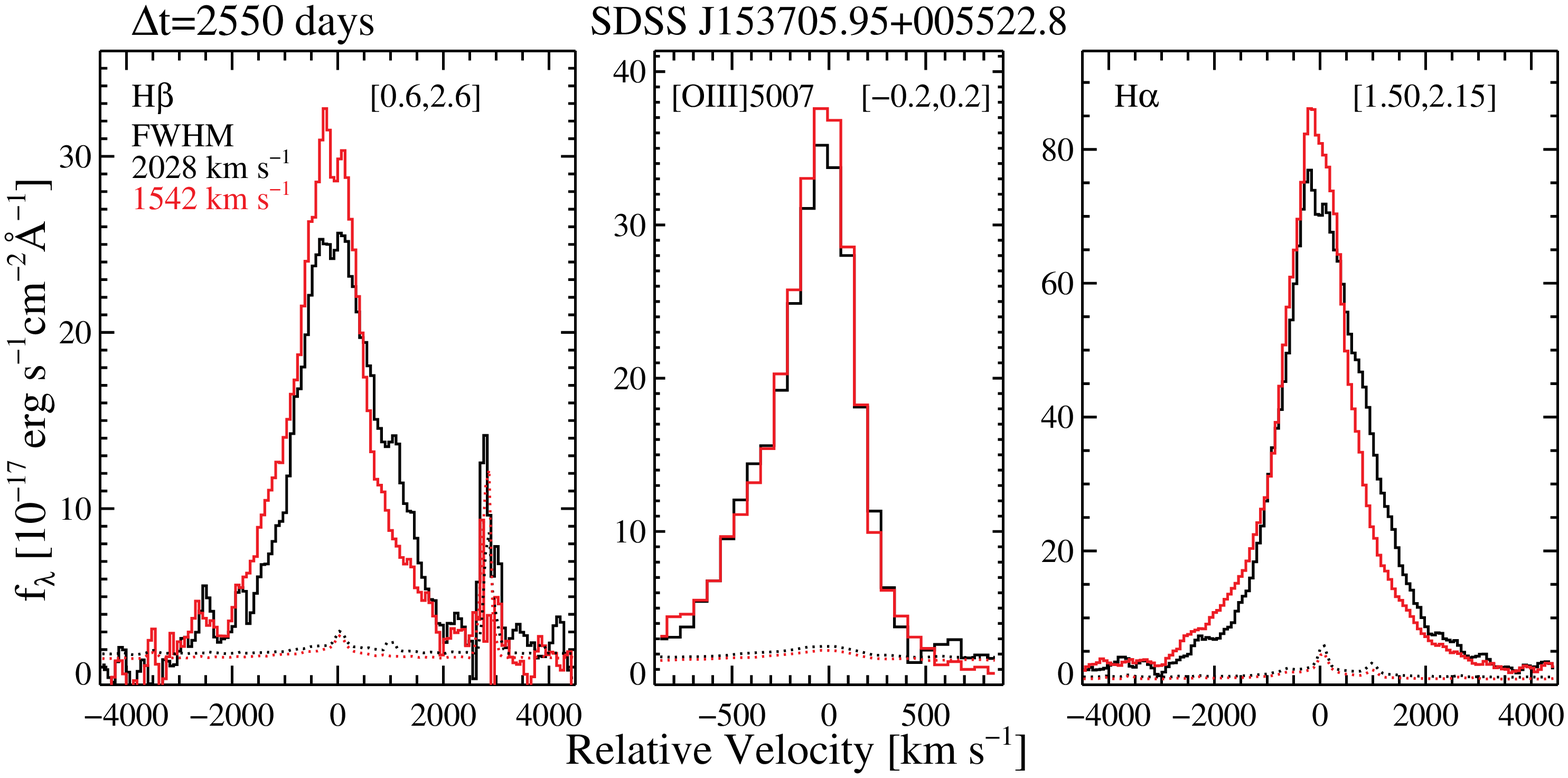}
 \caption{Same as Fig.\ \ref{fig:det_obj1}, for another set of detections. 
 }
 \label{fig:det_obj2}
\end{figure*}

\begin{figure*}
 \centering
 \includegraphics[width=0.45\textwidth]{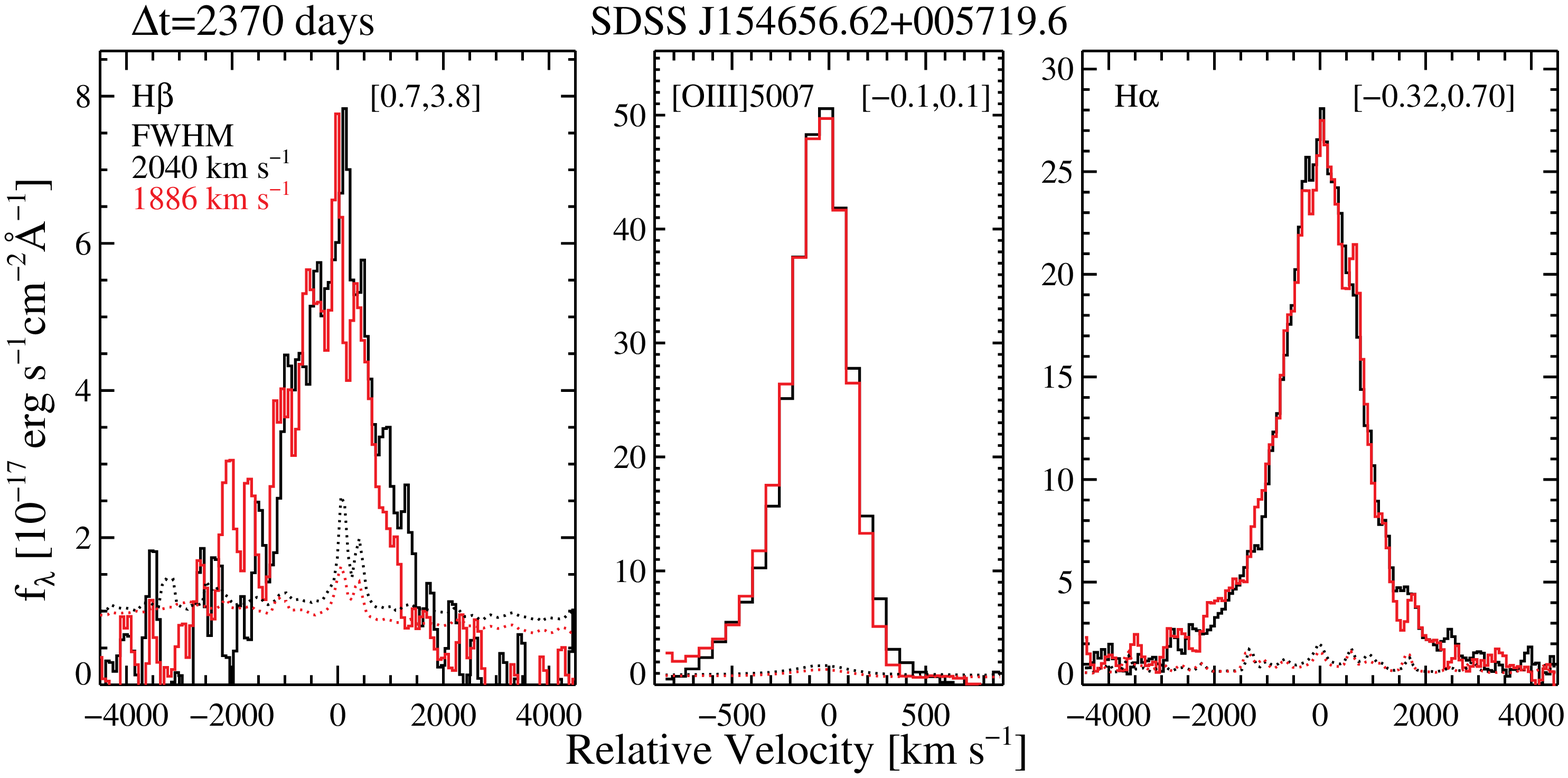}
 \includegraphics[width=0.45\textwidth]{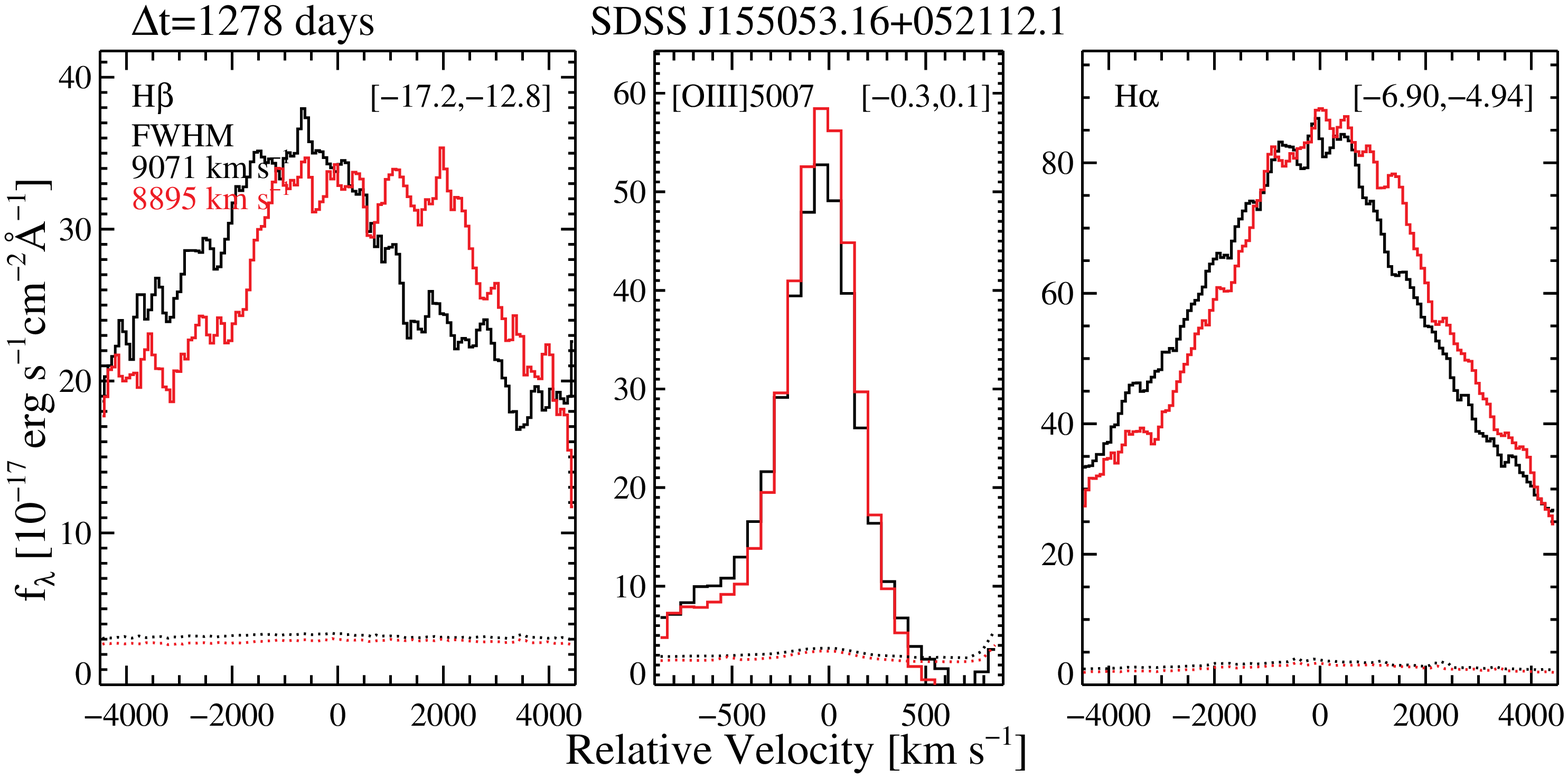}
 \includegraphics[width=0.45\textwidth]{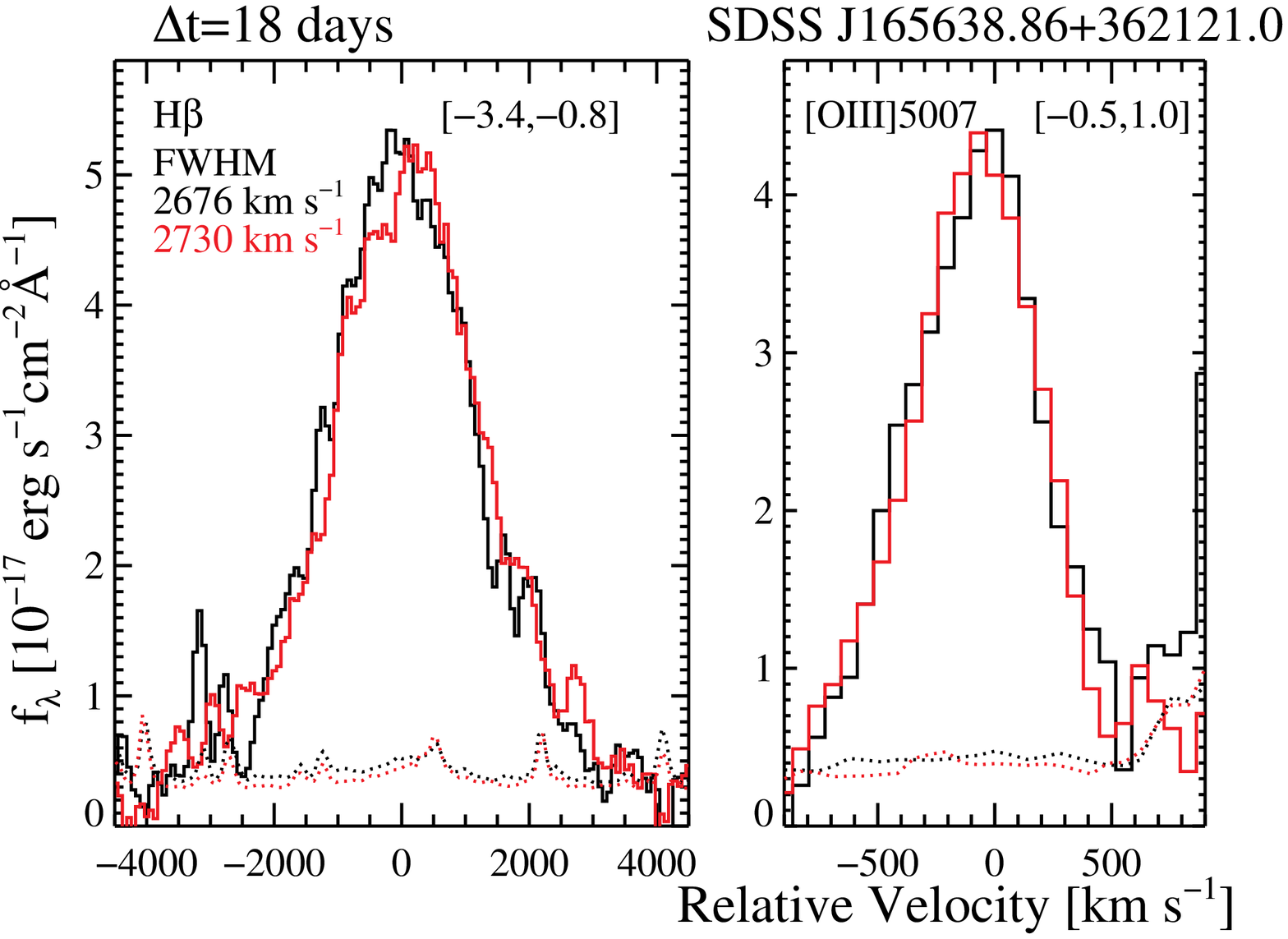}
 \includegraphics[width=0.45\textwidth]{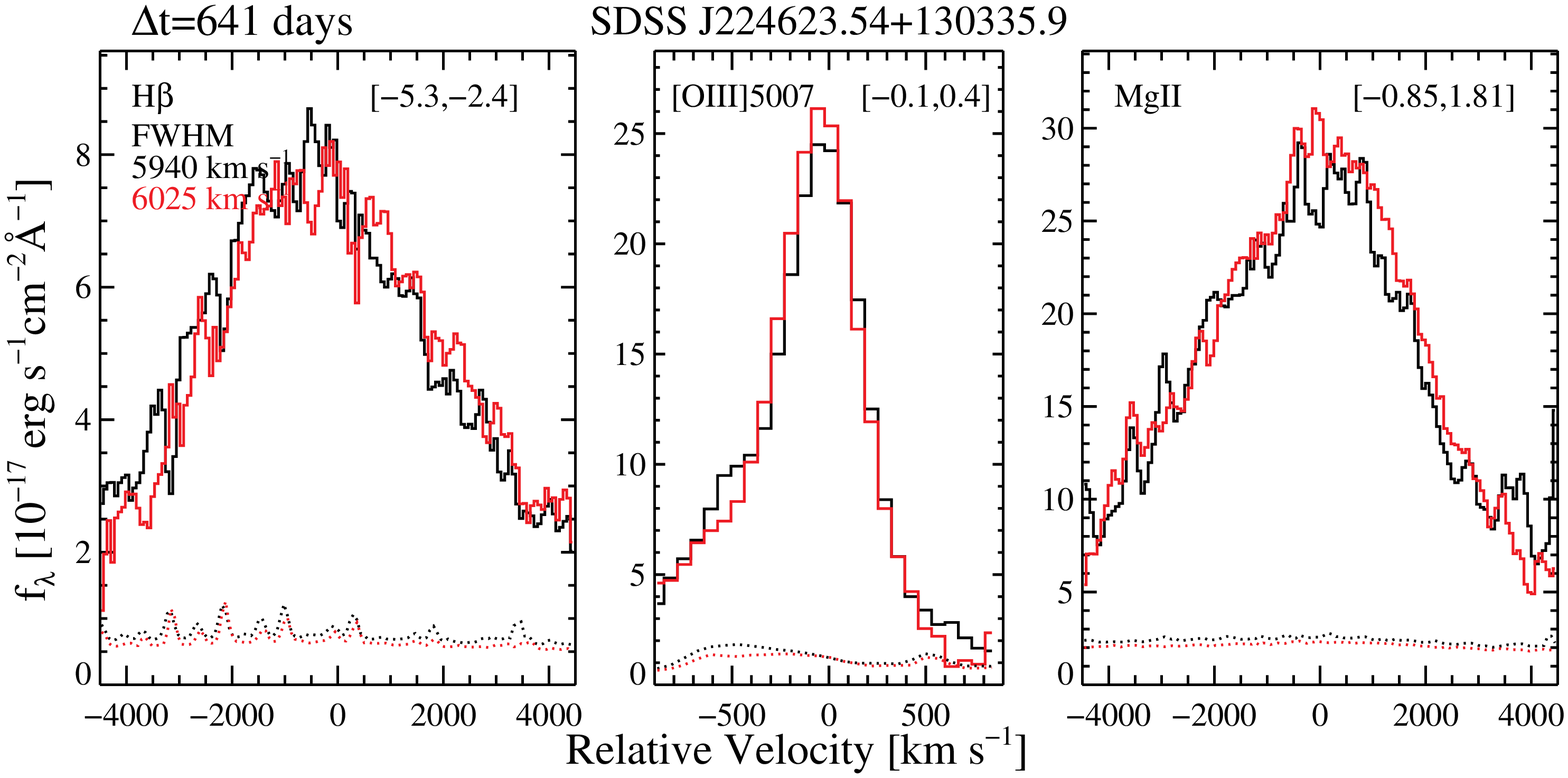}
 \includegraphics[width=0.45\textwidth]{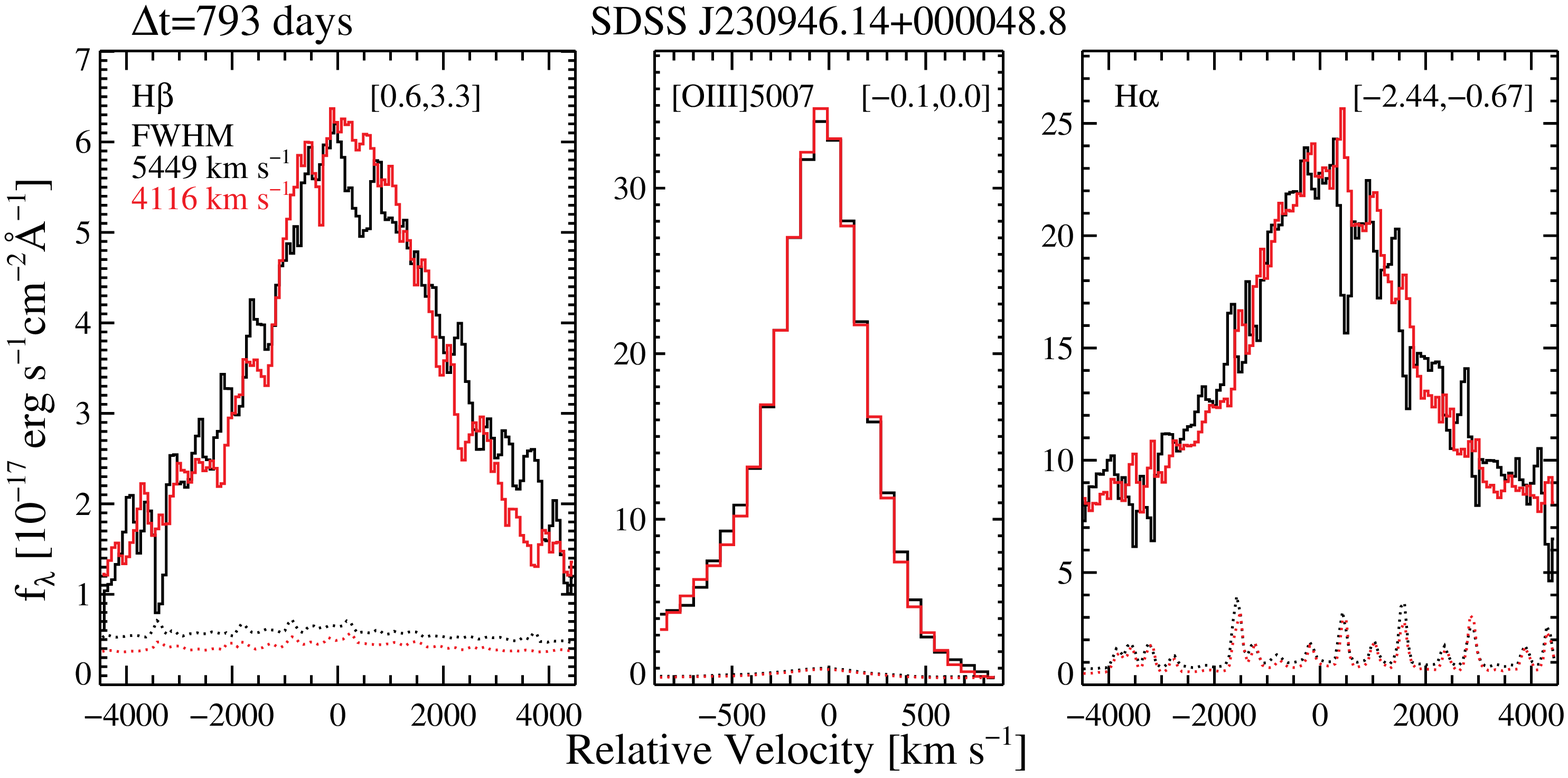}
 \includegraphics[width=0.45\textwidth]{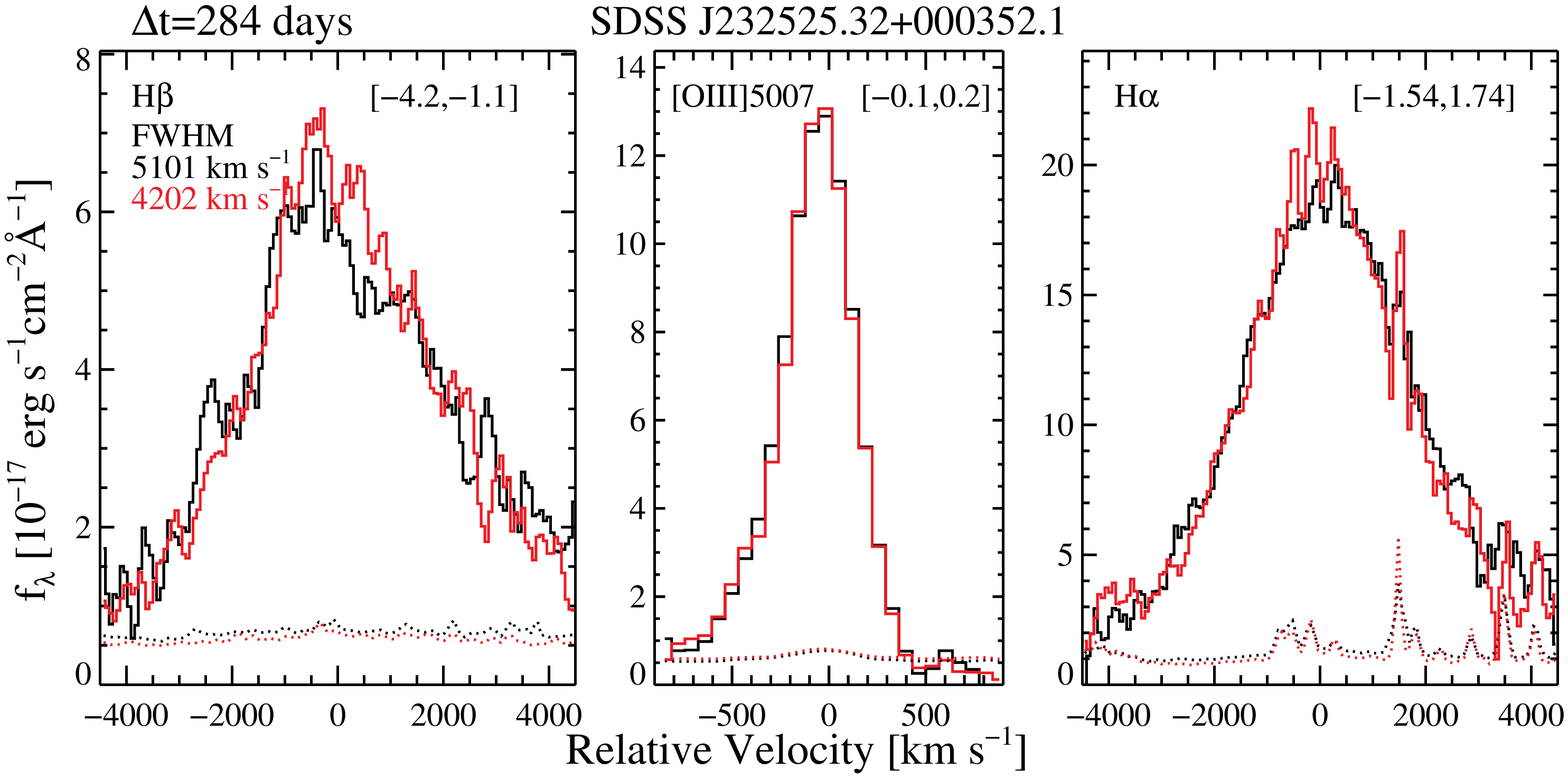}
 \includegraphics[width=0.45\textwidth]{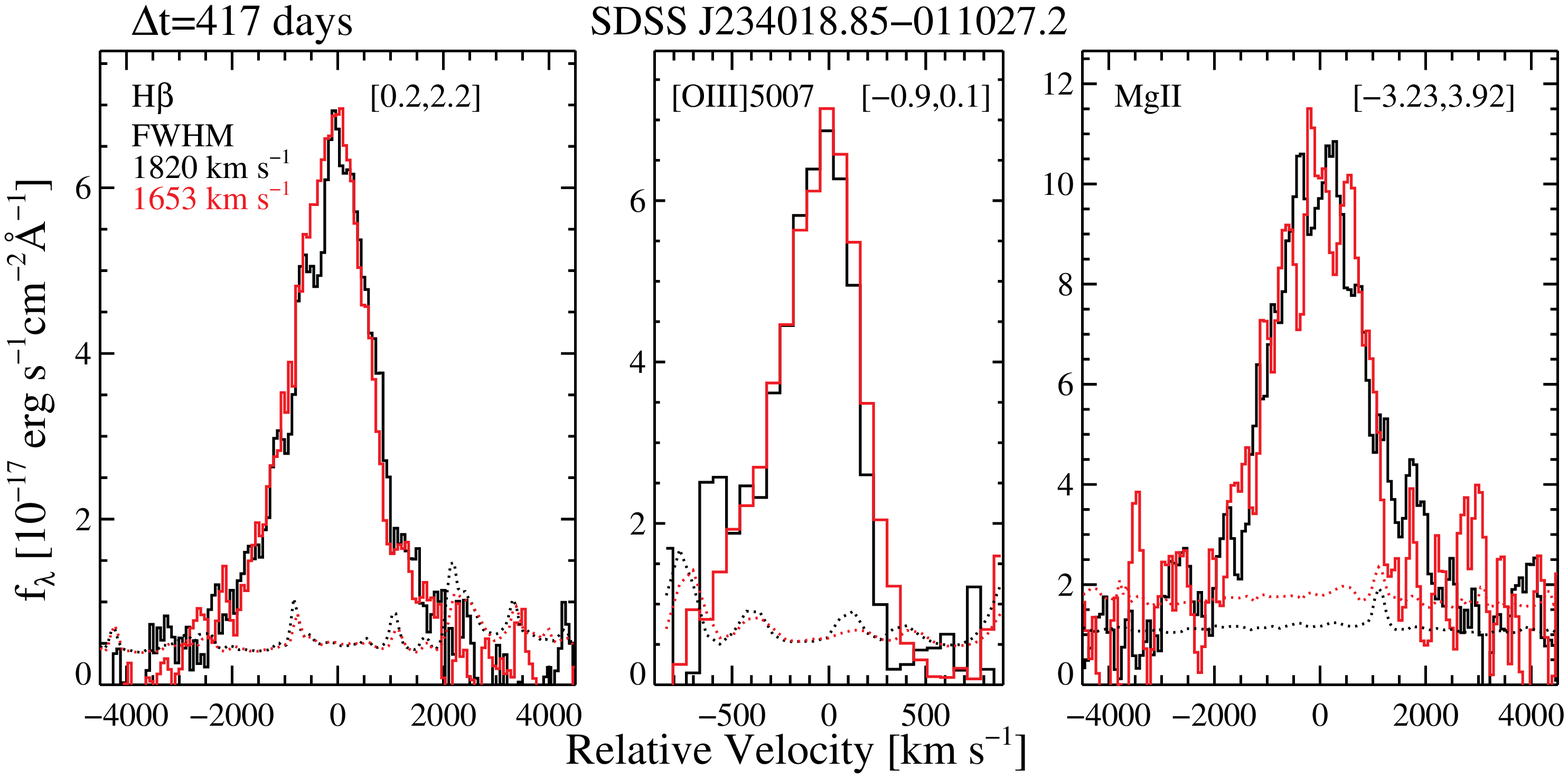}
 \includegraphics[width=0.45\textwidth]{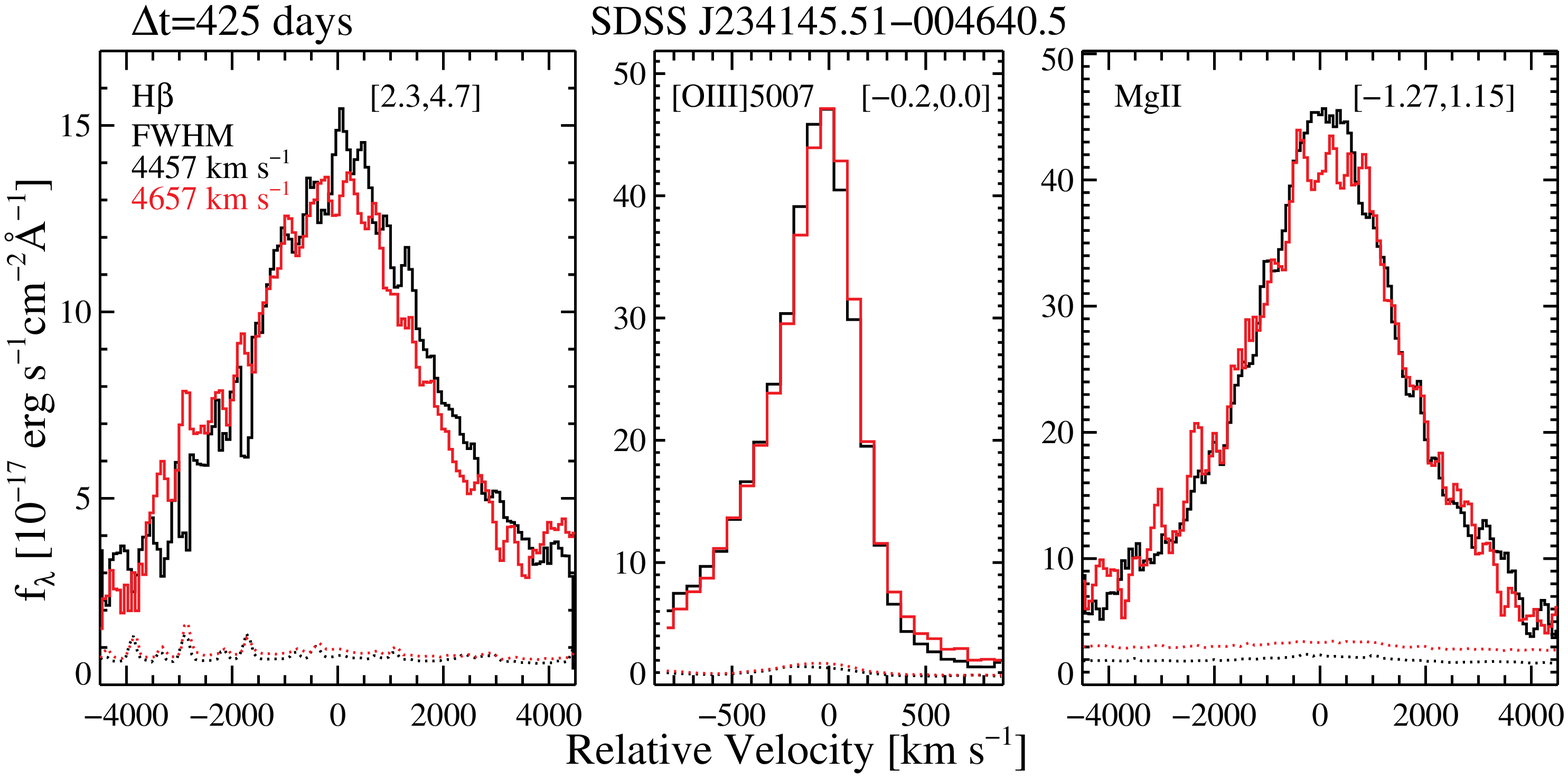}
 \includegraphics[width=0.45\textwidth]{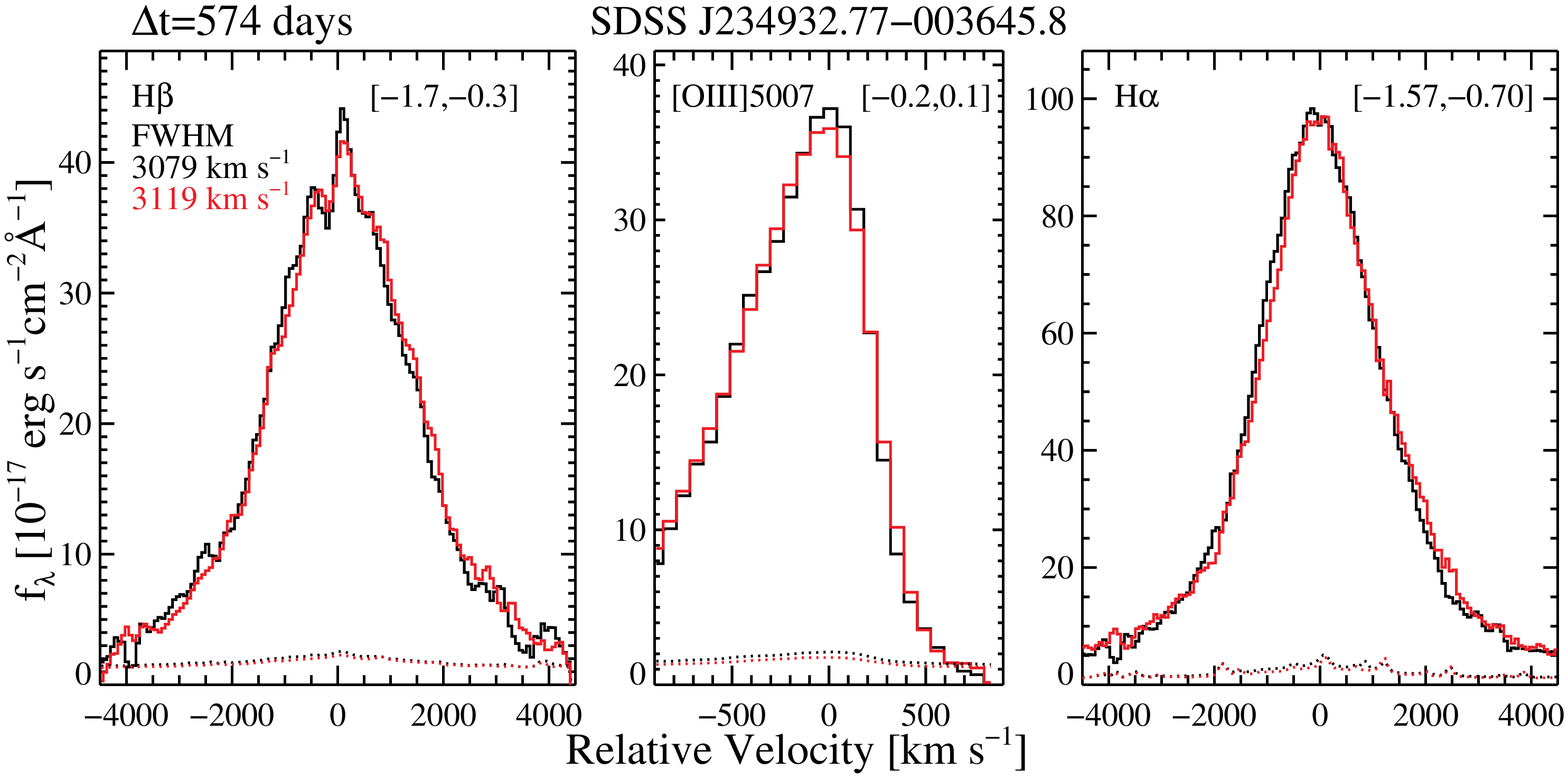}
 \includegraphics[width=0.45\textwidth]{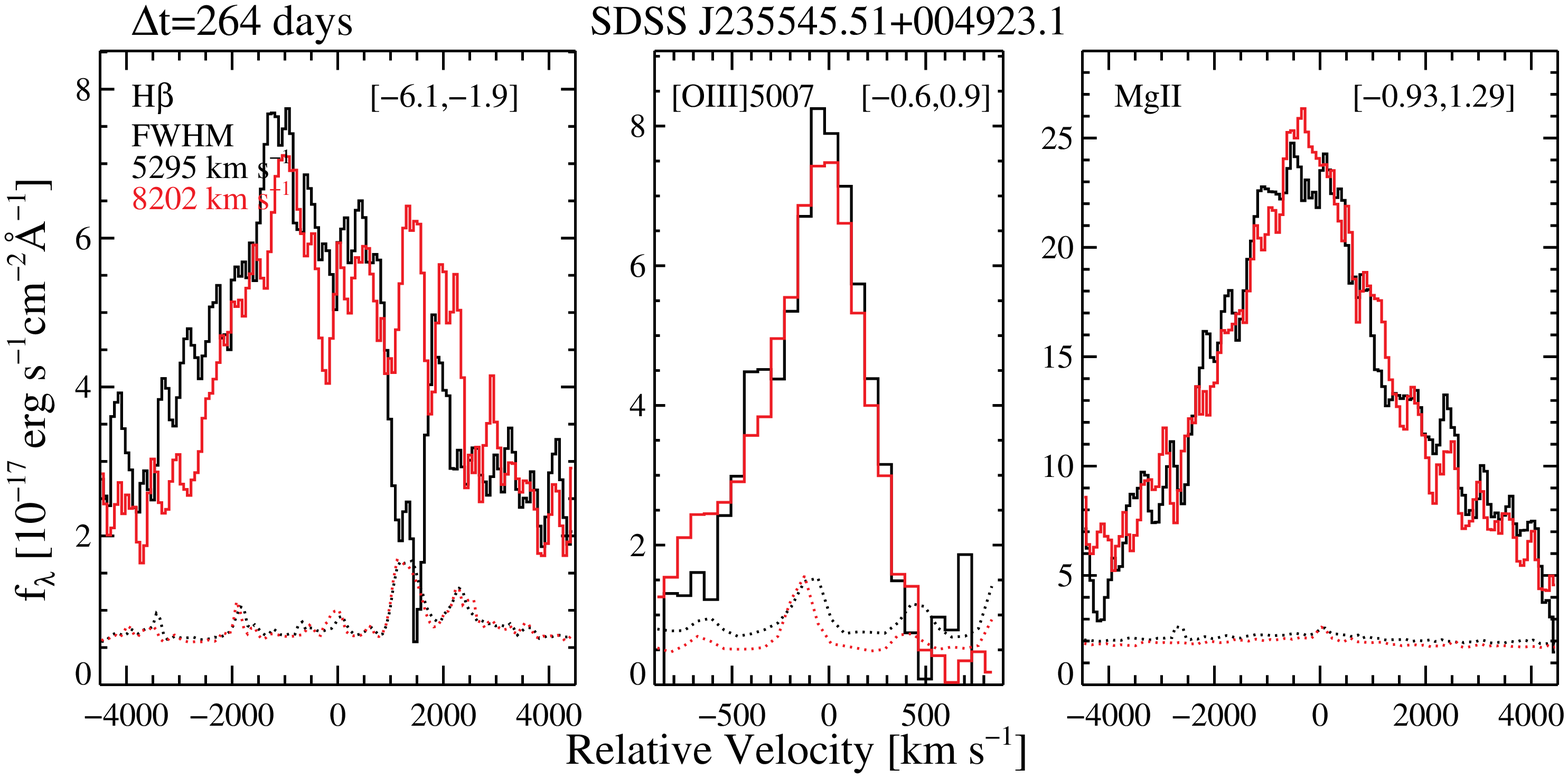}
 \caption{Same as Fig.\ \ref{fig:det_obj1}, for another set of detections. Note that for J2355$+$0049 there is significant absorption in the Epoch 1 spectrum near $+1500\ {\rm km\,s^{-1}}$ of \hbeta; this is not due to our narrow-line subtraction, but likely due to incorrect data reduction. Masking out these affected pixels in our cross-correlation still yields a consistent, negative shift.  
 }
 \label{fig:det_obj3}
\end{figure*}

Now we consider the general case with different binary fractions and binary separations. We determine the KS probability that the expected acceleration distribution is drawn from the same distribution as the observed one, for a series of values of $d/R_{\rm BLR}$, the binary fraction, and the mass of the companion BH, $M_2$. As discussed earlier, we adopt a constant mass $M_1=1.8\times 10^8\ M_\odot$ for BH 1. Each evaluation is performed for a mock sample of $10^5$ systems. 

We choose to parameterize the KS probability in terms of $d/R_{\rm BLR}$ instead of $d$, because objects in our sample spread over a range of BLR sizes. So for a fixed value of $d$, some objects will not satisfy the criterion $d>R_{\rm BLR}$ required in our binary scenario (see \S\ref{sec:prelim}), which will then complicate the expected distribution. 

If we consider the more restrictive condition that the BLR size must be smaller than the Roche radius, then we must have $d>f_r^{-1}R_{\rm BLR}$, where $f_r$ is given in Eqn.\ (\ref{eqn:roche}).

%If the companion BH is more massive, i.e., $M_2>M_1$, we further require that the BLR of BH 1 must be inside its Hill sphere, $(M_1/3M_2)^{1/3}d$. This imposes a higher threshold for the allowable binary separation, $d>(3M_2/M_1)^{1/3}R_{\rm BLR}$. 

Fig.\ \ref{fig:ks_prob} shows the contours of the KS probability as functions of the binary fraction and separation, at several different values of $M_2$, using the ``superior'' sample with our fiducial error estimates in the measured acceleration. The cyan lines show the $5\%$ probability contour, and darker regions are more probable. The dashed vertical lines in each panel mark the minimum separation of the hypothetical BBHs in units of the BLR size (i.e., the Roche limit, Eqn.\ \ref{eqn:roche}), in order for the broad line diagnostics to work. In each panel, regions with larger binary separations and lower binary fractions are disfavored, because they contribute negligible acceleration and thus cannot explain the observation that the observed distribution of accelerations is larger than expected from the measurement errors. Thus the observed acceleration distribution (error broadened) favors islands in the two-dimensional parameter space with high KS probabilities, bounded by the minimum separation (the vertical line). Since the mass for the active BH is $1.8\times 10^8\ M_\odot$ in our sample, these results suggest that the companion inactive BH must be several times more massive than the active BH to account for the observed acceleration distribution. In addition, we require a high binary fraction for the companion mass range probed ($\sim 5-10\times 10^8\ M_\odot$) and the companion BH must be orbiting not far outside the BLR. 

%But the binary separation is normally $\gtrsim$ several BLR radii, i.e., greater than the Roche radius. 

The above conclusions are based on the assumption that all the excess variance in acceleration we see is from binary motion. However, as argued below, apparent acceleration between two epochs can also be induced by BLR variability in single BHs. If some, or most, of the observed excess variance is caused by broad line variability, the constraints on the BBH population will be significantly weakened. 

In the extreme case where the observed acceleration distribution can be entirely explained by intrinsic broad line variability and measurement errors, our method only provides upper limits to the inactive BH mass and binary fraction, and lower limits to the separation. To illustrate this, we take the observed distribution as is, and randomly add intrinsic acceleration due to a BBH population as we did before, and see when the simulated distribution becomes significantly different from the observed one. 

Fig.\ \ref{fig:ks_prob2} shows the resulting KS contours for two values of companion BH mass, $q\equiv M_2/M_1=1$ and $q=5$, where $M_1=1.8\times 10^8\ M_\odot$ is the mass for the active BH in our sample. Since all the variance has already been accounted for, we no longer require a BBH population to contribute to the excess variance in the observed accelerations. As a result, there is essentially no constraint on the allowable binary parameters when the companion BH is less massive than $M_1$ (left panel). On the other hand, for more massive companion masses, certain regions in the two-dimensional parameter space are ruled out, as the resulting accelerations will exceed the observed variance (right panel). The constraints become stronger for more massive companion BHs. With a companion mass $M_2=9\times 10^8\ M_\odot$, the scenario of a $\gtrsim 90\%$ binary fraction with separations $d/R_{\rm BLR}<5$ can be ruled out at $\gtrsim 95\%$ confidence. 

The real situation is probably in between the two extreme cases considered in Figs.\ \ref{fig:ks_prob} and \ref{fig:ks_prob2}, with the excess acceleration due both to a BBH population and to broad line variability in single BHs. Future spectroscopic surveys may provide the prior information on the (stochastic) variance due to the latter effect, thus improving the statistical constraints on the BBH population with our method (see discussion in \S\ref{sec:future}).

%The KS probability decreases when $M_2$ increases, the binary fraction increases, and/or the binary separation decrease, as the predicted acceleration distribution becomes broader and hence more incompatible with the observed distribution. If all binaries are equal-mass binary, then the case where more than $\sim 50\%$ of quasars in our sample host a binary with separations below twice the BLR size can be ruled out at $95\%$ confidence. 

\subsection{Individual detections}\label{sec:binary_candidate} 

We show the 30 detections (28 unique systems) in Figs.\ \ref{fig:det_obj1}-\ref{fig:det_obj3}, where we plot the two-epoch spectra around the broad \hbeta\ line and the \OIIIb\ line in velocity space. We also show the \halpha\ or \MgII\ line of these objects, if it is covered in
SDSS spectra and the line is detected at $>6\sigma$. We follow the same $\chi^2$ cross-correlation approach to measure the velocity shift between two epochs based on \MgII\ or \halpha. But we note that the inferred velocity shift from \MgII\ or \halpha\ should be interpreted with caution, for the reasons discussed in \S\ref{sec:data}. The wavelength range for the $\chi^2$ cross-correlation is $[2750,2850]$ \AA\ for \MgII, and $[6450,6650]$ \AA\ for \halpha. For \halpha\ we carefully subtracted the narrow line components, fixing the relative flux ratios of all narrow lines at both epochs. In contrast, for \MgII, we use the full line flux (broad and narrow \MgII\ lines) in the $\chi^2$ cross-correlation, because the narrow \MgII\ component cannot be reliably constrained. 

\begin{figure}
 \centering
 \includegraphics[width=0.48\textwidth]{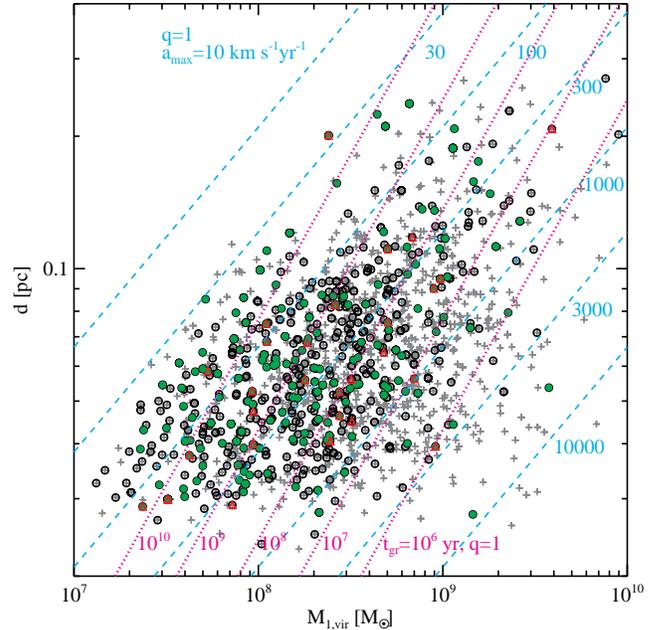}
 \caption{Distributions of quasars in the BH 1 mass versus binary separation plane, for the various samples defined earlier. Symbols are the same as in Fig.\ \ref{fig:Lbol_z}. We use the virial BH mass estimates for the active BH 1 mass. We also use the BLR size estimates for the binary separation, and the real binary separations are certainly larger, possibly by a large factor. The cyan dashed lines indicate constant maximum acceleration as in Eqn.\ (\ref{eqn:v_a}), assuming equal-mass binaries, $q=1$. The magenta dotted lines indicate constant orbital decay time due to gravitational radiation, $t_{\rm gr}$, for circular binaries and a mass ratio $q=1$, as defined in Eqn.\ (\ref{eqn:tgr}). }
 \label{fig:Mvir_d}
\end{figure}

In Fig.\ \ref{fig:Mvir_d} we show the distribution of the detected quasars in the BH 1 mass and binary orbit separation plane, along with other samples defined earlier. Again, we use the virial BH mass estimates for the active BH 1 mass and we caution that there are large uncertainties in these mass estimates. We also use the measured BLR sizes for the binary separation. As discussed in \S\ref{sec:prelim} and \S\ref{sec:general_pop}, the orbit separation of the hypothetical binary must be larger than the BLR size, i.e., all points in Fig.\ \ref{fig:Mvir_d} would move up if the true binary separations were known. The cyan dashed lines indicate constant maximum acceleration, as defined in Eqn.\ (\ref{eqn:v_a}), where we assume a constant binary mass ratio $q=1$. The magenta dotted lines indicate constant orbital decay time due to gravitational radiation in a circular binary with a mass ratio of $q=1$ \citep[][]{Peters_1964}
\begin{equation}\label{eqn:tgr}
t_{\rm gr}=\frac{5}{256}\frac{c^5}{G^3}\frac{d^4}{q(1+q)M_1^3}\ .
\end{equation}
Therefore most quasars in our sample are in the regime where the orbit decays from gravitational radiation in less than a Hubble time (if the virial mass estimate is correct and the orbital separation is not much bigger than the size of the BLR). Obviously, we expect binary BHs to be rare in the lower right region of the diagram, where the orbital decay time is much less than the Hubble time. 

The biggest concern with our detections is that the broad line profile could change over multi-year timescales, which may lead to an artificial line shift with the $\chi^2$ cross-correlation method. While visual inspection can lend some confidence to the interpretation as a genuine shift rather than a profile change (or vice versa), the correct interpretation is still ambiguous in some cases.

Below we give an object-by-object discussion of individual detections, and
comment on the robustness of the detection, as well as possible causes of the
detected velocity shift. To guide our discussion we divide these objects into three categories: 1) binary candidates; 2) broad \hbeta\ line variability, and 3) ambiguous cases. These categories are our best efforts to assign the ``most likely'' scenario to individual detections, and are not a rigorous classification of these systems. 
%The classifications of individual objects are listed in Table 2. 

A useful constraint on the expected acceleration from a BBH can be derived using the equations in \S\ref{sec:prelim}:
\begin{equation}\label{eqn:amax}
|a_1|<44qf_r^2\left(\frac{M_1}{10^8\ M_\odot}\right)\left(\frac{R_{\rm BLR}}{0.1\ {\rm pc}}\right)^{-2}\ {\rm km\,s^{-1}yr^{-1}}\ ,
\end{equation}
where $f_r(q)<1$ is given by Eqn.\ (\ref{eqn:roche}). The product $qf_r^2<0.6$ is a monotonic increasing function of $q$ for $0.05<q<20$.

\subsubsection{Binary candidates}\label{sec:class1}

\textbf{J030100.23$+$000429.3}\ For this object consistent velocity shifts between two epochs are detected for both \hbeta\ and \MgII\ over 261 days, despite the apparently noisy broad line spectrum for \MgII. The shape of the broad \hbeta\ is not significantly different between the two epochs. The inferred acceleration is $\sim 220-760\ {\rm km\,s^{-1}\,yr^{-1}}$ ($2.5\sigma$).

\textbf{J032213.89$+$005513.4}\ The S/N for this object is very high for both epochs. Consistent velocity shifts are detected for both \hbeta\ and \halpha\ over $\sim 3.4$ yrs. There is a slight change in the broad line profile between the two epochs. The inferred acceleration is $\sim 6-23\ {\rm km\,s^{-1}\,yr^{-1}}$ ($2.5\sigma$). 

\textbf{J122909.52$-$003530.0}\ For this object both \hbeta\ and \MgII\ show consistent velocity shifts over a time period of $\sim 4$ yrs. The change in line shape is insignificant, although there is some hint that the red-side flux changes a little between the two epochs. In the binary scenario, the inferred radial acceleration based on \hbeta\ is $50-110\ {\rm km\,s^{-1}\,yr^{-1}}$ ($2.5\sigma$). 

\textbf{J141020.57$+$364322.7} For this object we detected a velocity shift for \hbeta\ but not for \MgII. However, the limits we obtained from \MgII\ are fully consistent with the velocity shift detected in \hbeta. Upon visual inspection (see Fig.\ \ref{fig:det_obj2}), both \hbeta\ and \MgII\ show similar profile changes between the two epochs, with extra flux on the red-side in Epoch 1. Therefore we classify this object as a good BBH candidate. The inferred radial acceleration based on \hbeta\ is $50-110\ {\rm km\,s^{-1}\,yr^{-1}}$ ($2.5\sigma$). 

\textbf{J153705.95$+$005522.8}\ For this object consistent velocity shifts between two epochs are detected for both \hbeta\ and \halpha, over $\sim 7$ yrs. The broad line profile also changed dramatically between the two epochs: the line is substantially narrower, and the shape is more symmetric, in Epoch 2 than in Epoch 1. It is possible that there is only a portion of the broad line emission moving along with BH 1 in the hypothetical binary system, and the remaining broad line emission comes from a circumbinary region, which does not accelerate between the two epochs. It is also possible that some of the broad line emission comes from BH 2, and BH 2 is much more massive than BH 1. The inferred acceleration is $\sim 6-26\ {\rm km\,s^{-1}\,yr^{-1}}$ ($2.5\sigma$).

\textbf{J155053.16$+$052112.1}\ For this object we detected the largest velocity shift in our sample, for both \hbeta\ and \halpha, over $3.5$ yrs. Interestingly, the velocity shift for \hbeta\ is larger than that for \halpha\ (significant at the $>3\sigma$ level). At face value, this may argue against the binary hypothesis if the \hbeta\ and \halpha\ broad line emission comes from the same BLR. However, if the \hbeta\ BLR is mostly confined to the active BH, while the \halpha\ BLR also contains a circumbinary component (which does not accelerate), a larger velocity shift in \hbeta\ is expected. The width of the broad \hbeta\ is substantially larger than that of the broad \halpha, which is consistent with this scenario. The inferred acceleration based on \hbeta\ is then $\sim 250-340\ {\rm km\,s^{-1}\,yr^{-1}}$ ($2.5\sigma$). 

\textbf{J234932.77$-$003645.8}\ For this object both \hbeta\ and \halpha\ show consistent velocity shifts over a time period of $\sim 1.6$ yrs. The shape of the broad line profile remains more or less constant with consistent FWHM at both epochs. The narrow \hbeta\ subtraction is not perfect for both epochs, but does not affect the detected velocity shift. The inferred radial acceleration is in the range of $11-75\ {\rm km\,s^{-1}\,yr^{-1}}$ ($2.5\sigma$). 

Since the last spectroscopic observations of these candidates were a few years ago, taking a third-epoch spectrum now would be very helpful to confirm (or rule out) the proposed binary hypothesis. 

\begin{figure}
 \centering
 \includegraphics[width=0.48\textwidth]{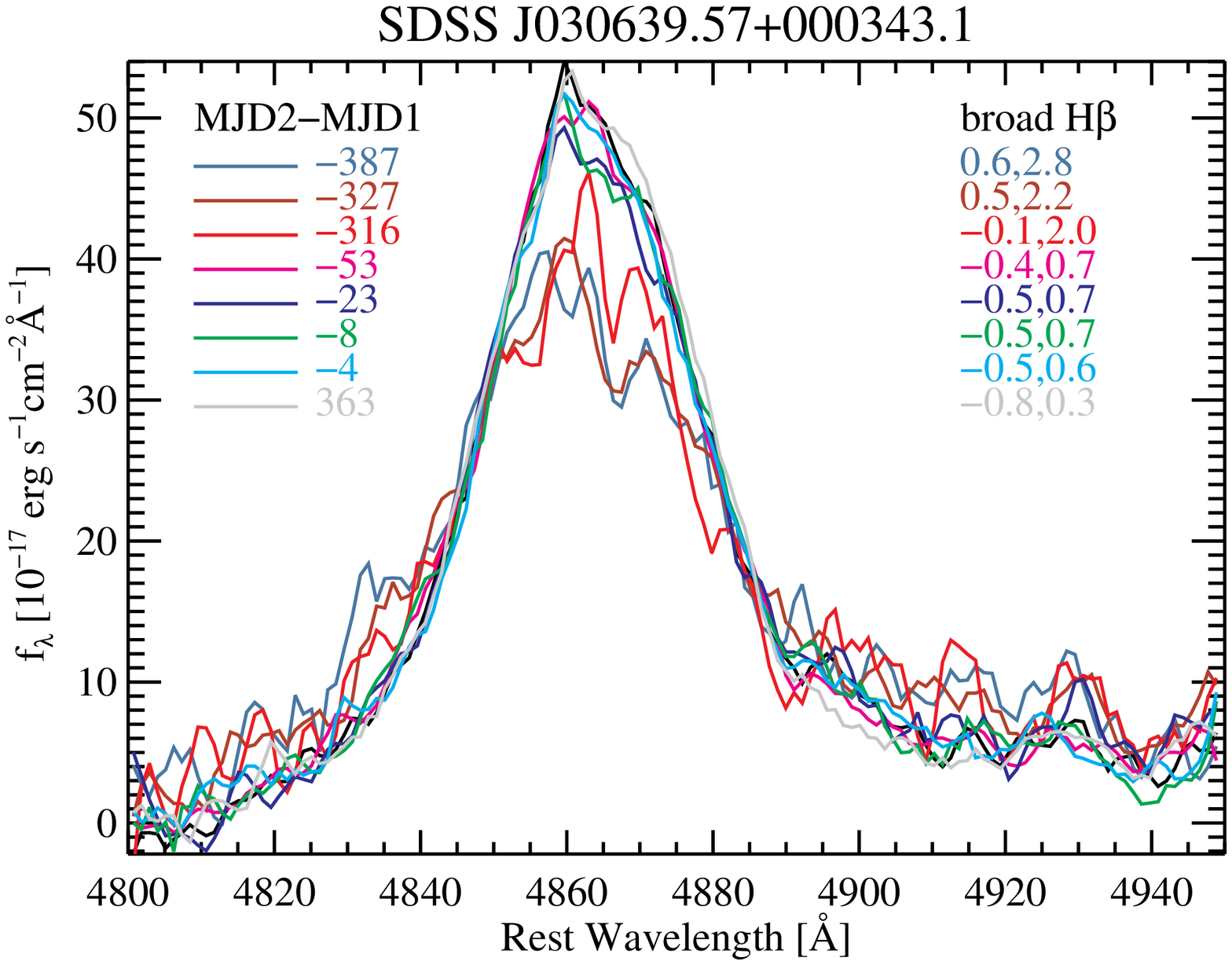}
 \includegraphics[width=0.48\textwidth]{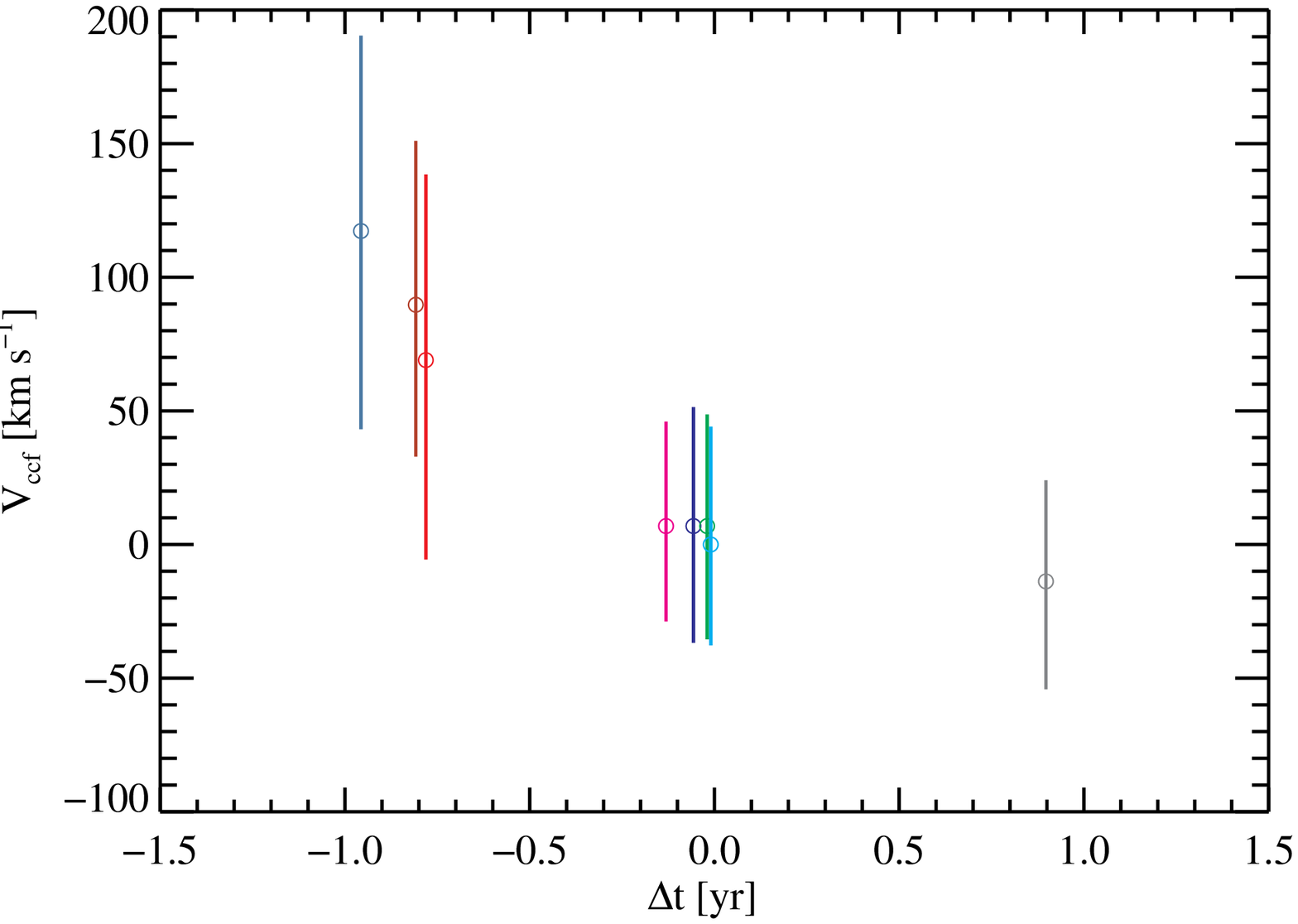}
 \caption{{\em Top:} broad \hbeta\ spectra for the quasar SDSS J030639.57$+$000343.1 at 9 epochs: the default Epoch 1 (the black line), and 8 different secondary epochs as listed in the legends. The time difference in MJD from the Epoch 1 spectrum and the measured pixel shifts ($2.5\sigma$) are marked on the left and right in the figure. The three spectra taken $\sim 1$ yr earlier than Epoch 1 have different broad line shape, and for the first two of them we have detected a velocity shift from Epoch 1 (as shown in Fig.\ \ref{fig:det_obj1}). However, the last epoch (the gray line) does not show the expected continuing velocity shift from the acceleration inferred from the earlier spectra. Thus the apparent velocity shifts between the first two spectra and the reference Epoch 1 are likely due to broad line variability rather than binary motion. {\em Bottom:} measured velocity shifts between different epochs and the reference Epoch 1, plotted against the restframe time separation. A straight line is expected from constant acceleration within such short time intervals. }
 \label{fig:0306}
\end{figure}

\subsubsection{Broad \hbeta\ line variability}\label{sec:class2}

\textbf{J030639.57$+$000343.1}\ For this object we have 8 different secondary epochs to cross-correlate with the default Epoch 1; all are in our good sample with good measurement precision. In Fig.\ \ref{fig:0306} (top) we show the broad line spectra at different epochs, along with the measured velocity shift with respect to the reference Epoch 1. We see a narrowing of the line profile from the first three spectra taken $\sim 10$ months earlier than Epoch 1, then the line profile stays more or less constant. In the bottom panel we show the measured radial velocity curve, again using Epoch 1 as reference. The time baseline is only $\sim 1.8$ yr in the quasar restframe. The expected binary period is $P> 86(1+q)^{-1/2}f_r^{-3/2}$ yr given the constraints of the binary separation (from the BLR size) and the virial BH mass estimate, where $q\equiv M_2/M_1$ and $f_r<1$ is a function of $q$ as defined in Eqn.\ (\ref{eqn:roche}). For $q>0.05$ we have $(1+q)^{-1/2}f_r^{-3/2}>1.9$ hence $P>160$ yr. Thus within such short time interval the expected BBH velocity should approximately have constant acceleration. If this is a binary system with a los acceleration inferred from the observations up to and including Epoch 1, we should expect to see a velocity shift within $\sim [-2.5,-0.5]$ (pixels) between the last epoch (gray spectrum/point in Fig.\ \ref{fig:0306}) and Epoch 1. While this is still consistent with what we measure, the fact that the last epoch has the same line profile as Epoch 1 while the first two epochs have very different profile strongly suggest that the detected velocity shift is more likely due to broad line variability.

\textbf{J135829.58$+$010908.6}\ This object is an unambiguous disk emitter with double-peaked broad line profiles. While we detected velocity shifts between the three epochs we have in the good sample, it is hard to ascribe this average velocity shift to binary motion. It appears that the relative strength of the two peaks changed over this period (seen in both \hbeta\ and \halpha). Variability of this kind has been seen in long-term monitoring of disk emitters \citep[e.g.,][]{Gezari_etal_2007,Lewis_etal_2010}, which usually invokes a non-binary interpretation \citep[but see, e.g.,][for an alternative BBH interpretation]{Bon_etal_2012}. We therefore classify this object as exhibiting broad line variability, although the BBH scenario cannot be completely ruled out given only three epochs. Similar objects were reported in \citet{Eracleous_etal_2012}.

\textbf{J074700.19$+$285608.5 and J165638.86$+$362121.0}\ These two objects are detections with the shortest time separation ($\Delta t<30$ days) between the two epochs, which implies rather large accelerations. Using Eqn.\ (\ref{eqn:amax}) and assuming the BLR sizes are accurate, it is difficult to get such large acceleration unless the BH mass $M_1$ is $\sim 10$ times larger than the virial BH mass estimate (see Table \ref{table:detection}) and the unseen BH is at least as massive for both objects. While this is not entirely ruled out, it is more likely that the velocity shifts detected for these two objects are due to short-term broad line variability. In addition, for J0747 the velocity shift based on \hbeta\ is {\em inconsistent} with that based on \halpha, which also favors the BLR variability scenario. 

\subsubsection{Ambiguous cases}\label{sec:class3}

We classify the remaining detections as ambiguous cases. This category includes more than half of our detections, which reflects the general ambiguity in interpreting the observed broad line velocity shifts. They have neither a second broad line (\MgII\ or \halpha) showing consistent velocity shifts, nor more than two epochs with sufficient quality to rule out the hypothetical binary acceleration. Thus, they could still be binary candidates, although the binary probability is perhaps lower than those presented in \S\ref{sec:class1}. Additional high-quality spectroscopic observations are needed to further investigate these systems.

%\begin{figure*}
% \centering
% \includegraphics[width=0.35\textwidth]{2045-417-53350_multi_epoch}
% \includegraphics[width=0.35\textwidth]{0411-265-51817_multi_epoch}
% \includegraphics[width=0.35\textwidth]{0413-598-51929_multi_epoch}
% \includegraphics[width=0.35\textwidth]{0414-341-51901_multi_epoch}
% %\includegraphics[width=0.35\textwidth]{0531-207-52028_multi_epoch}
% \includegraphics[width=0.35\textwidth]{0819-618-52409_multi_epoch}
% \includegraphics[width=0.35\textwidth]{0383-518-51818_multi_epoch}
% \includegraphics[width=0.35\textwidth]{0385-207-51877_multi_epoch}
% \includegraphics[width=0.35\textwidth]{0385-139-51877_multi_epoch}
% \includegraphics[width=0.35\textwidth]{0386-215-51788_multi_epoch}
% \caption{Multi-epoch spectra for the detections in \S\ref{sec:binary_candidate}. The non-detections are generally consistent with predictions from the acceleration inferred from the detections. 
% }
% \label{fig:det_obj_dup}
%\end{figure*}

%\section{Modeling}\label{sec:modeling}

\section{Discussion}\label{sec:disc}

%\textbf{Discuss the caveats of this broad line shift diagnostic, in particular the assumptions made in our working scenario, such as the active BH dominates the entire BLR dynamically, and that the companion BH orbits outside the BLR. The worse case scenario is that the BLR is so messy that it can never be used as a clear indicator for sub-pc binaries. This possibility deserves a serious discussion. }

\subsection{Caveats with the broad line diagnostics}\label{sec:caveat}

The basic assumption required to utilize the broad line shift as an indicator for the BBH orbital motion is that the BLR must move along with its associated BH. This means this method is only applicable to BBHs with separations $d\gtrsim f_r^{-1}R_{\rm BLR}$ \citep[see \S\ref{sec:prelim} and][]{Shen_Loeb_2010}. Fortunately, the typical BLR size for our sample is $\sim 0.06$ pc, so this method is still probing the interesting sub-pc regime where presumably the BBH has passed the final parsec barrier. 

The primary uncertainty is that the BLR is unlikely to be in a steady state, and changes in the BLR may produce artificial broad line velocity shifts. Velocity-resolved reverberation mapping results for a handful of objects have shown diverse kinematics of the BLR gas \citep[e.g.,][]{Sergeev_etal_1999,Denney_etal_2009,Grier_etal_2013} involving inflows and outflows. Changes of the kinematic structure of the BLR will occur on the dynamical time $t_{\rm dyn}\sim 24(R_{\rm BLR,0.1}/{\rm FWHM}_{4000})$ yr, i.e., the same timescale over which we search for the BBH signature. Broad line shifts without line shape changes are perhaps better explained by the BBH scenario, but the broad line shape may vary even in a BBH system. Occasionally an apparent velocity shift might be induced due to excess flux on one side of the broad line. This may indicate a BLR variability origin rather than a BBH; however, the BBH scenario cannot be completely ruled out if the BLR is composed of spatially distinct components, and the ``excess flux'' is indeed associated with the motion of one BH. In a similar spirit, the broad line profile could also change in the BBH scenario, if both BHs are active and the broad line is composed of two spectrally unresolved components \citep[e.g.,][]{Shen_Loeb_2010}. Invoking multiple broad lines sometimes helps confirm the detected velocity shift, but cannot be used to confirm or rule out the BBH scenario given the possible differences in the BLRs for different lines with various ionization potentials. 

It is practically impossible to distinguish between the BBH case and the BLR variability case for the observed broad line velocity shifts with only two epochs. The best (perhaps the only) way to differentiate these two cases is continued spectral monitoring of the broad line. BLR variability will lead to stochastic velocity shifts, while the BBH scenario will lead to coherent velocity shifts, either Keplerian (over long timescales comparable to the binary orbit period) or linear (over timescales much shorter than the binary orbit period). Of course, one also needs high S/N to make the measurements meaningful for this purpose. Currently there is no adequate spectroscopic monitoring data to provide priors on the stochastic velocity shifts induced by BLR variability, but future time-domain spectroscopic surveys may provide such information (see \S\ref{sec:future}).

Finally, although we have focused on circular binary orbits, the methodology developed in this work can be easily applied to a binary population with a distribution of eccentricities. Such an application will be appropriate once we have a good understanding of the velocity shifts due to broad line variability. 

\subsection{Uncertainty in the BLR size estimation}\label{sec:blr_size}

Throughout this work we have used BLR sizes estimated from the BLR $R-L$ relation found in reverberation mapping studies \citep[e.g.,][]{Bentz_etal_2009}. This is of course an approximation, as the BLR size measured in reverberation mapping represents the emissivity-weighted average radius of the BLR, and the actual BLR could extend beyond this typical size. It is nevertheless reasonable to expect that the bulk of the line emission comes from within this typical BLR size. If we have significantly underestimated the BLR sizes, the allowable binary parameter space shown in Fig.\ \ref{fig:ks_prob} will diminish, because then the allowable binary orbit separations (i.e., $d>f_r^{-1}R_{\rm BLR}$) will be too large to account for the observed excess variance in broad line velocity shifts; in this case the observed excess variance must be due to BLR variability.

An alternative way to estimate the BLR size is to use theoretical predictions based on photoionization calculations \citep[e.g.,][]{Dumont_1990}: $R_{\rm BLR}\sim 10^4R_G$, where $R_G\equiv 2GM_{1}/c^2$ is the gravitational radius of the active BH. In our opinion, such theoretical estimates are more uncertain than the BLR sizes derived empirically from reverberation mapping. In addition, such estimates have an explicit dependence on the highly uncertain BH
mass estimates \citep[e.g.,][]{Shen_2013}. Despite these concerns we have constructed models using $R_{\rm BLR}=10^4R_G$. Following our earlier discussion in \S\ref{sec:general_pop}, we assume a constant mass of $1.8\times 10^8\,M_\odot$ for BH 1, and use the new BLR size estimates (which are then the same for all quasars in our sample) to repeat the analysis in Fig.\ \ref{fig:ks_prob}. The new BLR size is $10^4R_G\sim 0.17\,$pc, larger than our fiducial BLR size estimates (with a median value of $\sim 0.06\,$pc) by almost a factor of 3. Therefore the allowable binary parameter space as shown in Fig.\ \ref{fig:ks_prob} disappears, and BLR variability is needed to explain the observed excess variance in broad line velocity shifts.

\subsection{Comparisons with other work}\label{sec:comp}

Most theoretical work on BBHs with a gaseous disk does not make specific predictions for the broad lines \citep[e.g.,][]{Armitage_Natarajan_2002,Escala_etal_2005,Dotti_etal_2006,Dotti_etal_2007,Mayer_etal_2007,Haiman_etal_2009,Kocsis_etal_2012,Rafikov_2012} with few exceptions \citep[e.g.,][]{Bogdanovic_etal_2008}. In addition, their models often involve a circumbinary accretion disk, which presumably complicates the applicability of the broad line diagnostics used in this work, if this circumbinary disk also contributes to the broad line emission. Using SPH simulation coupled with photoionization calculations, \citet{Bogdanovic_etal_2008} studied the temporal behavior of the broad \halpha\ line from the circumbinary accretion disk. The line profile is often double-peaked, showing the characteristics of a disk, and evolves with time, similar to the results in \citet{Shen_Loeb_2010} with toy models for the BLR. However, it is not yet feasible to make detailed comparisons between the simulation predictions and observations of quasars with double-peaked broad line profiles. \citet{Montuori_etal_2011} considered a BBH scenario where the BLR has a stratified structure for different line species with the higher-ionization lines deeper within the gravitational potential of the BH. Tidal erosion from the companion BH then reduces the flux ratio of the lower-ionization lines to higher-ionization lines. While this process may be at work for very close BBHs, it is difficult to differentiate this scenario from single BHs with abnormal broad line flux ratios, thus difficult to confirm the BBH nature with this feature alone. Finally, recent predictions of sub-pc BBH fractions in quasars from hierarchical cosmological merger models \citep[e.g.,][]{Volonteri_etal_2009,Kulkarni_Loeb_2012} still have large uncertainties, and a clear pathway connecting these predictions to direct observables is currently absent. Clearly more theoretical work is needed to probe the full potential of various broad line diagnostics, and to provide more theoretical guidance on the expected binary frequency. 

There are three recent observational studies that are closely related to the current work. \citet{Eracleous_etal_2012} and \citet{Decarli_etal_2013} both obtained second-epoch spectra for samples ($\sim 70$ and $\sim 30$ objects, respectively) of $z\lesssim 0.7$ SDSS quasars selected to have peculiar broad line characteristics, such as large offsets ($\gtrsim 1000\ {\rm km\,s^{-1}}$) between the broad \hbeta\ and narrow lines, and/or double-peaked or asymmetric line profiles. They used the two spectra (one from the SDSS) to measure broad \hbeta\ velocity shifts between the two epochs and reported several detections. The detection fraction in their samples is roughly consistent with ours if we restrict to the same restframe time separation (see Fig.\ \ref{fig:hbeta_hist}). But we caution that our statistics are poor (we only have 7 detections for the largest $\Delta t$ bin), and there are considerable differences in the sample and the quality of the spectroscopic data, in the approaches to measuring the velocity shift and in the associated error analysis among the three studies. The most significant difference is that our sample is typical of normal quasars, while theirs were selected to be more likely to host a BBH. Therefore they could not provide constraints on the general population as we do in \S\ref{sec:general_pop}. A more appropriate comparison will be provided in Paper II, where we also target quasars selected to be more likely BBHs. 

The third study, by \citet{Ju_etal_2013}, also used repeated spectroscopy from the SDSS, and focused on the general quasar population, as we do here. However, there are several key differences between the two studies: (1) their sample probes quasars at $0.4\lesssim z<2$ (although their reported detections are mostly at $z<0.7$) while our sample is at $z<0.8$ and has lower quasar luminosities; (2) they used the \MgII\ line instead of \hbeta. We did not choose the \MgII\ line as our primary line for the reasons discussed in \S\ref{sec:data}; (3) they used the several ``detections'' to constrain BBH models, while we believe it is more informative, and more appropriate, to use the full distribution in constraining the general BBH population; and (4) they did not utilize individual measurement errors for their sample, and used a typical ensemble value for the measurement errors instead. This could be problematic, as the several ``detections'' in their study could be due to large individual measurement errors, and their modeling is sensitive to this detected number. All these issues may explain the different conclusions (in particular, Ju et~al. find a much lower detection fraction) reached in the two studies, but one should also bear in mind the different assumptions that went into both analyses. We defer a more careful comparison to future work when we perform our own analysis on the \MgII\ line. 

\subsection{Future perspectives}\label{sec:future}

We have demonstrated the potential of using the broad line diagnostics, i.e., velocity shifts between epochs, to identify and characterize the BBH population in broad line quasars. However, to improve the power of this technique, there is both observational and theoretical need to understand the broad line emission in single and binary BHs. 

From an observational point of view, we need long-term spectroscopic monitoring of large samples of quasars to construct the power spectrum of the stochastic velocity shifts due to BLR variability in single BHs. Analogous to studying the photometric quasar continuum variability, we can construct a stochastic velocity shift ``structure function'', i.e., the random fluctuation of broad line shifts between epochs as a function of time separation. This must be done for a large sample to capture the diversity in the quasar population, and must be done over multi-year timescales and many epochs for the same objects, and with adequate S/N, to confirm the stochastic nature of the velocity shifts. A by-product of these programs would be potential discoveries of coherent velocity shifts, which would be the best candidates for BBHs. There are also general benefits from improved spectral quality, increased time baseline and more epochs for our statistical constraints. For instance, merely increasing the time baseline can shrink the measurement errors in acceleration. Parallel observational efforts are needed to better understand the BLR structure with reverberation mapping. 

From a theoretical perspective, we need BBH simulations with recipes for the BLR gas to predict the temporal evolution of the broad line emission and its correlation with the orbital motion of the BBH, as well as the effects of the (postulated) circumbinary disk on the broad lines (e.g., how much broad line emission comes from this circumbinary disk), with empirical constraints from observations, such as the BLR size, geometry and kinematics.

\section{Conclusions}\label{sec:con}

We have performed a systematic search for sub-parsec BBH candidates in optically selected, broad line quasars with multi-epoch (mostly two-epoch) SDSS spectroscopy. The spectra are separated by up to several years in the quasar restframe. Our sample is composed of normal quasars, which is different from the samples in other studies \citep[e.g.,][]{Eracleous_etal_2012,Decarli_etal_2013}. We have shown that targeting these normal quasars is not only as efficient as targeting offset or double-peaked broad line quasars in terms of finding BBH candidates, but also provides much more robust constraints on the BBH population in quasars. 

Our working hypothesis was that only one BH is active and that this BH dynamically dominates its own BLR, and that the companion BH is orbiting at a large enough distance such that the BLR is not tidally distorted or disrupted. Our sample is composed of $z<0.8$ quasars, with typical virial BH mass estimates $1.8\times 10^8\ M_\odot$ (for the active BH) and BLR sizes $\sim 0.01-0.1$ pc.  We searched for the expected acceleration (velocity shift) in the broad \hbeta\ line from the binary motion between the two epochs, using a cross-correlation technique. Our main findings and their implications are summarized below:

\begin{itemize}

\item[$\bullet$] The broad \hbeta\ line shape (FWHM, skewness, etc) can change over multi-year timescales, but dramatic changes in FWHM ($>20\%$) are rare.

\item[$\bullet$] Out of $688$ pairs of observations for which we have successfully measured the velocity shift between two epochs, we detected 28 systems with significant acceleration at the $\sim 2.5\sigma$ confidence level. Object-by-object inspections suggest that 7 of them are the best examples of sub-parsec BBH candidates, 4 of them are best explained by broad line variability in single BHs, and the rest are ambiguous. 

\item[$\bullet$] The full distribution of acceleration (mostly non-detections) for a subset from the 688 pairs with better-quality measurements and $\Delta t>0.4$ yr (restframe) was used to constrain the BBH population among general quasars. After accounting for measurement errors, excess variance is required to explain the observed distribution. If all the excess variance in velocity shifts is due to a BBH population, then (1) a large binary fraction is required; (2) the companion BH must be orbiting within a few times the BLR radius (details depending on the companion mass, see \S\ref{sec:general_pop}); (3) the companion BH must be more massive than the active one. On the other hand, if the excess variance is largely due to stochastic BLR variability in single BHs, then the requirement for a BBH population is substantially weakened, and even disfavored in the case of a very massive companion -- otherwise the predicted acceleration distribution will be broader than observed (after convolving with errors). 

\item[$\bullet$] Long-term spectroscopic monitoring of the detected BBH candidates with more epochs are required to confirm or rule out the BBH hypothesis. In addition, long-term spectroscopic monitoring of normal quasars will provide vital information on the stochastic velocity shift induced by BLR variability, and thus improve statistical constraints on the BBH population in broad line quasars with the methodology developed in this paper. 

\end{itemize}

\acknowledgements

We thank the anonymous referee and Mike Eracleous for useful comments on the manuscript. Support for the work of Y.S. and X.L. was provided by NASA
through Hubble Fellowship grant number HST-HF-51314.01 and HST-HF-51307.01,
respectively, awarded by the Space Telescope Science Institute,
which is operated by the Association of Universities for
Research in Astronomy, Inc., for NASA, under contract NAS
5-26555. This work was supported in part by NSF grant AST-0907890 and NASA grant NNA09DB30A (A.L.), and NASA grant NNX11AF29G (S.T.)

Funding for the SDSS and SDSS-II has been provided by the Alfred P. Sloan
Foundation, the Participating Institutions, the National Science Foundation,
the U.S. Department of Energy, the National Aeronautics and Space
Administration, the Japanese Monbukagakusho, the Max Planck Society, and the
Higher Education Funding Council for England. The SDSS Web Site is
http://www.sdss.org/.

\bibliographystyle{apj}

\end{document}